\documentclass[twocolumn,twocolappendix]{aastex631}

\let\tablenum\relax
\usepackage{siunitx}
\DeclareSIUnit\year{yr}
\DeclareSIUnit\bar{bar}
\DeclareSIUnit\molar{\mole\per\cubic\deci\metre}
\DeclareSIUnit\Molar{M}
\DeclareSIUnit\HTwo{\ce{H2}}
\DeclareSIUnit\HCN{\ce{HCN}}
\DeclareSIUnit\carbon{\ce{C}}
\DeclareSIUnit\molecules{molecules}
\DeclareSIUnit\photon{photon}
\usepackage[version=4]{mhchem}
\usepackage{booktabs}
\usepackage{multirow}
\usepackage{tabularx}
\usepackage{array}
\usepackage{soul}
\usepackage{placeins}
\usepackage{layouts}
\usepackage{float}
\usepackage{comment}

\accepted{March 19, 2025}

\acceptjournal{\textit{The Astrophysical Journal}}

\shorttitle{Options for Biomolecule Formation on an Oxidized Early Earth}
\shortauthors{Paschek et al.}

\begin{document}

\title{Deep Mantle-Atmosphere Coupling and Carbonaceous Bombardment: Options for Biomolecule Formation on an Oxidized Early Earth}

\correspondingauthor{Klaus Paschek}
\email{paschek@mpia.de}

\author[0000-0003-2603-4236]{Klaus Paschek}
\affiliation{Max Planck Institute for Astronomy, K{\"o}nigstuhl 17, 69117 Heidelberg, Germany}

\author[0000-0002-1493-300X]{Thomas K. Henning}
\affiliation{Max Planck Institute for Astronomy, K{\"o}nigstuhl 17, 69117 Heidelberg, Germany}

\author[0000-0002-0502-0428]{Karan Molaverdikhani}
\affiliation{Fakultät für Physik, Universitäts-Sternwarte, Ludwig-Maximilians-Universität München, Scheinerstr. 1, 81679 München, Germany}
\affiliation{Exzellenzcluster Origins, Boltzmannstr. 2, 85748 Garching, Germany}

\author[0000-0001-8325-8549]{Yoshinori Miyazaki}
\affiliation{Department of Earth and Planetary Sciences, Rutgers University, 610 Taylor Road, Piscataway, NJ 08854, USA}

\author[0000-0002-5449-4195]{Ben K. D. Pearce}
\affiliation{Department of Earth, Atmospheric, and Planetary Sciences, Purdue University, West Lafayette, IN 47907, USA}

\author[0000-0002-7605-2961]{Ralph E. Pudritz}
\affiliation{Origins Institute and Department of Physics and Astronomy, McMaster University, ABB 241, 1280 Main Street, Hamilton, ON L8S 4M1, Canada}
\affiliation{Max Planck Institute for Astronomy, K{\"o}nigstuhl 17, 69117 Heidelberg, Germany}

\author[0000-0002-3913-7114]{Dmitry A. Semenov}
\affiliation{Max Planck Institute for Astronomy, K{\"o}nigstuhl 17, 69117 Heidelberg, Germany}
\affiliation{Department of Chemistry, Ludwig Maximilian University of Munich, Butenandtstraße 5-13, House F, 81377 Munich, Germany}

\begin{abstract}\noindent
Understanding what environmental conditions prevailed on early Earth during the Hadean eon, and how this set the stage for the origins of life, remains a challenge. Geologic processes such as serpentinization and bombardment by chondritic material during the late veneer might have been very active, shaping an atmospheric composition reducing enough to allow efficient photochemical synthesis of \ce{HCN}, one of the key precursors of prebiotic molecules. \ce{HCN} can rain out and accumulate in warm little ponds (WLPs), forming prebiotic molecules such as nucleobases and the sugar ribose. These molecules could condense to nucleotides, the building blocks of RNA molecules, one of the ingredients of life. Here, we perform a systematic study of potential sources of reducing gases on Hadean Earth and calculate the concentrations of prebiotic molecules in WLPs based on a comprehensive geophysical and atmospheric model. We find that in a reduced \ce{H2}-dominated atmosphere, carbonaceous bombardment can produce enough \ce{HCN} to reach maximum WLP concentrations of $
\sim$\SIrange{1}{10}{\milli\Molar} for nucleobases and, in the absence of seepage, $\sim$\SIrange{10}{100}{\micro\Molar} for ribose. If the Hadean atmosphere was initially oxidized and \ce{CO2}-rich (\SI{90}{\percent}), we find serpentinization alone can reduce the atmosphere, resulting in WLP concentrations of an order of magnitude lower than the reducing carbonaceous bombardment case. In both cases, concentrations are sufficient for nucleotide synthesis, as shown in experimental studies. RNA could have appeared on Earth immediately after it became habitable (about \SI{100}{\mega\year} after the Moon-forming impact), or it could have (re)appeared later at any time up to the beginning of the Archean.
\end{abstract}

\keywords{Meteorites (1038) --- Carbonaceous chondrites (200) --- Earth atmosphere (437) --- Geological processes (2289) --- Pre-biotic astrochemistry  (2079) --- Computational astronomy (293) --- Chemical thermodynamics (2236) --- Complex organic molecules (2256) --- Interdisciplinary astronomy (804) --- Astrobiology (74)}

\section{Introduction}\label{sec:intro}

At the beginning of Earth's evolution, the surface of our nascent planet was a rather hostile environment, not hospitable to life. Volcanic activity and meteorite bombardment was likely high, and the hydrosphere still had to settle. The formation of the first primordial global ocean allowed liquid water, the basis for all life as we know it \citep{Westall2018}, to appear on the Earth's surface for the first time. However, this was likely disturbed several times by large impactors, causing the ocean to evaporate and the water to be lifted back into the atmosphere \citep{Chyba1990,Nisbet2001,Zahnle2006}.

The question of what the atmosphere above this ocean might have been like after it had finally settled at the end of the sterilizing giant impacts is an important one. Accretion during planet formation could have produced a primary \ce{H2}-rich atmosphere \citep{Oparin1938,Urey1951,Urey1952,Young2023}, which was eroded into space by the solar wind \citep[timescale of around \SI{100}{\mega\year},][]{Owen2017} and subsequently replaced by a \ce{CO2}-rich atmosphere of several hundred bars, which was outgassed by the magma ocean \citep{Zahnle2007,Miyazaki2022,Johansen2023,Johansen2024}. Rare earth element signatures in Hadean zircons at ${\sim\SI{4.35}{\giga\year}}$ indicate the presence of an already oxidized mantle \citep{Trail2011}. The redox state of the mantle was already close to the quartz-fayalite-magnetite mineral buffer, which describes the chemical state of reactions between minerals containing ferrous (\ce{Fe^{2+}}) and ferric (\ce{Fe^{3+}}) iron in the mantle. As on modern Earth, this state of the mantle results in the emission of mostly oxidized gases such as \ce{CO2} into the atmosphere during silicate volcanism.

The famous Urey-Miller experiments and many modern versions of them show that in a reducing atmosphere rich in \ce{H2}, \ce{CH4} is abundant and leads to the formation of \ce{HCN}, whereas a more neutral atmosphere dominated by \ce{CO2} is less favorable for the formation of organic molecules, including \ce{HCN} \citep{Haldane1929,Oparin1938,Urey1952,Miller1957a,Schlesinger1983,Stribling1987,Oro1990,Miyakawa2002,Cleaves2008,Benner2019a}. When dissolved in aqueous solution, this \ce{HCN} is able to react further to form many building blocks of life, such as amino acids, nucleobases, formaldehyde, sugars, and even nucleosides, the monomers of RNA \citep{Miller1953,Miller1955,Miller1957b,Miller1959,Oro1961a,Cleaves2008,Johnson2008,Powner2009,Bada2013,Sutherland2016,Becker2018,Benner2019a,Yadav2020}. RNA is of great interest for the origins of life, as its capabilities to store information and simultaneously self-catalyze its polymerization is one of the suggested starting points for chemical evolution and finally life in the so-called RNA world hypothesis \citep{Rich1962,Gilbert1986,Kruger1982,Guerrier-Takada1983,Guerrier-Takada1984,Zaug1986,Cech1986,Johnston2001,Vaidya2012,Attwater2018,Cojocaru2021,Kristoffersen2022}. It could also have been formed while being encapsulated in a primitive cell membrane and interacted with peptides in a more inclusive RNA-peptide world hypothesis \citep{DiGiulio1997,Muller2022}.

This raises the seeming contradiction of how the ingredients for life could have been formed on Hadean Earth, which likely had an oxidized \ce{CO2}-rich atmosphere in the early Hadean, while reducing conditions are required for prebiotic synthesis. To solve this problem, additional sources of reducing gases have been suggested.

\begin{figure*}[t]
    \centering
    \includegraphics[width=\textwidth]{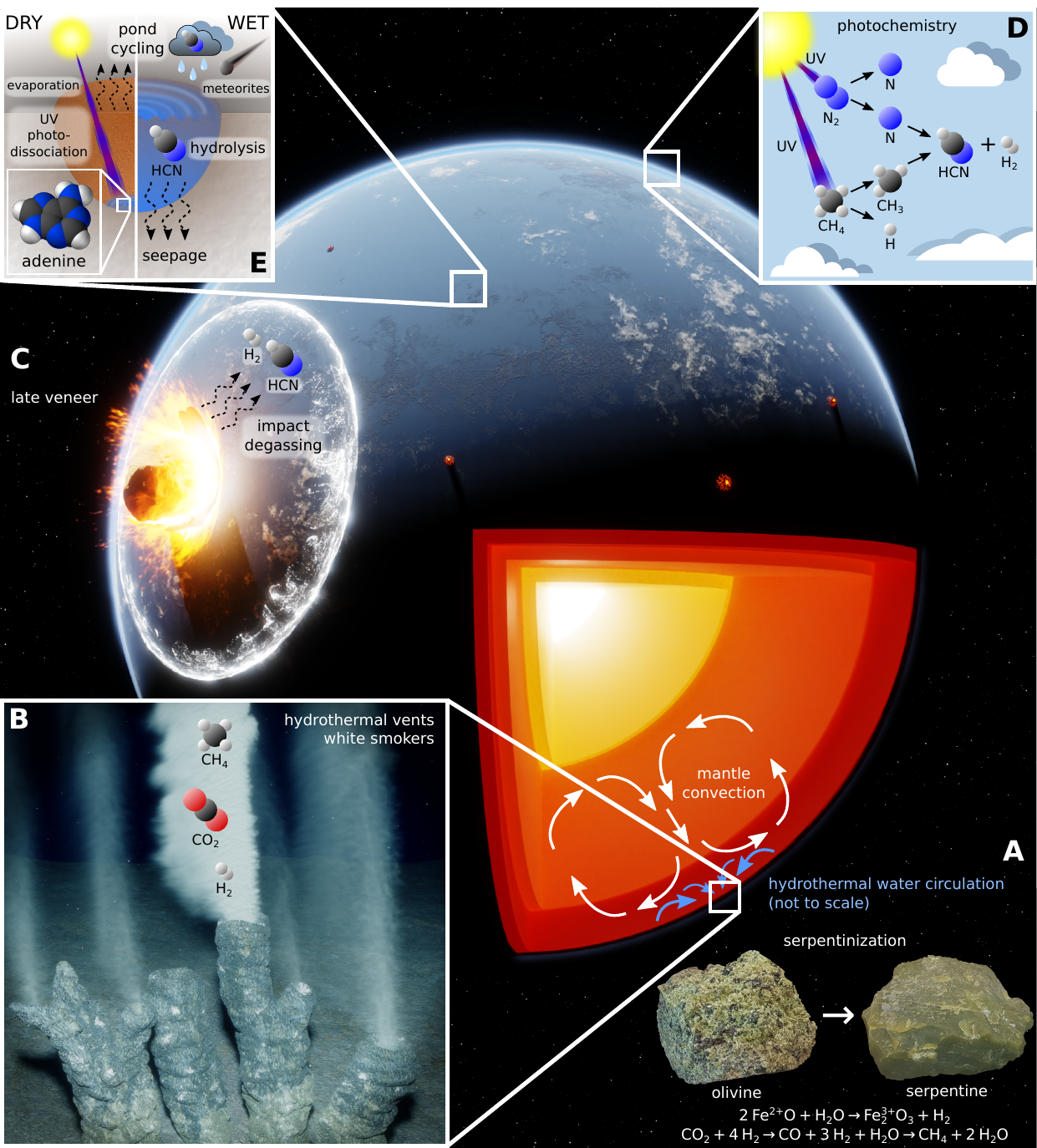}
    \caption{The ``\textbf{\textit{\ce{HCN} machine}}'': Geological, atmospheric, meteoritic, and chemical processes synthesizing the building blocks of life on the Hadean Earth (artist's impression, own creation, © Klaus Paschek). Panel \textbf{A} (lower right): Serpentinization and mantle processes lead to the efficient synthesis of \ce{H2} and \ce{CH4}, including reactions with water and \ce{CO2}. Panel \textbf{B} (lower left): Emission of \ce{H2}, \ce{CH4}, and \ce{CO2} from hydrothermal vents at volcanically active mid-ocean ridges. Panel \textbf{C} (center left): Degassing of \ce{H2} and \ce{HCN} by giant impacts. Panel \textbf{D} (top right): Synthesis of \ce{HCN} from \ce{CH4} and \ce{N2} by UV photochemistry in the atmosphere. Panel \textbf{E} (top left): Atmospheric \ce{HCN} rains out to the Earth's surface and enters lakes, ponds, and the ocean. In ponds, wet-dry cycling and aqueous chemistry convert \ce{HCN} into nucleobases, sugars, and ultimately RNA (oligo)nucleotides, key ingredients of life.}
    \label{fig:hadean_earth_scheme}
\end{figure*}

In Figure~\ref{fig:hadean_earth_scheme}, we provide an overview over these sources of reducing gases, and subsequent chemical processes operating together to form precursors of prebiotic molecules on the Hadean Earth. This begins with internal geological processes in the deep mantle, including the reaction between minerals derived from mantle magma and surface water (A), creating the emission of reducing and oxidizing gases from undersea volcanoes (B). These processes are joined by meteorites, which degas reducing species upon impact (C). It continues with photochemical reactions driven by UV irradiation in the proto-atmosphere (D). And it ends with the synthesis of biomolecules by wet-dry cycling in the first reservoirs of water on the emerging islands and land masses (E), which are fed by rain-out of the compounds formed in the atmosphere.

This Hadean Earth was likely dominated by a global ocean. Here, the first volcanic islands rise to the surface \citep{Bada2018,Korenaga2021,Chowdhury2023}. These are pushed together by plate tectonics to form the first basaltic land masses. These black land masses and island arcs populate the surface, visible on the day side of the Hadean Earth in the upper half of Figure~\ref{fig:hadean_earth_scheme}.

Because of this very active volcanism on this Hadean Earth, one source of reducing gases might haven been extensive serpentinization \citep{Russell2010,Holm2015,Preiner2018,Miyazaki2022} in the Earth's crust. After the magma ocean froze out, plate tectonics emerged, but it is still debated how and when it first appeared on Earth, and if it was present in the Hadean \citep{Chowdhury2023} or not \citep{Tarduno2023}. Before the onset of plate tectonics, the Earth may have been undergoing stagnant lid convection \citep{Debaille2013,Tosi2017,Tarduno2023}. In any case, material convecting in the mantle rises and melts, causing magma to rise up through cracks in the crust and participate in further crust formation. This created ridges where the magma erupts at the bottom of the oceans.

In Figure~\ref{fig:hadean_earth_scheme}(A), the geological cross-section of the Earth in the lower right shows that here the magma comes into contact with the ocean water, which enters the crust through fissures in a process called hydrothermal circulation. The Hadean crust was likely very different from today's crust, being thinner than today and undergoing rapid plate tectonic motion \citep{Sleep2001,Zahnle2007,Sleep2014,Miyazaki2019,Miyazaki2022}, influenced by early mantle differentiation. This early mantle was heterogeneous, containing iron-rich blobs that continuously supplied ferrous material to the surface \citep{Miyazaki2019,Miyazaki2022}. Such ferrous iron forms the iron-rich mineral group olivine, which comes into contact with ocean water through the hydrothermal circulation.

This olivine reacts with the water in the serpentinization reaction, converting the ferrous iron to more oxidized ferric iron in the mineral group serpentine, and reducing the water to \ce{H2} \citep{Klein2013}. One of the most important factors controlling the amount of \ce{H2} released is the hydrothermal circulation depth (HCD). With increasing depth, water penetrates further into the crust, leading to more extensive iron oxidation and consequently greater hydrogen production. This geochemical process is illustrated by the example rocks peridotite (containing olivine minerals) and serpentinite (containing serpentine minerals) and the corresponding generalized chemical reaction equations in the lower right of Figure~\ref{fig:hadean_earth_scheme}(A).

Serpentinization is the most stable and continuous source of reducing gases to be considered here, as mantle convection provides a continuous supply of ferrous iron to the reaction.

\ce{H2} further reacts with \ce{CO2} dissolved in the magma to form \ce{CH4} \citep{Kasting2005,McCollom2007,Thompson2022}. These gases erupt in white smokers, a type of hydrothermal vent, and rise to the ocean surface and enter the atmosphere, as shown in the zoomed-in inset (B) in the lower left of Figure~\ref{fig:hadean_earth_scheme}. Underwater volcanoes are driven purely by chemical reactions, not by direct eruption of silicate magma, creating a kind of ``chemical volcanism''.

Figure~\ref{fig:hadean_earth_scheme}(C), in the center left, shows one of the many meteorites that struck the early Earth during the late veneer \citep{Urey1952}. These impacts have been proposed as another source of reducing gases. The late veneer refers to the last layer of material, composed mainly of chondritic meteorites, that was late accreted to the Earth's mantle after core formation \citep{Morbidelli2015,Li2022}. Meteorites carry additional elements, such as metals and siderophiles (iron-loving elements), that could contribute to the release of reducing gases. This might explain why the Earth has an excess of highly siderophile elements (HSEs) in its crust and mantle, revealing the presence of the late veneer.

Enstatite chondrites, a type of iron-rich meteorites, and the siderophile fraction of ordinary and carbonaceous chondrites are expected to produce large amounts of \ce{H2} during impact, as the reduced iron in the meteorite reacts with water, similar to the serpentinization process discussed above \citep{Kasting1990,Hashimoto2007,Schaefer2007,Schaefer2010,Schaefer2017,Kuwahara2015,Pearce2022,Zahnle2020,Wogan2023}. 

There is, however, another important mechanism leading to direct \ce{HCN} production by meteoritic impact. Carbonaceous chondrites could have generated \ce{HCN} during impact, as the carbon reacts with an ambient \ce{N2} and water atmosphere in a reaction induced by the energy and heat released in the impact shock \citep{Kurosawa2013}. Even without the metal component in some types of carbonaceous chondrites, \ce{HCN} might be formed due to the vaporization and reaction of carbon alone.

The exact composition of the late veneer material, in particular the ratio between enstatite and carbonaceous impactors, has long been debated. Recent evidence from isotopic signatures of the primitive Earth's mantle and chondritic meteorite populations points to a mixed late veneer, although a pure enstatite or a pure carbonaceous bombardment remain valid possibilities \citep{Fischer-Godde2017,Varas-Reus2019,Budde2019,Hopp2020,Fischer-Godde2020,Bermingham2025}.

All of these source terms inject the gases \ce{H2}, \ce{CH4}, \ce{HCN}, and \ce{CO2} into the atmosphere. There they can react further to form prebiotic molecular precursors, such as \ce{HCN} and \ce{H2CO}.

The zoomed-in inset (D) in the upper right of Figure~\ref{fig:hadean_earth_scheme} shows that the molecules at the top of the atmosphere are exposed to UV radiation, which allows the molecules, e.g. \ce{N2} and \ce{CH4}, to split into radicals that can recombine to form new stable molecules. In this process, the reaction likely passes through \ce{H2CN} as an intermediate \citep[not shown]{Pearce2019}. In this photochemical reaction network \citep{Pearce2022}, there is a constant competition between reactions that produce oxidized gases such as \ce{CO2}, \ce{O2}, \ce{H2O}, etc.~and reduced gases such as \ce{H2}, \ce{CH4}, \ce{HCN}, etc. \ce{H2CO} can also be formed in weakly reducing atmospheres \citep{Pinto1980,Benner2019b}.

The outcome of the photochemical reaction network is primarily influenced by three key parameters. These parameters include: i) the general atmospheric composition at the start of the simulation, which may be either reducing or oxidizing, ii) the HCD of the primordial ocean penetrating into the Earth's crust, and iii) the composition of the late veneer material, which might consist of either iron-rich enstatite or carbon-rich carbonaceous chondrites, or both. We explore these various possibilities in a parameter study with several model scenarios.

The species that formed were rained out to the surface. This allowed these prebiotic precursors to accumulate on the Earth's surface, e.g., in the first water reservoirs, lakes and ponds on the first continental crust, which could have formed very early in the Earth's history, possibly as early as \SI{4.2}{\giga\year} ago \citep{McCulloch1993,Pearce2017,Chowdhury2023}. Charles Darwin suggested so-called warm little ponds (WLPs) as a possible location for the origin of life. The zoomed-in inset (E) in the upper left of Figure~\ref{fig:hadean_earth_scheme} gives an illustration of the various active processes that drive the chemical dynamics in these WLPs. \ce{HCN} in aqueous solution allows the synthesis of biomolecules such as amino acids and nucleobases as in the Urey-Miller experiments discussed above \citep{Johnson2008,Bada2013}. \ce{H2CO} can form sugars such as ribose in the formose reaction \citep{Breslow1959,Butlerow1861,Cleaves2015}. These WLPs can undergo wet-dry cycling, which allows for the synthesis of nucleotides \citep{Powner2009,Becker2016,Sutherland2016,Becker2018,Yadav2020} and their polymerization into RNA \citep{Benner2019a,DaSilva2015}. This involves condensation reactions that split off water, which is thermochemically inhibited in aqueous solution, but the dry phase of the WLP allows this process \citep{Ponnamperuma1963,Fuller1972,Powner2009,Saladino2017,Nam2018}. Prebiotic molecules can be further concentrated and nucleotides can be polymerized to RNA and DNA oligomers in cracks within rocks under the influence of geothermal heat flows \citep{Dirscherl2023,Matreux2024} caused by thermophoresis and convection processes corresponding to wet-dry cycling.

This wet-dry cycling synthesis includes source terms such as \ce{HCN} and \ce{H2CO} rain-out from the atmosphere and exogenous supply by carbonaceous chondrites, which contain a plethora of biomolecules \citep[see Figure~\ref{fig:hadean_earth_scheme}(E),][]{vanderVelden1977, Stoks1979,Stoks1981,Shimoyama1990,Callahan2011,Gilmour2003,Pizzarello2006,Smith2014,Furukawa2019,Oba2022,Paschek2022,Oba2023,Paschek2023,Paschek2024} that they might have released into the WLP \citep{Pearce2017}. Sinks are destruction of the biomolecules by UV photodissociation, hydrolysis, and seepage through pores at the bottom of the pond \citep{Pearce2017,Pearce2022}.

Here we build on the previous models by \citet{Pearce2022} and take a closer look at different sources of reducing gases appropriate in the context of the Hadean Earth. We take an agnostic approach and perform a systematic parameter study including different combinations of the contributing mechanisms outlined above to fully explore what might be feasible for prebiotic synthesis on the Hadean Earth.

The goal of this work is to bridge different scientific fields, ranging from geosciences to chemistry to astrophysics. Our idea and approach is to model several plausible scenarios and compare their results to get a more complete picture of what the Hadean Earth might have been like. Our goal is not to have a specific preference for any of the various hypotheses discussed in the scientific community, but to evaluate several possible scenarios, contributing mechanisms, and their interplay, given the scarce evidence available from the Hadean.

Section~\ref{sec:models} provides a short description of the models used in this paper. Details can be found in the Appendices. In Section~\ref{sec:methods}, we give a summary of the computational methods and the implementation of the models. The results across the whole parameter space, including photochemistry and resulting atmopheric compositions, rain-out, WLP cycling, and resulting biomolecule concentrations, are presented in Section~\ref{sec:results}. Discussion and conclusions follow in Sections~\ref{sec:discussion}~and~\ref{sec:conclusions}. In Appendix~\ref{sec:mantle}, we discuss the processes in the Hadean Earth's mantle and crust, including serpentinization, in more detail. Next, in Appendix~\ref{sec:late_veneer}, we outline the current evidence for the composition of the late veneer, and how an enstatite and/or carbonaceous bombardment might have influenced the Hadean atmosphere. In Appendix~\ref{sec:time}, we place the available evidence into the timeline of the Hadean, resulting in the two main environmental scenarios into which we will place our models. Then, in Appendix~\ref{sec:sources}, we discuss which surface gas fluxes emitted by the mantle and by impact degassing are feasible for the Hadean. These are summarized into two sets of fluxes representing the scenarios modeled in the present study in Appendix~\ref{sec:scenarios}. Supplementary results are presented in the Appendices~\ref{sec:supp_results}~and~\ref{sec:supp_results_high_bomb_rate}.

\section{Models}\label{sec:models}

In this Section, we provide an outline of our models. First, we fit the available evidence into the timeline of the Hadean eon to come up with appropriate environmental scenarios in which to place our simulations. Here, we give a brief summary of the available evidence and the assumptions made to arrive at these scenarios, and a more in-depth discussion of this can be found in Appendix~\ref{sec:time}.

The formation and evolution of Earth's early atmosphere during the Hadean eon are influenced by mantle composition, tectonic activity, and volcanic outgassing. Initially, the Earth's mantle might have released large amounts of \ce{CO2}, resulting in a dense, \ce{CO2}-rich atmosphere \citep{Zahnle2007,Miyazaki2022,Johansen2023,Johansen2024}. Tectonic processes may have rapidly sequestered much of this \ce{CO2} in the mantle, potentially causing a shift to a hydrogen-dominated atmosphere about \SI{4.4}{\giga\year} ago. Whether or not plate tectonics was active as the necessary process in the Hadean is still debated \citep{Chowdhury2023,Tarduno2023}. The presence of \ce{H2} and other reducing gases likely resulted from primordial gas accretion, serpentinization of mantle materials, and outgassing from hydrothermal vents that influenced early prebiotic chemistry.

In contrast, extrapolation of the Archean rock record toward the end of the Hadean (about \SI{4.0}{\giga\year} ago) suggests a shift toward more oxidizing conditions, as indicated by redox-sensitive elements \citep{Holland1984,Aulbach2016,Catling2017,Wogan2020a} and oxidation states of the mineral zircon \citep{Trail2011} at \SI{4.35}{\giga\year}. However, zircons crystallize at temperatures above \SI{600}{\celsius} \citep{Harrison2007}, reflecting deep mantle conditions, while serpentinization in the near-surface crust operates independently of this redox state. Hydrothermal vent outgassing from serpentinization could thus produce a reducing atmosphere, potentially out of equilibrium with the more oxidized state of the deep mantle.

Our simulations examine two primary scenarios, each capturing conditions at two critical Hadean epochs: the mid-Hadean (MH) at \SI{4.4}{\giga\year} ago and the end-Hadean (EH) at \SI{4.0}{\giga\year} ago. The MH scenario considers an initially reducing atmosphere (\SI{90}{\percent} \ce{H2}), while the EH scenario begins with an oxidizing atmosphere (\SI{90}{\percent} \ce{CO2}), reflecting the potential changes in redox state and atmospheric composition. These models test whether geological processes, such as serpentinization and impact degassing, could convert the atmosphere to a reduced state favorable for the synthesis of prebiotic molecules. The detailed parameters for these epochs are provided in Table~\ref{tab:epochs}. Here, $t$ is the time of the epoch in units of \si{\giga\year} from today, $p_\mathrm{surface}$ is the surface pressure in \si{\bar}, atmospheric gases are given in their initial molar mixing ratios (the molar mixing ratio is defined as the ratio between the amount of the respective atmospheric gas and the total amount of all gases, all in units of moles) at the beginning of the simulations in either \si{\percent} or parts per million (ppm), and $T$ is the surface temperature in \si{\celsius}. More discussions about these two epochs and how these parameter sets were calculated are presented in Appendix~\ref{sec:time}.

\begin{deluxetable}{llDD}[t]
   \tablecaption{Initial atmosphere compositions and parameters of the two considered epochs \citep[see also][]{Pearce2022}. Details can be found in Appendix~\ref{sec:time}.} 
   \label{tab:epochs}
    \tablehead{
        \twocolhead{} & \twocolhead{Mid-Hadean} & \twocolhead{End-Hadean} \\
        \twocolhead{Parameter} & \twocolhead{(MH)} & \twocolhead{(EH)}
    }
    \decimals
    \startdata
    $t$ & [\unit{\giga\year}] & 4.4 & 4.0 \\
    $p_{\mathrm{surface}}$ & [\unit{\bar}] & 1.5 & 2 \\
    \ce{H2} & [\%] & 90 & 0 \\
    \ce{CO2} & [\%] & 0 & 90 \\
    \ce{N2} & [\%] & 10 & 10 \\
    \ce{CH4} & [ppm] & 2 & 10 \\
    \ce{H2O} & [\%] (surface layer) & 1 & 1 \\
    $T$ & [\unit{\celsius}] & 78 & 51 \\
    \enddata
\end{deluxetable}

\vspace{-13pt}
As a next step, we explore a range of cases to simulate the atmospheric evolution in the MH and EH, distinguishing between the geological and impact-driven sources that contributed to the emission of reducing gases in the Hadean.

We consider geological sources alone and refer to them as the ``geology'' case G, where \ce{H2}, \ce{CO2}, and \ce{CH4} are released by volcanic activity and serpentinization driven by hydrothermal circulation. Recent models by \citet{Miyazaki2022} propose a heterogeneous Hadean mantle resulting in iron-rich upwelling and a thin crust. This fosters rapid plate tectonics, very active hydrothermal circulation from the surface ocean into the crust, extensive serpentinization \citep{Russell2010,Klein2013,Preiner2018}, and Fischer-Tropsch reactions \citep{Kasting2005,McCollom2007,Holm2015,Thompson2022} generating volcanic \ce{H2} and \ce{CH4} emissions. Detailed explanations of serpentinization and related mechanisms in the Hadean Earth's mantle and crust are provided in Appendix~\ref{sec:mantle}.

To assess how the HCD affects the release of these gases and influences atmospheric chemistry, we vary the HCD between \SI{0.5}{\kilo\meter}, \SI{1.0}{\kilo\meter}, and \SI{2.0}{\kilo\meter}, with cases named accordingly as G0.5, G1, and G2. The corresponding calculations, extending the models by \citet{Miyazaki2022} and determining the emission of these gases, and details of this modification of the HCD are presented in Appendix~\ref{sec:mantle_hcd}.

In scenarios focused on exogenous impacts, we consider both pure enstatite (referred to as case E) and pure carbonaceous (referred to as case C) bombardments.  Details on the late veneer and possible bombardment scenarios in the Hadean can be found in Appendix~\ref{sec:late_veneer}. In the main results in Section~\ref{sec:results}, we consider intermediate bombardment rates based on the lunar cratering record \citep{Pearce2017,Chyba1990}, and in the supplementary results in Appendix~\ref{sec:supp_results_high_bomb_rate}, we explore what would happen if there was a maxed-out and short-lived bombardment, and how this would affect the Hadean atmosphere and prebiotic synthesis in WLPs in both the MH and EH epochs.

An enstatite bombardment provides \ce{H2} by impact degassing, as the iron-rich impact ejecta react with the impact-vaporized ocean \citep{Sekine2003,Genda2017a,Genda2017b,Benner2019a,Zahnle2020,Citron2022,Itcovitz2022,Wogan2023}, which might counteract the oxidizing effects of volcanic \ce{CO2} outgassing. Meanwhile, a carbonaceous bombardment additionally generates \ce{HCN} emissions during impacts, as the impact shock allows the carbon-rich impact material to react with the surrounding \ce{N2}-\ce{H2O} atmosphere \citep{Kurosawa2013}. Appendix~\ref{sec:bombardment_model} contains detailed information on the considered bombardment rates in the Hadean and gas emission processes during impact events.

We also examine combinations of geological and impact sources by coupling the geological fluxes (using an HCD of \SI{2}{\kilo\meter} as a standard depth) with enstatite only, carbonaceous only, or mixed (\SI{50}{\percent} enstatite, \SI{50}{\percent} carbonaceous) bombardments, referred to as cases G2E, G2C, and G2EC, respectively. This combined approach allows us to explore the interplay of both endogenous and exogenous contributions. Details of all considered cases are available in Appendix~\ref{sec:scenarios}, and the corresponding surface gas fluxes resulting from the respective cases in both MH and EH epochs are given in Tables~\ref{tab:cases_red}~and~\ref{tab:cases_ox}.

\section{Computational Methods: Atmosphere Model and Warm Little Pond Cycling}\label{sec:methods}

To study the effect of the source fluxes of gases in the Hadean atmosphere, we use the 1D disequilibrium chemical kinetics model previously developed by \citet{Pearce2022}. This model combines the atmospheric chemistry code ChemKM \citep{Molaverdikhani2019,Molaverdikhani2020} with the chemical network CRAHCN-O \citep{Pearce2019,Pearce2020,Pearce2020a}, which comprises 259 one-, two-, and three-body reactions. It also includes the production of molecules by lightning in the lowest atmospheric layer of the model and the escape of hydrogen to space \citep{Zahnle2019}. The $P$-$T$ profiles of the atmosphere were calculated using the radiative transfer code petitRADTRANS \citep{Molliere2019}. It is important to note that the current model does not include the day/night cycle and its influence on atmospheric chemistry, as it is expected that some of the \ce{HCN} would be removed from the atmosphere by rain-out overnight. For a comprehensive breakdown of the atmospheric model used, see Appendix~A in \citet{Pearce2022}.

The atmosphere model calculates rain-out rates for the key prebiotic precursors \ce{HCN} and \ce{H2CO}, removing these products of the photochemical network from the atmosphere. This allows these prebiotic precursors to accumulate on the Earth's surface in the first reservoirs of water on land, e.g., WLPs. Through seasonal wet-dry cycling, the prebiotic precursors \ce{HCN} and \ce{H2CO} can form prebiotic molecules such as RNA building blocks and their precursors such as nucleobases, the sugar ribose, and 2-aminooxazole \citep{Butlerow1861,Breslow1959,Oro1961,LaRowe2008,Powner2009,Ferus2019,Yi2020}.

To explore this further, we follow the study by \citet{Pearce2022} and couple the atmospheric rain-out rates with the WLP cycling model developed by \citet{Pearce2017}. This WLP model combines experimentally determined yields of prebiotic molecules from \ce{HCN} and \ce{H2CO} with multiple sinks due to photodestruction by UV light, seepage through pores at the base of the WLPs, and hydrolysis. \ce{H2CO} can enter the ponds either directly by rain-out from the atmosphere or by aqueous synthesis from deposited \ce{HCN}. In the present study, we have added experimental yields for ribose from \ce{H2CO}. Using the hydroxide \ce{Ca(OH)2} as a catalyst in the formose reaction, the way to make sugars from aqueous mixtures of \ce{H2CO} and trace-amounts of glycolaldehyde \citep{Butlerow1861,Breslow1959}, the yield of ribose from \ce{H2CO} reached up to $1.22\times 10^{-3}$ \citep[K.~Kohler, O.~Trapp private communication;][]{Teichert2019,Paschek2022}. This allows estimates of the possible concentrations of these prebiotic molecules in WLPs on the Hadean Earth.

\section{Results}\label{sec:results}

\subsection{Atmosphere Compositions}

\subsubsection{Serpentinization and Volcanism}

\begin{figure*}[p]
    \centering
    \includegraphics[width=0.9\textwidth]{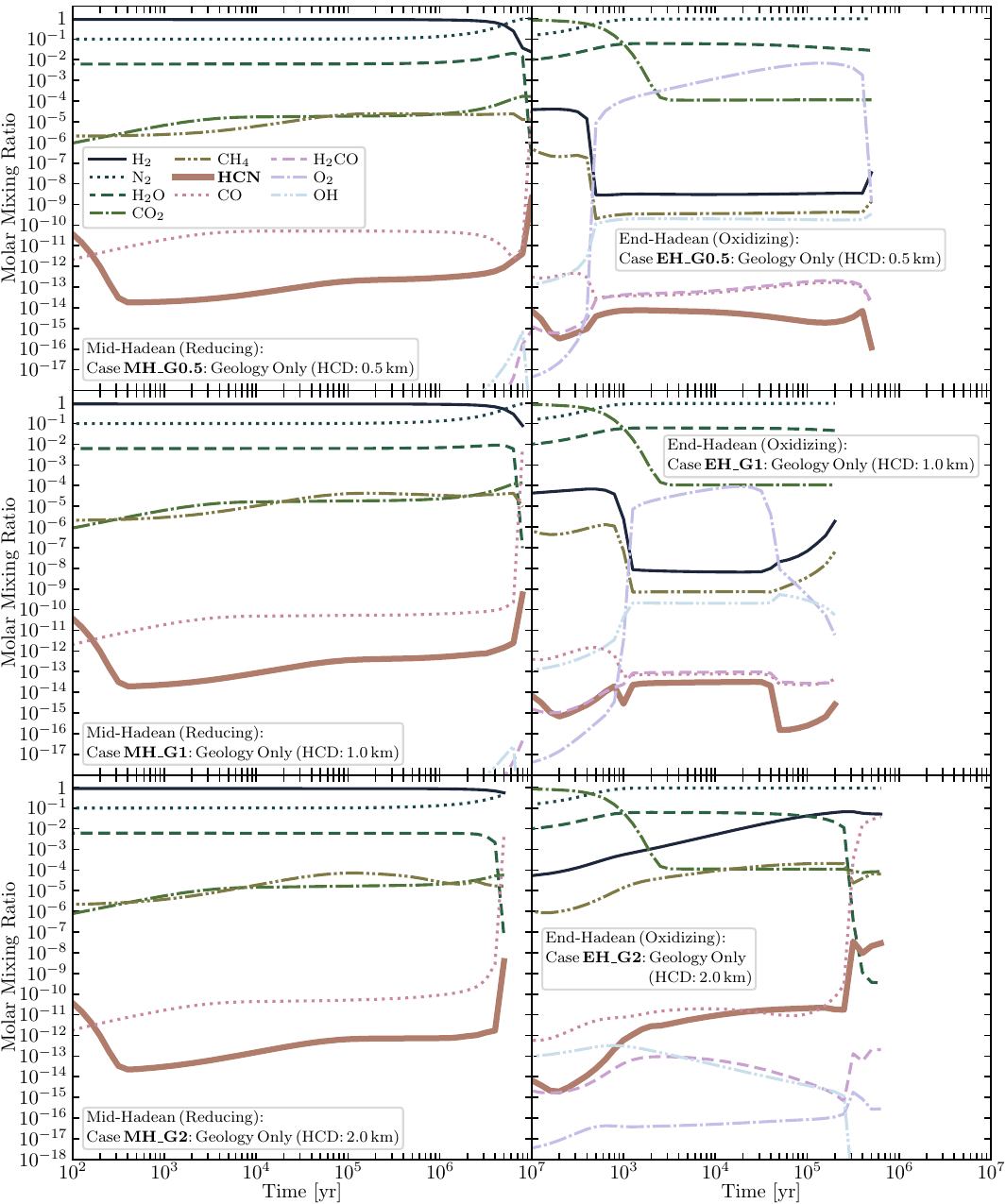}
    \caption{Effect of varying the hydrothermal circulation depth (HCD) during serpentinization on the atmospheric evolution. Shown is the simulated atmospheric composition of key species in the layer closest to the surface as a function of time. Only geological surface fluxes from extended models by \citet{Miyazaki2022} are considered as sources for \ce{H2}, \ce{CO2}, and \ce{CH4} (see cases MH/EH\_G0.5, MH/EH\_G1, and MH/EH\_G2 in Tables~\ref{tab:cases_red}~and~\ref{tab:cases_ox}). The two epochs of the mid-Hadean (MH) at \SI{4.4}{\giga\year} with reducing initial conditions and the end-Hadean (EH) at \SI{4.0}{\giga\year} with oxidizing initial conditions are compared (left vs.~right column; Panels \textbf{MH\_G0.5}, \textbf{MH\_G1}, \textbf{MH\_G2} vs.~\textbf{EH\_G0.5}, \textbf{EH\_G1}, \textbf{EH\_G2}). For each epoch, different HCDs of the primordial ocean penetrating the crustal rock are considered (top row (\textbf{MH\_G0.5}, \textbf{EH\_G0.5}): \SI{0.5}{\kilo\meter}, middle row (\textbf{MH\_G1}, \textbf{EH\_G1}): \SI{1.0}{\kilo\meter}, bottom row (\textbf{MH\_G2}, \textbf{EH\_G2}): \SI{2.0}{\kilo\meter}). The initial conditions for the reducing and oxidizing models are summarized in Table~\ref{tab:epochs}, closely following the established atmosphere models developed by \citet{Pearce2022}.}\label{fig:photochem_geo}
\end{figure*}

Figure~\ref{fig:photochem_geo} shows the effect of geological gas fluxes on the evolving atmosphere of the Hadean Earth. The surface fluxes for \ce{H2}, \ce{CO2}, and \ce{CH4} used in our atmosphere model come from the global upwelling of mantle material as simulated in newly calculated models that build on and extend the mantle model established by \citet{Miyazaki2022}. See Appendix~\ref{sec:mantle_hcd} and Figure~\ref{fig:H2_mantle_flux} for details. These mantle upwelling motions bring ferrous iron-rich magma and dissolved \ce{CO2} close to the surface. In contact with the surface hydrosphere, a combination of serpentinization, reverse gas-shift, and Fischer-Tropsch reactions (see Equations~\ref{eq:serp_simple}~and~\ref{eq:FT_eff}) results in the surface fluxes given in Tables~\ref{tab:cases_red}~and~\ref{tab:cases_ox}. Cases MH/EH\_G0.5, MH/EH\_G1, and MH/EH\_G2 represent these fluxes for HCDs of \SI{0.5}{\kilo\meter}, \SI{1.0}{\kilo\meter}, and \SI{2.0}{\kilo\meter}, respectively. See Appendix~\ref{sec:scenarios} for details.

Table~\ref{tab:results_atm} lists the maximum reached atmospheric molar mixing rations for key prebiotic precursors of interest achieved across all modeled scenarios in the present work. To determine these maximum values, only the time evolution of these molar mixing ratios after \SI{1000}{\year} is considered, due to the fact that many chemical species in the atmosphere initially show large fluctuations over many orders of magnitude, and by \SI{1000}{\year} at the latest begin to stabilize and reach their first plateau, resembling something close to steady state behavior (cf.~Figure~\ref{fig:photochem_geo} and following). We focus on these stabilized abundances because long-term atmospheric concentrations are most relevant for further implications of prebiotic synthesis in WLPs on geological time scales. We consider this a reasonable measure to ensure that Table~\ref{tab:results_atm} reflects characteristic values representative of geologically relevant time spans in the Hadean of millions of years and longer.

Maximum \ce{H2} levels are a good proxy for how much the atmosphere is reduced, especially for the initially oxidized models at \SI{4.0}{\giga\year}. \ce{CH4} is a major intermediate in the synthesis of \ce{HCN} in the computed photochemical reactions \citep{Pearce2022}. \ce{HCN} is our key prebiotic precursor molecule. Moreover, \ce{H2CO} is of interest as the key reactant in the formation of sugars.

\begingroup
\setlength{\tabcolsep}{3pt}
\begin{deluxetable*}{lDDcDDcDDcDD}[t]
    \tablecaption{Maximum resulting atmosphere concentrations of prebiotic precursors.}
    \label{tab:results_atm}
    \tabletypesize{\small}
        \tablehead{
            \colhead{} & \multicolumn{19}{c}{Max.~Molar Mixing Ratio}  \\
            \cmidrule{2-20}
            \colhead{} & \multicolumn{4}{c}{\ce{H2}} & \colhead{} & \multicolumn{4}{c}{\ce{CH4}} & \colhead{} & \multicolumn{4}{c}{\ce{HCN}} & \colhead{} & \multicolumn{4}{c}{\ce{H2CO}} \\
            \cmidrule{2-5}\cmidrule{7-10}\cmidrule{12-15}\cmidrule{17-20}
            \colhead{Case} & \twocolhead{MH (red.)} & \twocolhead{EH (ox.)} & \colhead{} & \twocolhead{MH (red.)} & \twocolhead{EH (ox.)} & \colhead{} & \twocolhead{MH (red.)} & \twocolhead{EH (ox.)} & \colhead{} & \twocolhead{MH (red.)} & \twocolhead{EH (ox.)} 
        }
        \decimals
        \startdata
        G0.5 & 8.94$\times 10^{-1}$ & 3.44$\times 10^{-8}$ && 2.49$\times 10^{-5}$ & 1.41$\times 10^{-9}$ && 2.38$\times 10^{-9}$ & 7.19$\times 10^{-15}$ && 1.13$\times 10^{-14}$ & 2.09$\times 10^{-13}$ \\
        G1 & 8.94$\times 10^{-1}$ & 1.86$\times 10^{-6}$ && 4.29$\times 10^{-5}$ & 6.25$\times 10^{-8}$ && 5.51$\times 10^{-10}$ & 3.19$\times 10^{-14}$ && 4.50$\times 10^{-17}$ & 9.61$\times 10^{-14}$ \\
        G2 & 8.94$\times 10^{-1}$ & 6.69$\times 10^{-2}$ && 7.22$\times 10^{-5}$ & 2.11$\times 10^{-4}$ && 3.96$\times 10^{-9}$ & 3.53$\times 10^{-8}$ && 3.10$\times 10^{-19}$ & 2.08$\times 10^{-13}$ \\
        E & 8.94$\times 10^{-1}$ & 3.06$\times 10^{-8}$ && 1.76$\times 10^{-6}$ & 5.36$\times 10^{-24}$ && 1.45$\times 10^{-14}$ & 6.97$\times 10^{-19}$ && 2.55$\times 10^{-23}$ & 8.89$\times 10^{-26}$ \\
        C & 8.94$\times 10^{-1}$ & 1.96$\times 10^{-7}$ && 1.76$\times 10^{-6}$ & 3.04$\times 10^{-24}$ && 1.23$\times 10^{-7}$ & 7.96$\times 10^{-19}$ && 2.55$\times 10^{-23}$ & 5.48$\times 10^{-26}$ \\
        G2E & 8.94$\times 10^{-1}$ & 8.03$\times 10^{-2}$ && 7.23$\times 10^{-5}$ & 2.16$\times 10^{-4}$ && 1.06$\times 10^{-8}$ & 3.23$\times 10^{-8}$ && 3.46$\times 10^{-17}$ & 1.66$\times 10^{-13}$ \\
        G2C & 8.94$\times 10^{-1}$ & 1.67$\times 10^{-2}$ && 7.22$\times 10^{-5}$ & 1.57$\times 10^{-4}$ && 7.54$\times 10^{-8}$ & 2.01$\times 10^{-10}$ && 3.10$\times 10^{-19}$ & 5.89$\times 10^{-12}$ \\
        G2EC & 8.94$\times 10^{-1}$ & 5.78$\times 10^{-2}$ && 7.43$\times 10^{-5}$ & 2.13$\times 10^{-4}$ && 4.35$\times 10^{-8}$ & 7.07$\times 10^{-9}$ && 2.49$\times 10^{-16}$ & 4.25$\times 10^{-13}$ \\
       \enddata
\end{deluxetable*}
\endgroup

\vspace{-24pt}
Case MH\_G0.5 with an HCD of \SI{0.5}{\kilo\meter} in an initially reducing atmosphere is presented in Figure~\ref{fig:photochem_geo}. The temporal evolution of key atmospheric species shows that the molar mixing ratios for the main prebiotic precursor \ce{HCN}, as well as \ce{CH4} as one of its main reactants, stabilize at $\sim\SI{500}{\year}$ at levels of \num{1.8e-14} and \num{2.2e-6}, respectively. Over the next million years, they rise steadily and moderately to levels of \num{4.7e-13} and \num{2.3e-5} at 3 million years, respectively. Finally, over the next 7 million years, the abundance of \ce{HCN} rises steeply to a maximum of \num{2.4e-9}. In addition, oxidizing species such as \ce{OH}, \ce{O2}, and the prebiotic precursor \ce{H2CO} begin to be produced, reaching a maximum of \num{1.1e-14} of \ce{H2CO}. 

In the process, water and \ce{H2} are consumed in the photochemical reactions in the atmosphere and decrease by several orders of magnitude. \ce{CO} levels rise sharply, while \ce{CH4} and \ce{H2O} levels decrease. This can be explained by the oxidation of methane by \ce{OH} radicals formed by photolysis of water to form \ce{H2CO}, which can be split by UV photolysis to form \ce{CO}, \ce{O2}, and other oxidized radicals. In addition, the decrease in \ce{CH4} can be further explained by the steep increase in \ce{HCN}, which is formed by reaction with atomic nitrogen from photolysis and likely passes through \ce{H2CN} as an intermediate \citep{Pearce2019}, as also visualized in simplified form in Figure~\ref{fig:hadean_earth_scheme}(D).

Similar trends can be identified in cases MH\_G1 and MH\_G2 in Figure~\ref{fig:photochem_geo} with increased HCDs. The increased surface fluxes of \ce{H2} and \ce{CH4} due to more productive serpentinization leave most of the reducing atmospheric species unchanged at similar levels as in the case of MH\_G0.5, but the synthesis of oxidizing species including \ce{H2CO} is strongly suppressed. In summary, the elevated surface fluxes of reducing gases do not significantly increase the atmospheric levels of reducing constituents, but substantially extend the time over which the atmosphere remains reducing from millions to at least 10 million years.

Simulations of an initially oxidizing atmosphere at \SI{4.0}{\giga\year} are presented in Figure~\ref{fig:photochem_geo}. In scenario EH\_G0.5 with an HCD of \SI{0.5}{\kilo\meter}, it can be seen that a moderately reduced state of the atmosphere cannot be sustained for the first hundred years. After about \SI{300}{\year}, the levels of reducing gases, e.g., \ce{H2} and \ce{CH4}, drop due to an abrupt rise of \ce{O2} and \ce{OH} in the atmosphere. This prevents effective formation of \ce{HCN}, leaving it at levels of \num{7.2e-15} and below, over 5 orders of magnitude less than the initially reducing model (see Figure~\ref{fig:photochem_geo}(MH\_G0.5)). This is not surprising, as oxidizing conditions are not a suitable environment for the effective synthesis of \ce{HCN} and other key reduced prebiotic molecules. Obviously, the supply of the necessary reducing reactants \ce{H2} and \ce{CH4} by serpentinization is insufficient at an HCD of \SI{0.5}{\kilo\meter}.

Increasing this HCD to \SI{1.0}{\kilo\meter}, as shown in Figure~\ref{fig:photochem_geo}(EH\_G1), allows to delay the growth of \ce{O2} in the atmosphere for several hundred years. The \ce{HCN} level reaches a maximum of \num{3.2e-14}, which is roughly 4 times higher than in case EH\_G0.5. After the steep rise of \ce{O2} at \SI{1000}{\year}, the period during which the atmosphere remains fully oxidized is shortened compared to case EH\_G0.5, and is partially reverted after about \SI{40000}{\year}. This allows the levels of \ce{H2} and \ce{CH4} to partially recover. Nevertheless, \ce{HCN} levels are not able to grow significantly and remain very low compared to the initially reducing models MH\_G0.5/1/2 at \SI{4.4}{\giga\year}.

This situation changes drastically when the HCD is increased to \SI{2.0}{\kilo\meter}, as shown in Figure~\ref{fig:photochem_geo}(EH\_G2). Because of increased serpentinization, the levels of \ce{H2} and \ce{CH4} grow over several hundred thousand years, suppressing and evading major oxidation of the atmosphere. This allows for peak \ce{HCN} abundances of \num{3.5e-8}, which is an order of magnitude higher than in the initially reducing scenario in Figure~\ref{fig:photochem_geo}(MH\_G2). The initial buildup of \ce{HCN} between \SIrange{200}{2000}{\year} coincides with a decline in \ce{CO2}, which is reduced in large amounts mainly by \ce{CH4} (and also by \ce{H2}) from serpentinization and acts as a carbon source for \ce{HCN}. \textit{This makes an initially oxidized atmosphere (i.e., high \ce{CO2}) the most suitable environment for \ce{HCN} synthesis in the presence of highly active serpentinization, a rather surprising finding.}

At the same time, the levels of \ce{O2} and \ce{OH} in the atmosphere are significantly suppressed in case EH\_G2. \ce{H2CO} levels are quite stable over all cases EH\_G0.5/1/2, with maximum levels around \num{e-14}. On the other hand, in the initially reducing MH models (Figures~\ref{fig:photochem_geo}(MH\_G0.5/1/2)), the \ce{H2CO} levels drop with increasing HCD due to suppressed availability of oxygen sources. An initially oxidized state of the atmosphere thus favors atmospheric \ce{H2CO} levels as another promising prebiotic precursor.

The final sharp increase of \ce{HCN} in case EH\_G2 at about \SI{300000}{\year} coincides with a steep increase of \ce{CO} and a decrease of \ce{CH4} and \ce{H2O}, which again can be explained by the formation process of \ce{H2CO} and the in parallel occurring reactions shown in Figure~\ref{fig:hadean_earth_scheme}(D), as already mentioned above when explaining the steep increases at the very end in cases MH\_G0.5/1/2. This means that this behavior in the initially oxidizing case EH\_G2 is equivalent to the behavior in the initially reducing cases, allowing a similarly extensive \ce{HCN} synthesis, and pushing the \ce{HCN} levels even above the initially reducing scenario. \textit{This underscores our striking finding that extensive serpentinization leads to the most \ce{HCN} in an initially oxidized atmosphere, not a reduced one.}

\subsubsection{Late Veneer in the Mid-Hadean}

\begin{figure*}[p]
    \centering
    \includegraphics[width=0.9\textwidth]{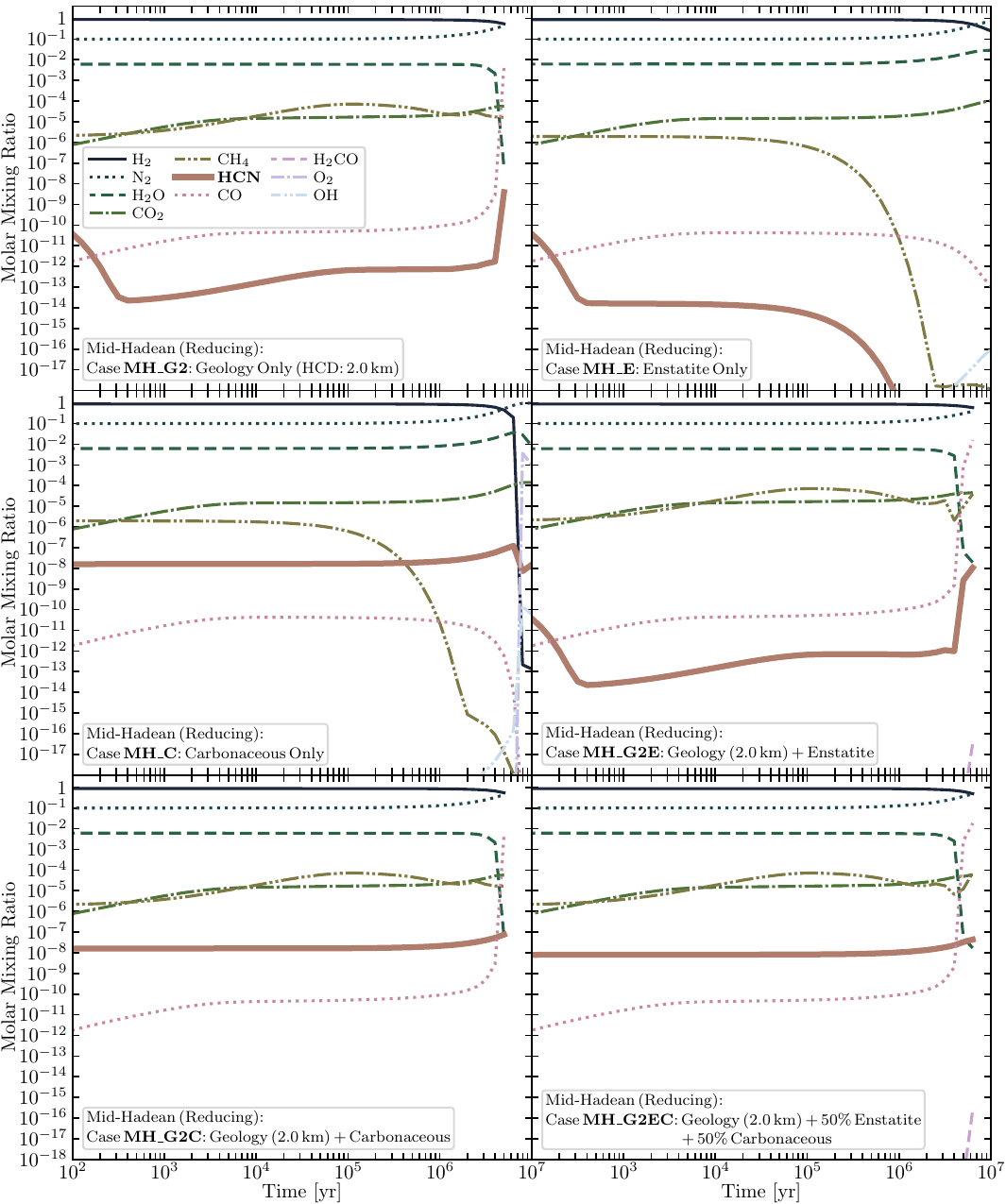}
    \caption{Testing whether exogenous chondritic bombardment is capable of keeping the atmosphere reduced while counteracting the oxidizing gases emitted by volcanism, as well as exploring its interplay with serpentinization. Shown is the simulated atmospheric composition of key species in the layer closest to the surface as a function of time starting at \SI{4.4}{\giga\year}. The atmosphere is initially set to a reducing state (\ce{H2}: \SI{90}{\percent}, \ce{N2}: \SI{10}{\percent}, \ce{CH4}: \SI{1}{ppm}) and a temperature of \SI{78}{\celsius} (Table~\ref{tab:epochs}), closely following the reducing case considered by \citet{Pearce2022}. Panel/case \textbf{MH\_G2} corresponds to a model driven by serpentinization alone (geology only) and is identical to Figure~\ref{fig:photochem_geo}(MH\_G2). It is shown here again for reference to facilitate comparison with the other cases. Cases E and C correspond to scenarios without any contribution from serpentinization, but with the reduction capacity of the late veneer alone competing with the \ce{CO2} flux emitted from the Earth's mantle. Panel \textbf{MH\_E} corresponds to a scenario with a bombardment of pure enstatite composition, and panel \textbf{MH\_C} instead corresponds to a pure carbonaceous composition. Panels \textbf{MH\_G2E}, \textbf{MH\_G2C}, and \textbf{MH\_G2EC} present the results corresponding to scenarios combining geological and late veneer source flux contributions, with case MH\_G2E combining the geology with an enstatite bombardment (cases MH\_G2 and MH\_E), case MH\_G2C combines the geology instead with a carbonaceous bombardment (MH\_G2 and MH\_C), and case MH\_G2EC combines all three with a half-half split bombardment composition (MH\_G2, \SI{50}{\percent} of MH\_E, and \SI{50}{\percent} of MH\_C).}\label{fig:photochem_red}
\end{figure*}

Figure~\ref{fig:photochem_red} shows and compares the results for cases MH\_G2/E/C/G2E/G2C/G2EC in Table~\ref{tab:cases_red} in the initially reducing scenario in the MH at \SI{4.4}{\giga\year}. Cases E and C correspond to scenarios including a bombardment of enstatite or carbonaceous composition, respectively. The geological contributions of \ce{H2} and \ce{CH4} are deactivated to examine the reducing potential of the late veneer alone, but the \ce{CO2} degassed from the mantle remains in the model. The purpose is to investigate whether the bombardment with meteorites of enstatite composition is capable of reducing the atmosphere while counteracting the oxidizing gases emitted by volcanism. The same holds for case C, where a purely carbonaceous bombardment and \ce{HCN} synthesis competes with the geological \ce{CO2} emission from volcanoes. See Appendix~\ref{sec:scenarios} for details. Finally, we examined the combination of these source terms by combining the geology with an HCD of \SI{2}{\kilo\meter} with the enstatite-only bombardment in case G2E, with the carbonaceous-only bombardment in case G2C, and perhaps the most agnostic assumption of a mixed bombardment of half and half composition in case G2EC. For comparison, case MH\_G2 (geology only) is again shown in Figure~\ref{fig:photochem_red}(MH\_G2). 

In case MH\_E, a late veneer of pure enstatite composition is the only source of reducing gases, in this case only \ce{H2}. Its reduction capacity competes with the geological source flux of \ce{CO2} emitted from the Earth's mantle. For about \SI{10000}{\year}, the enstatite bombardment is able to keep the \ce{CH4} and \ce{HCN} levels stable at \num{1.8e-6} and \num{1.5e-14}, respectively. For \ce{CH4} this is close to its initial abundance at the beginning of the simulation (see Table~\ref{tab:epochs}). After that, the levels of all reducing gases begin to drop steeply. This suggests that enstatite bombardment, as we have included it here in the model with a continuous intermediate impact rate (see Appendix~\ref{sec:bombardment_model}), is not able to drive significant reducing chemistry in the atmosphere. It is able to delay the oxidation of the atmosphere from the mantle \ce{CO2} for about \SI{10000}{\year}, but it cannot stop the inevitable decline of reducing gases and thus has no potential to enable effective prebiotic synthesis.

It is important to note that an increased supply of \ce{H2} alone is not sufficient to fuel an effective \ce{HCN} synthesis in the atmosphere. Comparing the \ce{H2} surface fluxes for serpentinization and enstatite bombardment in Table~\ref{tab:cases_red}, the \ce{H2} flux for enstatite bombardment in case MH\_E is actually higher than that generated by serpentinization in case MH\_G2. However, the resulting \ce{HCN} levels in the atmosphere are higher for serpentinization despite the lower \ce{H2} flux compared to enstatite bombardment. Instead, it is the additional \ce{CH4} flux resulting from serpentinization and subsequent reactions that makes the big difference here, raising \ce{HCN} levels significantly, while the enstatite bombardment is not a source of \ce{CH4}. The maximum \ce{HCN} levels reached in case MH\_G2 are more than five orders of magnitude higher than in case MH\_E (see Table~\ref{tab:results_atm}).

This does not mean that an enstatite bombardment with more singular cataclysmic events is not capable of providing significant amounts of prebiotic precursors in the atmosphere. For example, the models by \citet{Wogan2023,Zahnle2020} assume that a major fraction or even all of the HSE excess in the Earth's crust and mantle was delivered to Earth by one enstatite impact. They have shown that this allows for significant \ce{CH4} and \ce{HCN} synthesis in the post-impact atmosphere. We explore what would happen if our models experienced the same maxed-out bombardment and its effect on atmospheric photochemistry and prebiotic synthesis in WLPs in the supplementary results in Appendix~\ref{sec:supp_results_high_bomb_rate}.

However, in what is one of the most important findings of this paper, our results indicate that enstatite bombardment is clearly outcompeted by serpentinization and is not able to drive any reducing chemistry, but only delays the decay of reducing gases.

In case MH\_C, a pure carbonaceous bombardment, this becomes even clearer. Figure~\ref{fig:photochem_red}(MH\_C) shows that the absence of any \ce{H2} flux leads to a drop in atmospheric \ce{CH4} only slightly earlier than for the enstatite bombardment. This means that the \ce{H2} flux from the enstatite bombardment is nearly negligible in the formation of \ce{CH4} in this context.

The direct synthesis of \ce{HCN} during the impact of carbonaceous meteorites (see Figure~\ref{fig:photochem_red}(MH\_C)) is able to increase the \ce{HCN} levels in the atmosphere to a maximum of \num{1.2e-7}, which is almost 2 orders of magnitude more than serpentinization alone (case MH\_G2) and nearly seven orders of magnitude more than enstatite bombardment (case MH\_E). Furthermore, the \ce{HCN} abundance is stable over the entire time evolution at a molar mixing ratio in the atmosphere above \num{e-8}. 

As another major result of this study, this makes a carbonaceous bombardment the most promising candidate for contributing the most \ce{HCN} to the Hadean Earth's atmosphere.

Figure~\ref{fig:photochem_red}(MH\_G2E) explores the possibility of a combination of very active serpentinization (case MH\_G2) and pure enstatite bombardment (case MH\_E). When compared to Figure~\ref{fig:photochem_red}(MH\_G2), the evolution of reducing gases matches very closely. This confirms the findings above that serpentinization has a stronger potential to feed the production of reducing gases in the atmosphere than enstatite bombardment. Nevertheless, the enstatite bombardment pushes up the maximum \ce{HCN} molar mixing ratio by a factor of $\sim\num{2.7}$ compared to serpentinization alone (compare cases MH\_G2 and MH\_G2E in Table~\ref{tab:results_atm}).

The combination of serpentinization and carbonaceous bombardment (see Figure~\ref{fig:photochem_red}(MH\_G2C)) shows that the \ce{CH4} level in the atmosphere is dominated by serpentinization. In contrast, \ce{HCN} closely follows the same behavior as in case MH\_C with carbonaceous bombardment alone. In case MH\_G2C, the maximum \ce{CH4} level matches case MH\_G2, and the maximum \ce{HCN} level is close to case MH\_C (see Table~\ref{tab:results_atm}).

Finally, Figure~\ref{fig:photochem_red}(MH\_G2EC) shows the most agnostic case with all source terms active. Serpentinization is set to the highest efficiency with an HCD of \SI{2.0}{\kilo\meter}, and the late veneer consists of equal amounts of enstatite and carbonaceous impactors. Looking at the trends in all of the previous cases, one would expect the \ce{CH4} levels to be dominated by serpentinization and the \ce{HCN} levels by carbonaceous bombardment, and this is indeed the behavior seen in Figure~\ref{fig:photochem_red}(MH\_G2EC). It appears that the direct supply of \ce{CH4} and \ce{HCN} dominates over their synthesis by photochemistry. Nevertheless, the entire network of photochemical reactions helps to keep the abundances of oxidizing gases low.

\subsubsection{Late Veneer in the End-Hadean}

\begin{figure*}[p]
    \centering
    \includegraphics[width=0.9\textwidth]{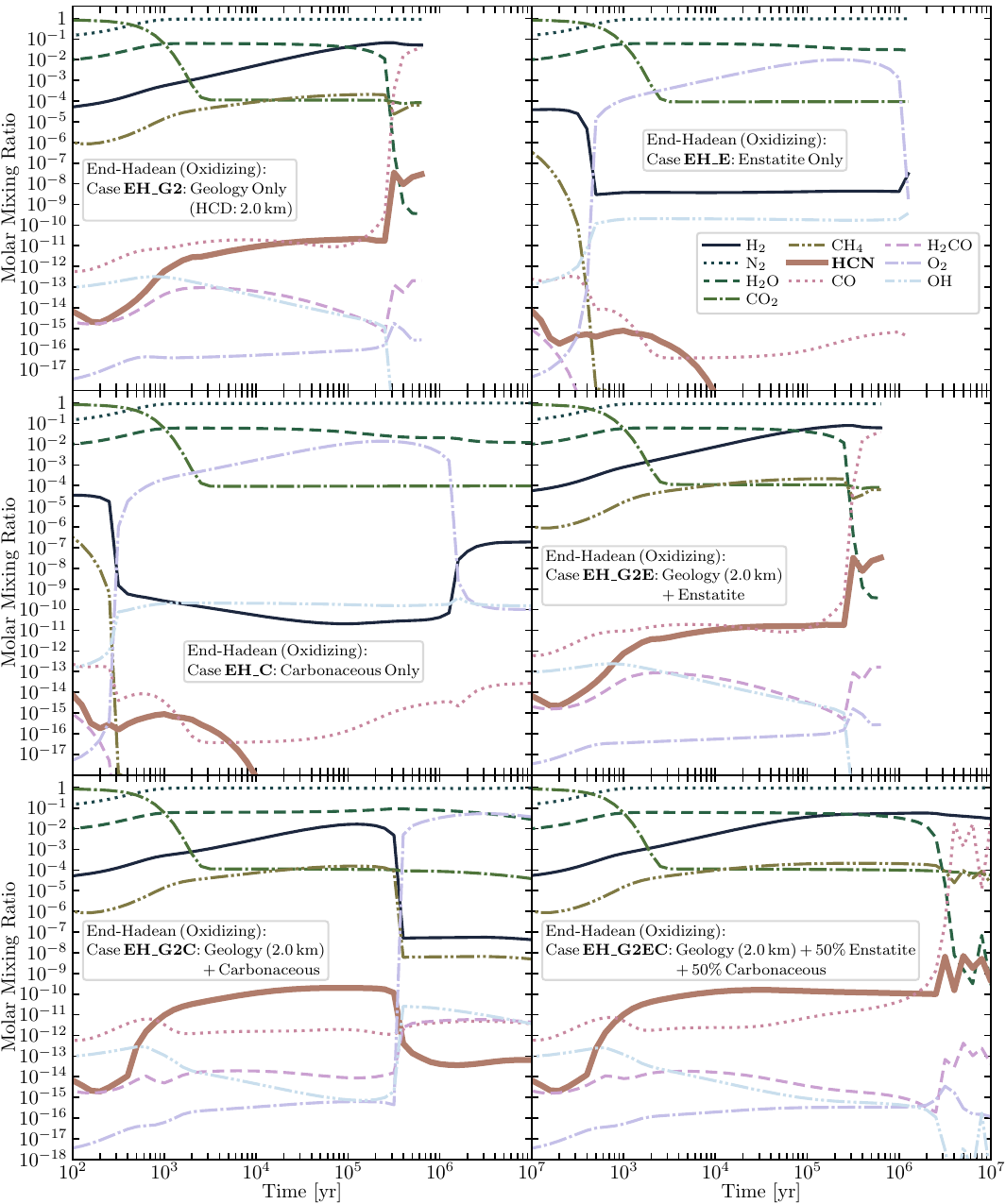}
    \caption{Testing whether exogenous chondritic bombardment is capable of reducing an initially oxidized atmopshere while counteracting the oxidizing gases emitted by volcanism, as well as exploring its interplay with serpentinization. Shown is the simulated atmospheric composition of key species in the layer closest to the surface as a function of time starting at \SI{4.0}{\giga\year}. The atmosphere is initially set to an oxidizing state (\ce{CO2}: \SI{90}{\percent}, \ce{N2}: \SI{10}{\percent}, \ce{CH4}: \SI{10}{ppm}) and a temperature of \SI{51}{\celsius} (Table~\ref{tab:epochs}), closely following the oxidizing case considered by \citet{Pearce2022}. Panel/case \textbf{EH\_G2} corresponds to a model driven by serpentinization alone (geology only) and is identical to Figure~\ref{fig:photochem_geo}(EH\_G2). It is shown here again for reference to facilitate comparison with the other cases. Cases E and C correspond to scenarios without any contribution from serpentinization, but with the reduction capacity of the late veneer alone competing with the \ce{CO2} flux emitted from the Earth's mantle. Panel \textbf{EH\_E} corresponds to a scenario with a bombardment of pure enstatite composition, and panel \textbf{EH\_C} instead corresponds to a pure carbonaceous composition. Panels \textbf{EH\_G2E}, \textbf{EH\_G2C}, and \textbf{EH\_G2EC} present the results corresponding to scenarios combining geological and late veneer source flux contributions, with case EH\_G2E combining the geology with an enstatite bombardment (cases EH\_G2 and EH\_E), case EH\_G2C combines the geology instead with a carbonaceous bombardment (EH\_G2 and EH\_C), and case EH\_G2EC combines all three with a half-half split bombardment composition (EH\_G2, \SI{50}{\percent} of EH\_E, and \SI{50}{\percent} of EH\_C).}\label{fig:photochem_ox}
\end{figure*}

Figure~\ref{fig:photochem_ox} gives an overview of the different cases and the influence of impacts on the initial oxidizing atmosphere in the EH at \SI{4.4}{\giga\year}. Figure~\ref{fig:photochem_ox}(EH\_G2) is identical to Figure~\ref{fig:photochem_geo}(EH\_G2) and is shown again for easy comparison with the other cases EH\_E/C/G2E/G2C/G2EC.

The pure enstatite bombardment with serpentinization turned off in Figure~\ref{fig:photochem_ox}(EH\_E) shows that the \ce{H2} flux from impact degassing is not sufficient to stabilize the \ce{H2} level in the atmosphere. After about \SI{300}{\year}, its abundance begins to drop sharply and remains at a level of \num{3.1e-8} for the next million years. The level of \ce{CH4} shows the same decline from about \SI{100}{\year} and \ce{HCN} never reaches a significant amount. This coincides with a sharp increase in \ce{O2} and \ce{OH}, indicating that the enstatite bombardment is not able to reduce the atmosphere, while its own \ce{H2} flux is even suppressed after the onset of the formation of oxidized species in the atmosphere.

The fact that \ce{HCN} is not stable and cannot be effectively produced in an oxidized atmosphere becomes clear in Figure~\ref{fig:photochem_ox}(EH\_C). The direct synthesis of \ce{HCN} by carbonaceous impacts is switched on after \SI{1000}{\year}, because at this time \ce{N2} starts to exceed the atmospheric abundance of \ce{CO2} and the ratio \ce{N2}/\ce{CO2} is high enough to allow the formation of \ce{HCN} \citep[see Appendix~\ref{sec:cc_impactors}]{Kurosawa2013}. Nevertheless, this is too late, as the full oxidation of the atmosphere already happens at about \SI{300}{\year} with a steep decrease of \ce{H2} and an increase of \ce{O2} in the atmosphere. Despite the high flux of \ce{HCN} due to a purely carbonaceous bombardment, it is not able to build up significantly in the atmosphere and remains at molar mixing ratios of \num{8.0e-19} and below. Between 1 and 2 million years, \ce{H2} levels increase and \ce{O2} levels decrease, partially reversing the oxidation of the atmosphere, but \ce{HCN} levels have already dropped drastically. It might take much more time than the 10 million years shown for them to recover.

Figure~\ref{fig:photochem_ox}(EH\_G2E) shows that the combination of serpentinization and enstatite bombardment increases the levels of \ce{H2} in the atmosphere, but not significantly for \ce{CH4} and \ce{HCN} (see Table~\ref{tab:results_atm}). Apparently, the \ce{H2} flux from enstatite degassing does not contribute substantially to the budget of formed \ce{HCN}, whereas \ce{CH4} emitted from hydrothermal vents does. The main trends in the temporal evolution here are dictated by serpentinization (cf. Figure~\ref{fig:photochem_ox}(EH\_G2)).

Yet enstatite bombardment does not destabilize the reducing effect of serpentinization, whereas carbonaceous bombardment does. For the first \SI{30000}{\year} in Figure~\ref{fig:photochem_ox}(EH\_G2C), the atmosphere is significantly reduced. In particular, the activation of direct \ce{HCN} synthesis by carbonaceous impactors around \SI{1000}{\year} pushes its atmospheric abundance significantly to maximum values of \num{2.0e-10}. However, around \SI{300000}{\year} there is a sudden increase in \ce{O2} and a sudden decrease in reducing gas levels. The high \ce{HCN} abundance leads to a relative decrease in the molar mixing ratio of \ce{H2}, keeping it below values of \num{1.7e-2}. This is slightly too low to prevent the growth of oxidizing gases in the photochemical network.

In a mixed bombardment scenario, as assumed in Figure~\ref{fig:photochem_ox}(EH\_G2EC), this stabilizing effect of the enstatite bombardment becomes very apparent, as the additional \ce{H2} flux from enstatite impactors keeps the reduced state of the atmosphere stable for tens of millions of years. The final increase in \ce{HCN}, \ce{H2CO}, and \ce{CO} at 3 million years again coincides with a decrease in \ce{CH4} and \ce{H2O}, which can be explained as above by the oxidation of methane to \ce{H2CO} and the formation of \ce{HCN} by methane and nitrogen radicals reacting through the \ce{H2CN} intermediate.

In summary, for both MH and EH models, if serpentinization is active according to the extended Hadean mantle models \citep[Appendix~\ref{sec:mantle_hcd},][]{Miyazaki2022}, it dominates the levels of \ce{CH4} in the atmosphere. Since \ce{HCN} levels are closely correlated with \ce{CH4} as its major chemical intermediate \citep[cf.][]{Pearce2022}, serpentinization is the most important and reliable driving force in \ce{HCN} synthesis. In addition, when present, carbonaceous bombardment clearly dominates the abundance of \ce{HCN} by direct synthesis during impacts \citep{Kurosawa2013}.

In EH models, however, carbonaceous bombardment requires support from serpentinization. Due to its delayed activation in an initially \ce{CO2}-rich atmosphere, it either emerges in a highly oxidized atmosphere (see case EH\_C) or destabilizes the \ce{H2} levels and thus the reduced state of the atmosphere (see case EH\_G2C). An enstatite component in the bombardment allows to prevent this (see case EH\_G2EC). However, strongly active serpentinization with its \ce{CH4} surface flux is by far the most effective way not only to reduce an initially oxidizing atmosphere, but also to exploit the initially high \ce{CO2} levels as a carbon source for an effective \ce{HCN} synthesis. The resulting \ce{HCN} yields are comparable in magnitude to its synthesis in an initially reducing atmosphere. 

After evaluating the various scenarios, we can summarize that serpentinization, and this is the most important finding in this study, might resolve one of the most widely debated issues in the origins of life research community, namely that an initially oxidized atmosphere on Hadean Earth would prevent sufficiently effective synthesis of prebiotic molecules.

\subsection{Rain-out}

Our atmospheric model can provide rain-out of chemical species from the lowest layer of the atmosphere closest to the surface. This effectively removes these molecules from the atmosphere at each time step and affects the balance between atmospheric gases in the photochemical network, which is already included in the calculation of the atmospheric mixing ratios as shown in the previous Sections. The rain can accumulate on the surface of the first volcanic islands and continental crust emerging from the global ocean. The calculated rain-out rates define an influx of these chemical species into the first small water bodies forming on these landmasses, e.g., small lakes and WLPs, which allow the concentration of key prebiotic precursors, e.g., the water-soluble molecules \ce{HCN} and \ce{H2CO}. These rain-out rates are directly correlated with and follow the time evolution of the molar mixing ratio of the respective species, as already seen in a previous study \citep{Pearce2022}.

\begin{figure}
    \centering
    \includegraphics[width=\columnwidth]{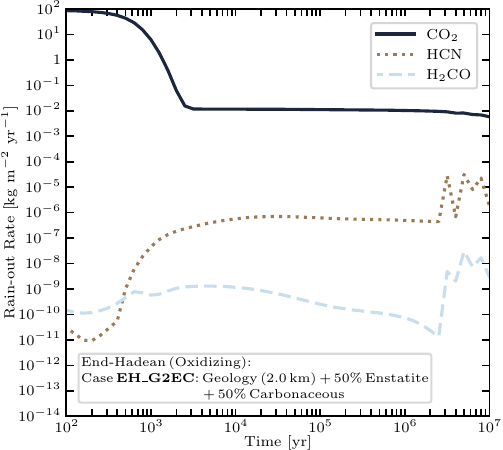}
    \caption{Rain-out rates of \ce{CO2}, \ce{HCN}, and \ce{H2CO} from the lowest atmospheric layer closest to the surface as a function of time for case EH\_G2EC.}
    \label{fig:rain-out}
\end{figure}

As an example, Figure~\ref{fig:rain-out} shows the time evolution of the rain-out rates for \ce{CO2}, \ce{HCN}, and \ce{H2CO} for case EH\_G2EC, incorporating all the different processes providing reducing gases as well as \ce{CO2} from volcanic outgassing on the Hadean Earth. It also corresponds to one of the highest rain-out rates reached across all simulations in the EH models (see Table~\ref{tab:results_rain}).

\begingroup
\setlength{\tabcolsep}{3pt}
\begin{deluxetable*}{lDDcDDcDD}[t]
    \tablecaption{Maximum resulting rain-out rates of prebiotic precursors.}
    \label{tab:results_rain}
    \tabletypesize{\small}
        \tablehead{
            \colhead{} & \multicolumn{14}{c}{Max.~Rain-out Rate [\unit{\kilo\gram\per\meter\squared\per\year}]} \\
            \cmidrule{2-15}
            \colhead{} & \multicolumn{4}{c}{\ce{CO2}} & \colhead{} & \multicolumn{4}{c}{\ce{HCN}} & \colhead{} & \multicolumn{4}{c}{\ce{H2CO}}\\
            \cmidrule{2-5}\cmidrule{7-10}\cmidrule{12-15}
            \colhead{Case} & \twocolhead{MH (red.)} & \twocolhead{EH (ox.)} & \colhead{} & \twocolhead{MH (red.)} & \twocolhead{EH (ox.)} & \colhead{} & \twocolhead{MH (red.)} & \twocolhead{EH (ox.)}
        }
        \decimals
        \startdata
        G0.5 & 1.86$\times 10^{-2}$ & 1.20$\times 10^{-2}$ && 1.11$\times 10^{-5}$ & 3.19$\times 10^{-11}$ && 8.38$\times 10^{-10}$ & 1.47$\times 10^{-8}$ \\
        G1 & 1.49$\times 10^{-2}$ & 1.12$\times 10^{-2}$ && 2.57$\times 10^{-6}$ & 1.41$\times 10^{-10}$ && 3.34$\times 10^{-12}$ & 6.77$\times 10^{-9}$ \\
        G2 & 6.26$\times 10^{-3}$ & 1.17$\times 10^{-2}$ && 1.85$\times 10^{-5}$ & 1.57$\times 10^{-4}$ && 2.30$\times 10^{-14}$ & 1.46$\times 10^{-8}$ \\
        E & 1.27$\times 10^{-2}$ & 9.81$\times 10^{-3}$ && 6.79$\times 10^{-11}$ & 3.09$\times 10^{-15}$ && 1.89$\times 10^{-18}$ & 6.26$\times 10^{-21}$ \\
        C & 1.56$\times 10^{-2}$ & 9.99$\times 10^{-3}$ && 5.76$\times 10^{-4}$ & 3.53$\times 10^{-15}$ && 1.89$\times 10^{-18}$ & 3.86$\times 10^{-21}$ \\
        G2E & 4.96$\times 10^{-3}$ & 1.17$\times 10^{-2}$ && 4.95$\times 10^{-5}$ & 1.43$\times 10^{-4}$ && 2.57$\times 10^{-12}$ & 1.17$\times 10^{-8}$ \\
        G2C & 6.25$\times 10^{-3}$ & 1.17$\times 10^{-2}$ && 3.53$\times 10^{-4}$ & 8.91$\times 10^{-7}$ && 2.30$\times 10^{-14}$ & 4.15$\times 10^{-7}$ \\
        G2EC & 6.53$\times 10^{-3}$ & 1.17$\times 10^{-2}$ && 2.03$\times 10^{-4}$ & 3.14$\times 10^{-5}$ && 1.84$\times 10^{-11}$ & 2.99$\times 10^{-8}$ \\
       \enddata
\end{deluxetable*}
\endgroup

\vspace{-24pt}
\ce{HCN} and \ce{H2CO} rain-out peaks around \SIrange{3}{5}{\mega\year} with rates of \SI{3.1e-5}{\kilo\gram\per\meter\squared\per\year} and \SI{3.0e-8}{\kilo\gram\per\meter\squared\per\year}, respectively. To get a rough feel of the magnitudes involved, the average global (water) precipitation from \numrange{1983}{2023} on Earth was \SI{2.81}{\milli\meter\per\day} \citep{Adler2024}, equivalent to $\sim\SI{1000}{\kilo\gram\per\meter\squared\per\year}$. In tropical coastal areas, even over \SI{3500}{\milli\meter\per\year} and more can be reached locally \citep{Ogino2016}, equivalent to \SI{3500}{\kilo\gram\per\meter\squared\per\year} and more. The shape of the time evolution of the rain-out rates in Figure~\ref{fig:rain-out} is very similar to the corresponding atmospheric molar mixing ratios in Figure~\ref{fig:photochem_ox}(EH\_G2EC).

Comparing the rain-out rates in Table~\ref{tab:results_rain} shows that the highest \ce{HCN} rain-out rate is reached in case MH\_C and the highest \ce{H2CO} rain-out rate is reached in case EH\_G2C. The initially reducing atmosphere does not always lead to the highest \ce{HCN} rain-out rates. In general it is advantageous to have initially reducing conditions for \ce{HCN} formation and rain-out, but for the cases including serpentinization at an HCD of \SI{2.0}{\kilo\meter} (cases G2, G2E, G2C, G2EC) the initially oxidizing conditions allow similar or even higher rates (cases G2 and G2E) compared to the reducing conditions. The reason is that the initially present \ce{CO2} can be exploited as an abundant carbon source for \ce{HCN} synthesis, as already discussed for the atmospheric abundances in the previous Sections. In addition, carbonaceous bombardment and subsequent direct \ce{HCN} synthesis during impacts significantly enhances \ce{HCN} rain-out if the atmosphere is either non-oxidized or only moderately oxidized.

When serpentinization is included in the model, \ce{H2CO} rain-out rates are systematically higher by about \numrange{12}{14} orders of magnitude in the initially oxidizing scenario, making atmospheric photochemistry a potentially significant source of \ce{H2CO} in EH WLP settings.

\subsection{Prebiotic Molecules in Warm Little Ponds}

One way for rained-out \ce{HCN} and \ce{H2CO} to accumulate and concentrate on Hadean Earth is to enter the first emerging WLPs, small reservoirs of water about a meter in diameter that formed on the first landmasses to emerge from the global ocean. Due to the seasonal cycles of the environment, these ponds can go through repeated states of desiccation and rewetting by precipitation. Rainwater containing the prebiotic precursors \ce{HCN} and \ce{H2CO} resulting from atmospheric photochemistry can accumulate in these WLPs. Experimental studies have shown that these precursor molecules are able to form complex prebiotic molecules such as RNA-building blocks during wet-dry cycling \citep{Oro1961,LaRowe2008,Ferus2019,Butlerow1861,Breslow1959,Yi2020}. To simulate this process in the context of our atmospheric models, we developed a WLP wet-dry cycling model \citep{Pearce2017,Pearce2022}, which uses experimentally determined yields of prebiotic molecules formed from \ce{HCN} and \ce{H2CO}, and used the determined rain-out rates in Table~\ref{tab:results_rain} as influx terms supplying the prebiotic synthesis.

\begin{figure*}[t]
    \centering
    \includegraphics[width=\textwidth]{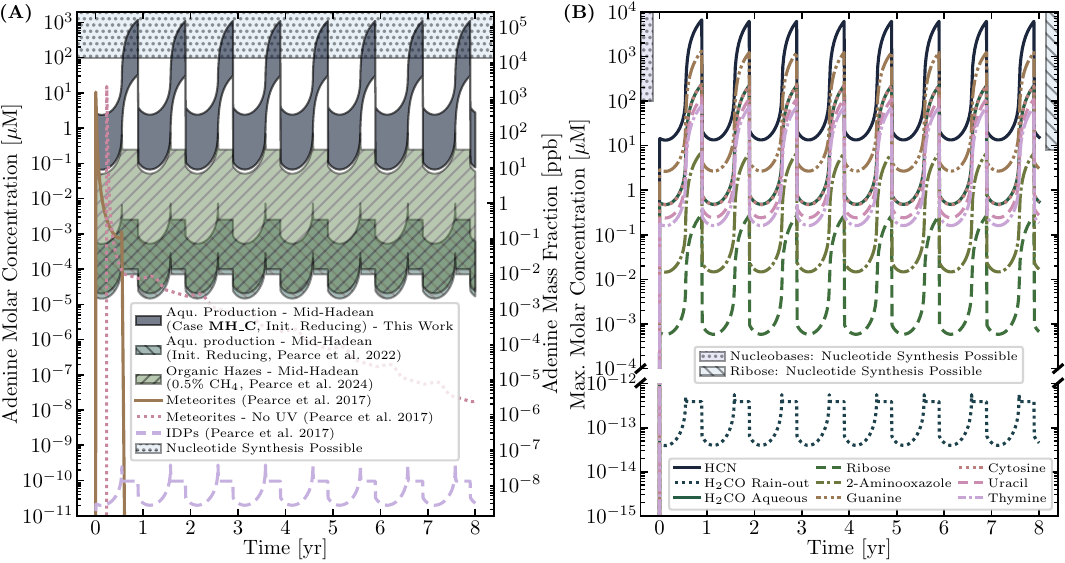}
    \caption{Concentrations of prebiotic molecules over time resulting from the warm little pond (WLP) cycling model \citep{Pearce2017,Pearce2022}. Inflow fluxes for \ce{HCN} and \ce{H2CO} are derived from atmospheric rain-out as a result of photochemistry in case MH\_C (Table~\ref{tab:results_rain}). Panel \textbf{A} shows a comparison of the resulting adenine concentrations in this work with the previous atmospheric model with some assumptions based on the present Earth that might be less suitable for the Hadean Earth \citep{Pearce2022}, an organic haze experiment with particles containing biomolecules formed in an atmosphere with \SI{5}{\percent} \ce{CH4}, which fall into the pond \citep{Pearce2024}, delivery from meteorites and interplanetary dust particles (IDPs) \citep{Pearce2017}, and the amounts required for nucleotide synthesis as determined by experimental studies \citep{Ponnamperuma1963,Fuller1972,Powner2009,Becker2016,Saladino2017,Becker2018,Nam2018,Teichert2019}. Aqueous production is sourced from atmospheric rain-out of \ce{HCN} multiplied by experimental yields of adenine \citep{Oro1961,Wakamatsu1966,Hill2002,Pearce2022}. The pond cycles through 6 months of wet and 6 months of dry conditions. Sinks for concentration are photodissociation by UV light, hydrolysis, and seepage through pores at the bottom of the pond. Panel \textbf{B} shows concentrations for several prebiotic molecules, including accumulating \ce{HCN} and \ce{H2CO} from rain-out in case MH\_C, and subsequently formed nucleobases, the sugar ribose, required amounts of nucleobases and ribose for nucleotide synthesis in experiments, and 2-aminooxazole, which is a key intermediate in the Powner-Sutherland pathway \citep{Powner2009}.}\label{fig:WLP}
\end{figure*}

Figure~\ref{fig:WLP} shows as an example the resulting abundances for key RNA building blocks and intermediates using the rain-out rates for case MH\_C. The left panel shows the evolution of the prebiotic molecule adenine, one of the genetic letters in RNA and DNA molecules, over time. One year corresponds to a full cycle of 6 months of filling the pond with rain and 6 months of dry conditions (no rain), allowing evaporation to dry the pond to a minimum of \SI{1}{\milli\meter} and concentrating the molecules in the process. This concentration process is what allows these complex prebiotic biomolecules to form from \ce{HCN} and \ce{H2CO}.

In the left panel (A), the shaded blue area shows the range between the minimum and maximum yields of adenine formed by prebiotic synthesis from \ce{HCN}. This \ce{HCN} was supplied by rain-out from the atmosphere, and the influx rates correspond to the values shown in Table~\ref{tab:results_rain}. The maximum yield for adenine is a concentration of \SI{1.2}{\milli\Molar} in the WLP. It is summarized together with the maximum reached concentrations of other prebiotic molecules relevant for the origins of life, including all considered cases, in Table~\ref{tab:results_WLP}. For comparison, several WLP adenine concentrations from previous studies are also plotted in Figure~\ref{fig:WLP}(A).

The dark green area with ``\textbackslash\textbackslash''-hatching shows the resulting concentrations for the initially reduced MH from the previous study by \citet{Pearce2022}. The computational methods and workflow used in that earlier study were very similar to those used in this work. Atmospheric photochemistry, rain-out rates, and prebiotic synthesis in the WLP were calculated using the same approach. It considered the same initially reducing atmosphere in the MH at \SI{4.4}{\giga\year} (see Table~\ref{tab:epochs}). This happened to be the most productive scenario for prebiotic molecule synthesis in the WLPs considered in this earlier study by \citet{Pearce2022}.

\begingroup
\setlength{\tabcolsep}{3pt}
\begin{splitdeluxetable*}{lDDcDDcDDcDDBlDDcDDcDDBlDDcDDcDD}
    \tablecaption{Maximum yields of prebiotic organic molecules in warm little ponds (WLPs).}
    \label{tab:results_WLP}
    \tabletypesize{\small}
        \tablehead{
            \colhead{} & \multicolumn{19}{c}{Max.~Warm Little Pond Concentration [\unit{\micro\Molar}]} & \colhead{} & \multicolumn{14}{c}{Max.~Warm Little Pond Concentration [\unit{\micro\Molar}]} & \colhead{} & \multicolumn{14}{c}{Max.~Warm Little Pond Concentration [\unit{\micro\Molar}]} \\
            \cmidrule{2-20}\cmidrule{22-35}\cmidrule{37-50}
            \colhead{} & \multicolumn{4}{c}{\ce{HCN} from Rain-out} & \colhead{} & \multicolumn{4}{c}{\ce{H2CO} from Rain-out} & \colhead{} & \multicolumn{4}{c}{\ce{H2CO} from Aqueous Synth.\tablenotemark{a}} & \colhead{} & \multicolumn{4}{c}{Ribose} & \colhead{} & \multicolumn{4}{c}{2-Aminooxazole} & \colhead{} & \multicolumn{4}{c}{Adenine} & \colhead{} & \multicolumn{4}{c}{Guanine}& \colhead{} & \multicolumn{4}{c}{Cytosine} & \colhead{} & \multicolumn{4}{c}{Uracil} & \colhead{} & \multicolumn{4}{c}{Thymine} \\
            \cmidrule{2-5}\cmidrule{7-10}\cmidrule{12-15}\cmidrule{17-20}\cmidrule{22-25}\cmidrule{27-30}\cmidrule{32-35}\cmidrule{37-40}\cmidrule{42-45}\cmidrule{47-50}
            \colhead{Case} & \twocolhead{MH (red.)} & \twocolhead{EH (ox.)} & \colhead{} & \twocolhead{MH (red.)} & \twocolhead{EH (ox.)} & \colhead{} & \twocolhead{MH (red.)} & \twocolhead{EH (ox.)} & \colhead{} & \twocolhead{MH (red.)} & \twocolhead{EH (ox.)} & \colhead{Case} & \twocolhead{MH (red.)} & \twocolhead{EH (ox.)} & \colhead{} & \twocolhead{MH (red.)} & \twocolhead{EH (ox.)} & \colhead{} & \twocolhead{MH (red.)} & \twocolhead{EH (ox.)} & \colhead{Case} & \twocolhead{MH (red.)} & \twocolhead{EH (ox.)} & \colhead{} & \twocolhead{MH (red.)} & \twocolhead{EH (ox.)} & \colhead{} & \twocolhead{MH (red.)} & \twocolhead{EH (ox.)}
        }
        \decimals
        \startdata
        G0.5 & 3.66 & 1.05$\times 10^{-5}$ && 2.49$\times 10^{-4}$ & 4.38$\times 10^{-3}$ && 1.32$\times 10^{-1}$ & 3.79$\times 10^{-7}$ && 1.61$\times 10^{-4}$\tablenotemark{b} & 5.35$\times 10^{-6}$\tablenotemark{c} & G0.5 & 3.99$\times 10^{-3}$ & 1.14$\times 10^{-8}$ && 6.47$\times 10^{-1}$ & 1.86$\times 10^{-6}$ && 7.24$\times 10^{-1}$ & 2.08$\times 10^{-6}$ & G0.5 & 1.26$\times 10^{-1}$ & 3.63$\times 10^{-7}$ && 6.49$\times 10^{-2}$ & 1.86$\times 10^{-7}$ && 4.31$\times 10^{-2}$ & 1.24$\times 10^{-7}$ \\
        G1 & 8.48$\times 10^{-1}$ & 4.66$\times 10^{-5}$ && 9.93$\times 10^{-7}$ & 2.01$\times 10^{-3}$ && 3.06$\times 10^{-2}$ & 1.68$\times 10^{-6}$ && 3.73$\times 10^{-5}$\tablenotemark{b} & 2.46$\times 10^{-6}$\tablenotemark{c} & G1 & 9.23$\times 10^{-4}$ & 5.07$\times 10^{-8}$ && 1.50$\times 10^{-1}$ & 8.23$\times 10^{-6}$ && 1.68$\times 10^{-1}$ & 9.21$\times 10^{-6}$ & G1 & 2.93$\times 10^{-2}$ & 1.61$\times 10^{-6}$ && 1.50$\times 10^{-2}$ & 8.26$\times 10^{-7}$ && 9.97$\times 10^{-3}$ & 5.48$\times 10^{-7}$ \\
        G2 & 6.11 & 1.21$\times 10^{3}$ && 6.85$\times 10^{-9}$ & 4.36$\times 10^{-3}$ && 2.20$\times 10^{-1}$ & 4.34$\times 10^{1}$ && 2.69$\times 10^{-4}$\tablenotemark{b} & 5.30$\times 10^{-2}$\tablenotemark{b} & G2 & 6.65$\times 10^{-3}$ & 1.33 && 1.08 & 2.17$\times 10^{2}$ && 1.21 & 2.41$\times 10^{2}$ & G2 & 2.11$\times 10^{-1}$ & 4.34$\times 10^{1}$ && 1.08$\times 10^{-1}$ & 2.17$\times 10^{1}$ && 7.18$\times 10^{-2}$ & 1.45$\times 10^{1}$ \\
        E & 2.24$\times 10^{-5}$ & 1.02$\times 10^{-9}$ && 5.62$\times 10^{-13}$ & 1.86$\times 10^{-15}$ && 8.07$\times 10^{-7}$ & 3.67$\times 10^{-11}$ && 9.84$\times 10^{-10}$\tablenotemark{b} & 4.48$\times 10^{-14}$\tablenotemark{b} & E & 2.44$\times 10^{-8}$ & 1.11$\times 10^{-12}$ && 3.95$\times 10^{-6}$ & 1.80$\times 10^{-10}$ && 4.43$\times 10^{-6}$ & 2.01$\times 10^{-10}$ & E & 7.73$\times 10^{-7}$ & 3.52$\times 10^{-11}$ && 3.97$\times 10^{-7}$ & 1.81$\times 10^{-11}$ && 2.63$\times 10^{-7}$ & 1.20$\times 10^{-11}$ \\
        C & 6.59$\times 10^{3}$ & 1.16$\times 10^{-9}$ && 5.64$\times 10^{-13}$ & 1.15$\times 10^{-15}$ && 2.37$\times 10^{2}$ & 4.19$\times 10^{-11}$ && 2.90$\times 10^{-1}$\tablenotemark{b} & 5.12$\times 10^{-14}$\tablenotemark{b} & C & 7.25 & 1.27$\times 10^{-12}$ && 1.19$\times 10^{3}$ & 2.05$\times 10^{-10}$ && 1.32$\times 10^{3}$ & 2.30$\times 10^{-10}$ & C & 2.37$\times 10^{2}$ & 4.02$\times 10^{-11}$ && 1.19$\times 10^{2}$ & 2.06$\times 10^{-11}$ && 7.91$\times 10^{1}$ & 1.37$\times 10^{-11}$ \\
        G2E & 1.64$\times 10^{1}$ & 1.03$\times 10^{3}$ && 7.64$\times 10^{-7}$ & 3.48$\times 10^{-3}$ && 5.90$\times 10^{-1}$ & 3.72$\times 10^{1}$ && 7.20$\times 10^{-4}$\tablenotemark{b} & 4.54$\times 10^{-2}$\tablenotemark{b} & G2E & 1.80$\times 10^{-2}$ & 1.14 && 2.95 & 1.86$\times 10^{2}$ && 3.27 & 2.07$\times 10^{2}$ & G2E & 5.72$\times 10^{-1}$ & 3.72$\times 10^{1}$ && 2.95$\times 10^{-1}$ & 1.86$\times 10^{1}$ && 1.97$\times 10^{-1}$ & 1.24$\times 10^{1}$ \\
        G2C & 3.72$\times 10^{3}$ & 2.94$\times 10^{-1}$ && 6.84$\times 10^{-9}$ & 1.23$\times 10^{-1}$ && 1.34$\times 10^{2}$ & 1.06$\times 10^{-2}$ && 1.64$\times 10^{-1}$\tablenotemark{b} & 1.63$\times 10^{-4}$\tablenotemark{c} & G2C & 4.10 & 3.20$\times 10^{-4}$ && 6.70$\times 10^{2}$ & 5.19$\times 10^{-2}$ && 7.45$\times 10^{2}$ & 5.81$\times 10^{-2}$ & G2C & 1.34$\times 10^{2}$ & 1.01$\times 10^{-2}$ && 6.70$\times 10^{1}$ & 5.21$\times 10^{-3}$ && 4.47$\times 10^{1}$ & 3.45$\times 10^{-3}$ \\
        G2EC & 1.81$\times 10^{3}$ & 1.03$\times 10^{1}$ && 5.49$\times 10^{-6}$ & 8.91$\times 10^{-3}$ && 6.51$\times 10^{1}$ & 3.73$\times 10^{-1}$ && 7.94$\times 10^{-2}$\tablenotemark{b} & 4.66$\times 10^{-4}$\tablenotemark{b} & G2EC & 1.99 & 1.13$\times 10^{-2}$ && 3.25$\times 10^{2}$ & 1.85 && 3.62$\times 10^{2}$ & 2.06 & G2EC & 6.50$\times 10^{1}$ & 3.59$\times 10^{-1}$ && 3.25$\times 10^{1}$ & 1.85$\times 10^{-1}$ && 2.17$\times 10^{1}$ & 1.24$\times 10^{-1}$ \\
       \enddata
    \tablenotetext{a}{Formaldehyde synthesized aqueously from rained-out \ce{HCN}.}
    \tablenotetext{b}{Most ribose is synthesized in formose reaction starting from formaldehyde, which in turn is synthesized aqueously from rained-out \ce{HCN}.} 
    \tablenotetext{c}{Most ribose is synthesized in formose reaction starting from formaldehyde rained-out directly from the atmosphere.}
\end{splitdeluxetable*}
\endgroup

The only difference is that \citet{Pearce2022} used different fluxes for the source terms of reducing gases. In particular, the \ce{CH4} fluxes from serpentinization are more representative of the present-day Earth \citep{Guzman-Marmolejo2013}, and \ce{H2} emitted by serpentinization was omitted. Additionally, a carbonaceous late veneer was not considered, excluding the possibility of direct \ce{HCN} synthesis during impact \citep{Kurosawa2013}. The maximum adenine yield in the present work is more than five orders of magnitude higher than the yield of \SI{7.3}{\nano\Molar} calculated by \citet{Pearce2022}.

The light green area with ``//''-hatching shows the resulting concentrations from an organic haze experiment with particles containing biomolecules formed in an atmosphere with \SI{5}{\percent} \ce{CH4}, which fall into the pond \citep{Pearce2024}. The maximum adenine yield in the present work was more than three orders of magnitude higher than the yield of \SI{0.7}{\micro\Molar} obtained by \citet{Pearce2024}.

Delivery of biomolecules by carbonaceous chondrites and interplanetary dust particles (IDPs) was considered using the same wet-dry cycling model by \citet{Pearce2017}. In the present study, the maximum adenine yield was about two orders of magnitude higher than the yield of \SI{10.6}{\micro\Molar} in this previous work. The exogenous delivery of biomolecules by meteorites might still be a way to enhance the concentrations of prebiotic molecules for a limited time during the wet phase of the pond, facilitating the synthesis of RNA building blocks.

The light blue area with dotted hatching shows the range of nucleobase abundances that was required to successfully synthesize nucleotides in experiments \citep{Ponnamperuma1963,Fuller1972,Powner2009,Becker2016,Saladino2017,Becker2018,Nam2018,Teichert2019}. Only our models (here, e.g., case MH\_C) were able to generate the required nucleobases concentrations and for the first time enter the regime of feasible nucleotide synthesis in WLPs, unlike the models by \citet{Pearce2017,Pearce2022,Pearce2024}.

Figure~\ref{fig:WLP}(B) shows concentrations for several prebiotic molecules, including accumulating \ce{HCN} and \ce{H2CO} from rain-out in case MH\_C, and the subsequent formation of the nucleobases guanine, cytosine, uracil, and thymine, the sugar ribose, and 2-aminooxazole as a key intermediate in the Powner-Sutherland pathway \citep{Powner2009}. The required nucleobase concentrations for nucleotide synthesis in experiments are shown again, here as a vertical bar with dotted hatching in the top left corner. In addition, the required ribose concentrations are indicated by a vertical bar with ``\textbackslash\textbackslash''-hatching in the top right corner. The simulated ribose abundances do not reach high enough concentrations to allow nucleotide synthesis in experiments. However, by turning off the seepage at the bottom of the WLP, the required ribose concentrations can be reached, see the following Sections~\ref{sec:seepage_off}~and~\ref{sec:discussion}.

\begin{figure*}[t]
    \centering
    \includegraphics[width=\textwidth]{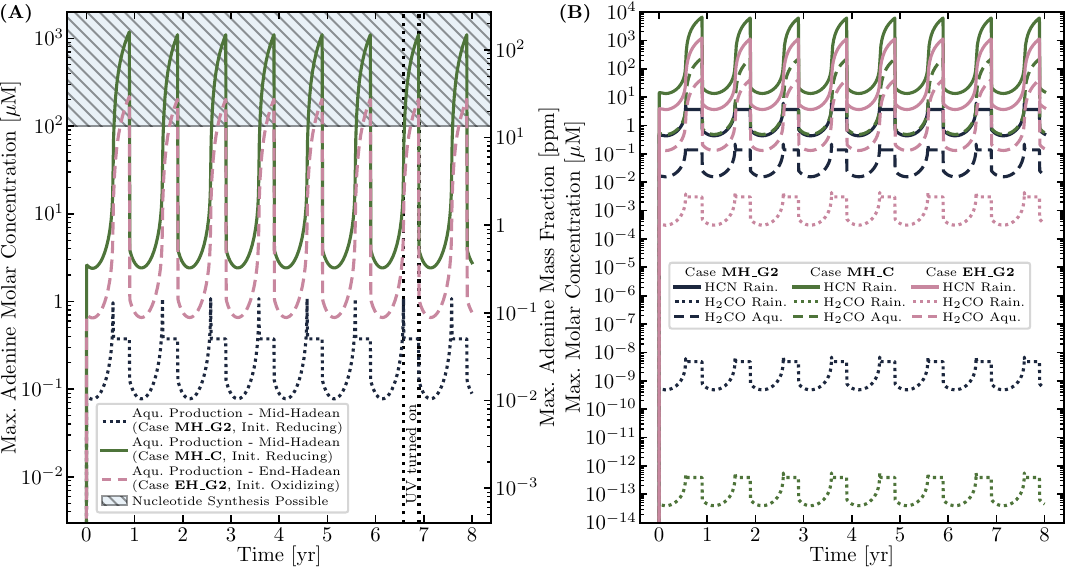}
    \caption{Concentrations of prebiotic molecules over time resulting from the warm little pond (WLP) cycling model \citep{Pearce2017,Pearce2022}. Inflow fluxes for \ce{HCN} and \ce{H2CO} are derived from atmospheric rain-out as a result of the photochemistry model. Panel \textbf{A} shows a comparison of the maximum resulting adenine concentrations for some of the most productive cases considered in this work, MH\_G2, MH\_C, and EH\_G2. In several of these cases, sufficient amounts of adenine are reached for nucleotide synthesis, as shown by experimental studies \citep{Ponnamperuma1963,Fuller1972,Powner2009,Becker2016,Saladino2017,Becker2018,Nam2018,Teichert2019}. Aqueous production is sourced from atmospheric rain-out of \ce{HCN} multiplied by experimental yields of adenine \citep{Oro1961,Wakamatsu1966,Hill2002,Pearce2022}. The pond cycles through 6 months of wet and 6 months of dry conditions. Sinks for the concentration are photodissociation by UV light, hydrolysis, and seepage through pores at the bottom of the pond. Between the two vertical dotted lines is the phase of the cycles where UV irradiation is turned on while the pond is dried out. Panel \textbf{B} shows concentrations for several prebiotic molecules including accumulating \ce{HCN} and \ce{H2CO} from rain-out and subsequent aqueous synthesis of \ce{H2CO} for the same three cases. Please note that the maximum pond concentrations of prebiotic biomolecules in case EH\_G2 in an initially oxidizing atmosphere are comparable to the other two cases shown with initially reducing conditions.}\label{fig:WLP_compare}
\end{figure*}

Figure~\ref{fig:WLP_compare} gives a concise overview of the most productive scenarios considered in this study. It can be seen that significantly high concentrations of prebiotic biomolecules in WLPs are possible even in an initially oxidizing environment. In case EH\_G2, serpentinization provides high enough fluxes of \ce{H2} and \ce{CH4} to exploit the initially \ce{CO2}-rich atmosphere for its carbon and subsequent \ce{HCN} synthesis. This highlights that an initially oxidizing atmosphere is a favorable scenario for highly active serpentinization in the Hadean. The resulting pond concentrations are more than two orders of magnitude higher than an initially reducing scenario (case MH\_G2) and less than one order of magnitude lower than the most effective scenario overall, a purely carbonaceous bombardment in the MH (case MH\_C).

Another interesting phenomenon can be clearly observed in Figure~\ref{fig:WLP_compare}(A) if one takes a closer look at the shape of the curves in cases MH\_C and EH\_G2 and compares them with case MH\_G2. The vertical dotted lines indicate the phase in which UV irradiation is turned on as the pond has dried out. In case MH\_G2, at the beginning of this phase, the concentrations reach their maximum in a sharp peak as the pond dries down to its minimum level of \SI{1}{\milli\meter}. Then UV irradiation is turned on and the concentration drops to a plateau in a steady state where production from \ce{HCN} influx and destruction from UV dissociation equilibrate. This behavior was always observed in the previous studies \citep[see also Figure~\ref{fig:WLP}]{Pearce2017,Pearce2022,Pearce2024}.

In contrast, in cases MH\_C and EH\_G2, we reach for the first time concentrations high enough to prevent this equilibration of influx and dissociation, and the concentration continues to rise in the nearly dry pond. The rate of adenine photodestruction was measured to be \SI{1e-4}{\per\photon} \citep{Poch2015,Pearce2022}. This sink rate is exceeded by synthesis from \ce{HCN} in cases MH\_C and EH\_G2 (see Figure~\ref{fig:WLP_compare}(A)) as well as in cases MH\_G2C, MH\_G2EC, and EH\_G2E (not shown) in the present study. The remaining sink terms that cap biomolecule abundances are seepage through pores at the bottom of the pond and hydrolysis. This effectively renders destruction by UV light negligible in the context of prebiotic synthesis in WLPs if high enough synthesis rates can be achieved.

\subsubsection{No Seepage}\label{sec:seepage_off}

It might be possible that after some time the pores at the bottom of the pond became clogged due to adsorption of biomolecules on the mineral surfaces or deposition of amphiphiles and mineral gels \citep{Hazen2010,Deamer2017,Damer2020}. In the wet-dry pond cycling model, this situation can be represented by turning off the seepage sink term. Table ~\ref{tab:results_WLP_no_seepage} summarizes the resulting maximum yields of prebiotic biomolecules following from such a model after allowing the molecules to accumulate for \SI{10000}{\year}.

For adenine, guanine, and cytosine, pond concentrations do not increase significantly. This is due to their relatively high rates of hydrolysis, which has become the dominant sink term without seepage \citep[for details, see Table~A6 in][]{Pearce2022}. On the other hand, in the most productive case MH\_C, uracil reaches a maximum concentration of \SI{0.89}{\milli\Molar} (less than an order of magnitude increase over the scenario with seepage), and thymine reaches \SI{1.8}{\milli\Molar} (more than an order of magnitude increase). Since hydrolysis rates for 2-aminooxazole and ribose are not provided in the \citet{Pearce2022} model, their more than an order of magnitude increased concentrations of \SI{9.6}{\micro\Molar} and \SI{0.24}{\milli\Molar} in comparison to the simulations with seepage, respectively, represent potential maximum values in case MH\_C without seepage.

\section{Discussion}\label{sec:discussion}

When trying to estimate whether the Hadean Earth as modeled here was able to generate conditions suitable for the origins of life, the key question to answer is whether the concentrations of biomolecules such as nucleobases, ribose, and 2-aminooxazole are high enough to promote the synthesis of nucleotides, the monomers of RNA molecules. Experimental studies that have successfully demonstrated the formation of nucleotides from solutions of nucleobases, ribose, and phosphates required minimum nucleobase concentrations of $\sim\SI{100}{\micro\Molar}-\SI{100}{\milli\Molar}$ for nucleobases and $\sim\SI{8}{\micro\Molar}-\SI{15}{\milli\Molar}$ for ribose \citep{Ponnamperuma1963,Fuller1972,Powner2009,Becker2016,Saladino2017,Becker2018,Nam2018,Teichert2019}.

The maximum abundances for the most productive case MH\_C in Table~\ref{tab:results_WLP} reach the \si{\milli\Molar} range for the purines adenine and guanine, and reach the \SI{100}{\micro\Molar} range for the pyrimidines cytosine, uracil, and thymine. Furthermore, in the case of EH\_G2 in an initially oxidizing scenario, purines reach the \SI{100}{\micro\Molar} range and pyrimidines reach the \SI{10}{\micro\Molar} range. In the absence of seepage, the concentrations for the pyrimidines uracil and thymine can even reach the \SI{100}{\micro\Molar} range in case EH\_G2 (see Table~\ref{tab:results_WLP_no_seepage}). These concentrations are well within the range required for successful nucleotide synthesis in the experiments discussed above. Assuming maximum bombardment rates as presented in the supplementary results in Appendix~\ref{sec:supp_results_high_bomb_rate}, these maximum nucleobase concentrations can be enhanced by about an order of magnitude (see Table~\ref{tab:results_WLP_high_bomb_rate}).

With seepage turned on, ribose concentrations reach the \SI{100}{\nano\Molar} range in case MH\_C and the \SI{10}{\nano\Molar} range in case EH\_G2 (see Table~\ref{tab:results_WLP}), which is one to two orders of magnitude below the concentrations required for successful nucleotide synthesis. Without seepage, however, the concentrations reach almost the \SI{10}{\micro\Molar} range in case MH\_C and the \si{\micro\Molar} range in case EH\_G2 (see Table~\ref{tab:results_WLP_no_seepage}). These concentrations are sufficient to allow nucleotide synthesis, as shown in the experiments by \citet{Saladino2017}, but it must be noted that in these experiments ribose was not dissolved in water but in formamide.

At maxed-out bombardment rates and without seepage, as assumed for the supplementary results in Appendix~\ref{sec:supp_results_high_bomb_rate}, ribose concentrations reach the \SI{100}{\micro\Molar} range (see Table~\ref{tab:results_WLP_high_bomb_rate_no_seepage}), which is close to, but still falls short of, the \unit{\milli\Molar} concentrations required for nucleotide synthesis in \textit{aqueous} solution as performed in laboratory experiments \citep{Ponnamperuma1963,Fuller1972,Nam2018,Powner2009}. It should also be noted that reaching these \SI{100}{\micro\Molar} ribose concentrations would require a bombardment intensity comparable to impacts the size of present-day dwarf planets (\SI{2e25}{\gram}, \SI{2300}{\kilo\meter} diameter), as suggested by \citet{Zahnle2020,Wogan2023}.

This is the first model to show that the nucleobase and, without seepage, also the ribose concentrations generated by aqueous \textit{in situ} synthesis are potentially sufficient for nucleotide synthesis in WLPs.

\subsection{A New Hope: \ce{HCN} Production in Initially Oxidizing Environments}

We demonstrated for the first time that the synthesis of sufficiently high concentrations of nucleoside building blocks is possible even in an initially oxidized atmosphere rich in \ce{CO2}. Furthermore, in case EH\_G2, these concentrations are replenished by serpentinization and subsequent photochemistry in a continuous and stable manner, rather than relying on singular cataclysmic events as in the studies by \citet{Wogan2023,Zahnle2020} and others, who considered very large enstatite impacts as the source of reducing gases on the Hadean Earth. Further increases in ribose concentrations might be necessary, however, as it is not yet certain that nucleotide synthesis in aqueous rather than formamide solution in WLPs would be successful with the current results.

Comparing the initially reducing and oxidizing models in the geology only scenario in Figures~\ref{fig:photochem_geo}(MH\_G2)~and~(EH\_G2) leads to the conclusion that an initially oxidized and therefore \ce{CO2}-rich atmosphere might be necessary to fuel an effective \ce{HCN} synthesis. The critical point is when the Earth begins to develop an effective upwelling of ferrous iron-rich mantle material and simultaneously the first hydrosphere settles on the surface, forming liquid water oceans. The contact of this liquid water with the ferrous iron from the mantle provides a large reducing potential, which in turn can utilize the carbon in atmospheric \ce{CO2} when reacting with \ce{H2} and \ce{CH4}, promoting the formation of \ce{HCN}. This results in an order of magnitude higher \ce{HCN} level compared to the initially reducing situation. In cases MH\_G0.5/1/2 there is a lack of \ce{CO2}, and the \ce{HCN} level is directly correlated with the \ce{CH4} level as the main precursor and carbon source. This is consistent with previous models and the findings by \citet{Pearce2022}.

This paper has elucidated a new and hitherto unknown regime that allows \ce{CO2} to be exploited for its carbon and \ce{HCN} synthesis to reach new levels. We have discovered that it is serpentinization that produces the required high \ce{H2} and \ce{CH4} fluxes that make this possible.

These high fluxes might also more realistically represent the situation prevailing in the Hadean. Over most of the time evolution in model EH\_G2, \ce{CH4} and \ce{HCN} levels are still loosely correlated. However, in the time span of $\sim$\SIrange{200}{2000}{\year}, \ce{HCN} levels experience a brief but strong boost, as much of the carbon in the initial \ce{CO2} atmosphere is converted directly into \ce{HCN} via \ce{CH4}.

As a novel insight, our results suggest that a primary \ce{CO2}-rich atmosphere might even be beneficial as a carbon source if hydrothermal circulation reached deep enough into the crust in the Hadean. This might open the possibility of shifting the scenario in case EH\_G2 from the EH at \SI{4.0}{\giga\year} to the very beginning of the Hadean even before \SI{4.4}{\giga\year}, but only if the hydrosphere was already (at least temporarily) present then. Strongly active serpentinization might have already occurred in the earliest Hadean, when the very first primordial \ce{CO2}-rich atmosphere was still present. This would create an additional scenario, not considered in the present work, that predates even the MH models. The very first atmosphere on Earth might have been a primordial \ce{CO2}-dominated one, which formed by \ce{CO2} outgassing from the magma ocean of the forming Earth \citep{Zahnle2007,Miyazaki2022,Johansen2023,Johansen2024}.

The time window of the most effective \ce{HCN} production would then be limited by the time when liquid water forms on the surface and the time scale of the subduction of atmospheric \ce{CO2} into the Earth's mantle. According to \citet{Miyazaki2022} this can be as short as 160 million years after the onset of plate tectonics. In extended models with HCD up to \SI{2.0}{\kilo\meter} (see Figure~\ref{fig:H2_mantle_flux}), \ce{CO2} was flushed into the mantle in only \SI{82.75}{\mega\year}. If most of the prebiotic precursors for the origins of life in a scenario analogous to case EH\_G2 were prepared before \SI{4.4}{\giga\year}, a rapid settling of the hydrosphere \citep{Abe1993,Korenaga2021} after the onset of plate tectonics is required. This would allow a \ce{CO2} atmosphere to be exploited for its carbon before it is flushed into the mantle. In contrast, the new models by \citet{Guo2025} employ a different scenario with a linearly decreasing plate velocity throughout the Hadean, which would allow for only \SI{90}{\bar} of \ce{CO2} to be removed from the atmosphere into the mantle \citep[this specific information is only available in the preprint version of the article:][]{Guo2024}.

There is no need to wait until the \ce{CO2}-rich atmosphere is gone to have an efficient prebiotic synthesis, as it might even be a favorable environment now to set the stage and the right conditions for the origins of life. This might make it possible to set the time for the emergence of the RNA world to only hundreds of millions of years or even much less after the formation of the Earth or $>\SI{4.4}{\giga\year}$ from today. It might be interesting to explore this idea further with in-depth modeling in future work.

Assuming a HCD of \SI{2}{\kilo\meter} in the Hadean, serpentinization of upwelling mantle material at mid-ocean ridge spreading zones results in a \ce{CH4} flux of \SI{1.12e10}{\molecules\per\centi\meter\squared\per\second} (see Table~\ref{tab:source_terms}), which corresponds to a global \ce{CH4} flux of $\sim\SI{3}{\tera\mole\per\year}$. \citet{Thompson2022} compiled an extensive list of abiotic \ce{CH4} sources and compared them to the present-day flux caused by biological activity to assess the potential of \ce{CH4} to serve as a biosignature if detected in exoplanet atmospheres using the James Webb Space Telescope. \citet{Kasting2005} estimated for serpentinization at spreading zones a global \ce{CH4} flux of \SI{1.5}{\tera\mol\per\year} during the Hadean, and \citet{Catling2017} a flux of \SI{0.03}{\tera\mol\per\year} for the present-day Earth, using a similar estimation method as in the present study (see Appendix~\ref{sec:mantle_model_assumptions}), starting with an expected \ce{H2} flux and combining it with a measured \ce{CH4}/\ce{H2} ratio in vent fluids.

For the slightly different scenario of serpentinization located at subduction zones, \citet{Fiebig2007} estimated an Archean flux of \SIrange{40}{80}{\mega\tonne\per\year}, equivalent to \SIrange{2.5}{5}{\tera\mol\per\year}. It is expected that subduction-related serpentinization will result in higher \ce{CH4} fluxes than spreading zone-related serpentinization, since the subducting oceanic plate drags a lot of water deep into the mantle, and serpentine has been found to be present at depths of $\sim\SIrange{5}{200}{\kilo\meter}$ by seismic velocity measurements on present-day Earth \citep{Reynard2013,Guillot2015}. However, since the presence and extent of plate tectonics and related subduction activity as found on the present-day Earth is still in question for the Hadean Earth \citep{Chowdhury2023,Tarduno2023}, we consider spreading zone serpentinization to be a more robust scenario for estimating \ce{CH4} fluxes on the Hadean Earth.

Our considered flux of $\sim\SI{3}{\tera\mole\per\year}$, about twice the flux estimated by \citet{Kasting2005}, fits well within the order of magnitude of the estimated global \ce{CH4} fluxes in the other studies, and thus appears to be a robust estimate valid for the Hadean Earth. Biological activity on Earth today results in \SI{30}{\tera\mole\per\year} of \ce{CH4} emission \citep{Jackson2020,Thompson2022}, and therefore \ce{CH4} remains a valid biosignature even for a Hadean Earth with a rather large HCD as considered here.

It must be acknowledged that the new models by \citet{Guo2025} \citep[more comprehensive information can be found in the preprint version of the article:][]{Guo2024} introduce a linearly decreasing plate velocity throughout the Hadean, in contrast to the \citet{Miyazaki2022} models used in the present work, which use a constantly high plate velocity throughout the Hadean. \citet{Guo2024} argue that a decreasing plate velocity results from a three times higher surface heat flux, and consider it more reasonable because convective mixing gradually destroys the initial chemical heterogeneities, allowing the rapid and constant plate velocities as used by \citet{Miyazaki2022}. This would imply that \ce{CO2} would be flushed into the mantle more slowly by seafloor weathering, and that ferrous iron would well up at a slower rate at mid-ocean ridges, possibly leading to lower surface fluxes of \ce{H2}, and consequently \ce{CH4}, in the Hadean. Due to the lost rock record, there is almost no way to validate which plate velocities better represent the Hadean and how the gas fluxes might have evolved. It could be interesting to model how these lower surface fluxes of reducing gases affect atmospheric photochemistry, rain-out, and biomolecule concentrations in WLPs, and see if there is a difference from current models in future studies.

The \citet{Miyazaki2022} model suggests that a heterogeneous mantle composition and active serpentinization in the Hadean could provide abundant \ce{H2}. This in turn might create a strong reducing potential and enable the formation of prebiotic molecules inside hydrothermal vents \citep[see, e.g.,][]{Martin2006}. However, in the context of atmospheric \ce{HCN} synthesis and rain-out, oceanic dilution makes it difficult to achieve sufficiently high concentrations of \ce{HCN} and \ce{H2CO} from atmospheric sources for significant prebiotic synthesis in hydrothermal vent environments.

Serpentinization is the most stable and continuous source of reducing gases considered in our models. It is fed by a steady stream of iron-rich mantle material. Our model assumes steady fluxes and continuous evolution of the atmosphere. Therefore, the effect of serpentinization should be well captured in our simulations. We consider it to be the most continuous and reliable way to supply reducing gases to the Hadean atmosphere.

\subsection{Comparison with Impact Scenarios}

Impacts, on the other hand, although we have treated them as continuous in our models for simplicity, might actually represent singular cataclysmic events that cause sudden bursts of prebiotic synthesis in the atmosphere and ponds. Furthermore, they can completely reshape the composition of the atmosphere and even evaporate the hydrosphere.

Comprehensive and detailed models considering single impacts and their aftermath in the Hadean Earth's atmosphere have been published, e.g., by \citet{Zahnle2020,Wogan2023}, \citet{Citron2022,Itcovitz2022}, \citet{Benner2019a}, \citet{Genda2017a,Genda2017b}, and \citet{Sekine2003}. These studies all focus on an enstatite bombardment and are able to generate significant amounts of prebiotic molecules in the atmosphere and \ce{HCN} rain-out. The models by \citet{Wogan2023} resulted in maximum \ce{HCN} rain-out rates of \SI{e9}{\molecules\per\centi\meter\squared\per\second}, which is equivalent to $\sim\SI{1.4e-5}{\kilo\gram\per\meter\squared\per\year}$ of \ce{HCN}. This required an impactor larger than \SI{5e21}{\kilo\gram} (diameter of $\sim\SI{1330}{\kilo\meter}$). This would mean that most of the HSE excess was delivered to the Hadean Earth's crust and mantle in just this single impact.

In the cases MH\_C (reducing, carbonaceous bombardment) and EH\_G2 (oxidizing, serpentinization with \SI{2}{\kilo\meter} of HCD), our model reached maximum rain-out rates of \SI{5.8e-4}{\kilo\gram\per\meter\squared\per\year} and \SI{1.6e-4}{\kilo\gram\per\meter\squared\per\year}, respectively (see Table~\ref{tab:results_rain}). With a very concentrated bombardment in case EH\_G2EC\_HSE\_max (oxidizing, maxed-out bombardment rates) in the supplementary results in Appendix~\ref{sec:supp_results_high_bomb_rate}, even \SI{1.4e-2}{\kilo\gram\per\meter\squared\per\year} of rain-out were reached. This is one to three orders of magnitude higher than the models by \citet{Wogan2023}. However, the enstatite bombardment cases MH/EH\_E resulted in rain-out rates of \SI{6.8e-11}{\kilo\gram\per\meter\squared\per\year} and \SI{3.1e-15}{\kilo\gram\per\meter\squared\per\year}, respectively, which is over \numrange{5}{9} orders of magnitude lower than in the models by \citet{Wogan2023}. This is to be expected, since our model treats this bombardment only as a continuous background source of \ce{H2}, and we have chosen only a moderate bombardment intensity \citep[see Appendix~\ref{sec:bombardment_model},][]{Pearce2022} in the main results.

\citet{Wogan2023} specifically point out that these high \ce{HCN} rain-out rates in their model require high levels of \ce{H2} and \ce{CH4} in the atmosphere, leading to strong greenhouse warming reaching surface temperatures $>\SI{360}{\kelvin}$. This might be a problem for the longevity of the first forming RNA molecules. In the present work, the carbonaceous bombardment might be a less drastic alternative to avoid too high atmospheric \ce{H2} and \ce{CH4} concentrations while still providing the highest \ce{HCN} rain-out of all modeled scenarios, even in the initially oxidizing scenario EH\_G2EC\_HSE\_max in the supplementary results (Appendix~\ref{sec:supp_results_high_bomb_rate}).

\section{Conclusions}\label{sec:conclusions}

\begin{figure*}[t]
    \centering
    \includegraphics[width=\textwidth]{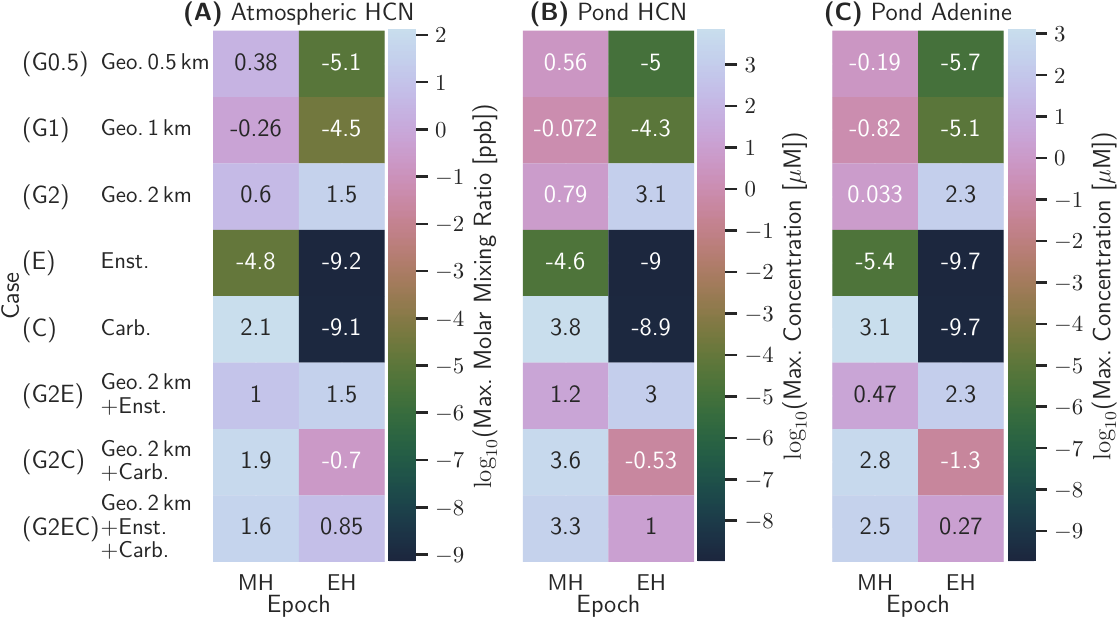}
    \caption{Heatmap showing a visual representation of the maximum simulated \ce{HCN} atmospheric mixing ratios and biomolecule pond concentrations in Tables~\ref{tab:results_atm}~and~\ref{tab:results_WLP}. Panel \textbf{A} shows the logarithm of the maximum molar mixing ratio of \ce{HCN} reached in the Hadean atmosphere over all considered cases and both epochs of the mid-Hadean (MH) with an initially reducing atmosphere at \SI{4.4}{\giga\year} as well as the end-Hadean (EH) with an initially oxidizing atmosphere at \SI{4.0}{\giga\year}. Panel \textbf{B} shows the logarithm of the maximum concentration of \ce{HCN} and panel \textbf{C} shows the logarithm of the maximum concentration of the biomolecule adenine reached in the pond cycling model over all scenarios. A heatmap plot is a graphical representation that uses color to show the magnitude of values. The representation helps visualize patterns, correlations, and distributions within our dataset across both epochs and all cases considered, where brighter colors represent higher values, and darker colors represent lower values.}\label{fig:heatmap_HCN}
\end{figure*}

To provide an overview of the potential for prebiotic synthesis in the many scenarios considered here, and to make this visually easy to grasp, Figure~\ref{fig:heatmap_HCN} shows the peak concentrations of \ce{HCN} in the atmosphere and WLPs, as well as the biomolecule adenine in WLPs, in a heatmap comparing all cases considered. It clearly shows that serpentinization with an HCD of \SI{2}{\kilo\meter} is productive in prebiotic synthesis regardless of the initial conditions. In the initially reducing conditions of scenario MH (mid-Hadean), carbonaceous bombardment is able to promote the most productive prebiotic synthesis, while enstatite bombardment alone, as modeled here, leads to significantly lower \ce{HCN} and biomolecule concentrations of up to five orders of magnitude less. In the initially oxidizing conditions of scenario EH (end-Hadean), serpentinization with an HCD of \SI{2}{\kilo\meter} is required to achieve comparable biomolecule concentrations as in the MH scenario.

For the first time, our model of the Hadean Earth has been able to justify sufficiently high abundances of nucleobases and the sugar ribose in WLPs, in the range used by experiments showing successful formation of nucleotides, the monomers of RNA. Nevertheless, the notoriously low yield of ribose in the formose reaction, even when using very effective catalysts such as \ce{Ca(OH)2}, remains a challenge, and further advanced theoretical modeling efforts and experimental studies are needed to make a confident statement that ribose abundances are sufficient, as this study finds that they are not yet.

We find that between 1.0 and \SI{2.0}{\kilo\meter} of HCD, serpentinization becomes capable of reducing even a highly oxidized atmosphere (initially \SI{90}{\percent} \ce{CO2}). This is the requirement to allow for an effective synthesis of the building blocks of life, i.e., to start from atmospheric \ce{HCN} and seed the early Earth with the ingredients for the origins of life. This eliminates the need for a primary reducing atmosphere, since serpentinization is a continuous and stable source of reducing gases over long timescales. An initial \ce{CO2}-rich atmosphere might even be advantageous, since it can be exploited for its carbon in photochemical \ce{HCN} synthesis in the atmosphere, as shown in the simulation of case EH\_G2. It is crucial to assume serpentinization in a manner appropriate for the Hadean Earth, as estimates of its activity based on the present-day Earth might severely underestimate its potential.

UV irradiation is usually considered a threat to the stability of biomolecules formed by prebiotic synthesis and an argument against considering WLPs as one of the possible sites for the origins of life. With this model, we are able for the first time to achieve rates of biomolecule formation by prebiotic synthesis in WLPs that are high enough to make the influence of UV photodissociation negligible. Despite the fact that water bodies on the first land masses on Earth are exposed to UV, it might not pose a significant threat to the synthesis of the building blocks of the RNA world in the most productive scenarios considered here.

Exogenous delivery of biomolecules by meteorites from space resulted in maximum pond concentrations about two orders of magnitude lower \citep{Pearce2017} than the most productive scenario considered in this work. A meteorite falling into a pond during its wet phase, or into ponds that do not completely dry out, might still be an interesting way to temporarily boost biomolecule concentrations and increase yields for nucleoside/-tide synthesis and polymerization, pushing toward the emergence of the RNA world.

Formation of atmospheric \ce{HCN} by lightning has been included in the model (see Section~\ref{sec:methods}), but produces negligible amounts compared to UV radiation \citep[as also shown in][]{Pearce2022}.

Furthermore, rain-out of \ce{H2CO} is negligible, and ribose is predominantly formed by \ce{H2CO} synthesized aqueously from \ce{HCN} in WLPs. These results confirm the findings of \citet{Pearce2022}.

Carbonaceous bombardment and immediate \ce{HCN} synthesis during impact \citep{Kurosawa2013} might be the most potent source of biomolecules and the building block of life in WLPs. Recent evidence that the HSE excess of the Earth's crust and mantle might have a significant contribution from carbonaceous chondrites \citep{Varas-Reus2019,Budde2019,Hopp2020,Fischer-Godde2020} brings this type of impactor back into the picture as a contributor to the reducing inventory of the Hadean atmosphere. In an initially reducing atmosphere (\SI{90}{\percent} \ce{H2}) carbonaceous impactors can fully unfold their potential to provide \ce{HCN} to the atmosphere and WLPs, whereas in an initially oxidizing atmosphere the \ce{N2}/\ce{CO2} ratio in the atmosphere rises too slowly and an already advanced oxidation with high \ce{O2} levels prevents this. Another advantage of the carbonaceous bombardment scenario is that high atmospheric \ce{H2} and \ce{CH4} levels are not required to achieve high biomolecule concentrations in WLPs, potentially preventing a strong greenhouse effect and allowing habitable temperatures on Hadean Earth, setting the stage for the emergence of the RNA world and the origins of life.

The authors would like to thank Kai Kohler and Oliver Trapp for performing formose sugar synthesis experiments for us and providing yields for ribose synthesis from \ce{H2CO}. We thank Wolfgang Bach and Mario Trieloff for very fruitful discussions on geological processes, which were instrumental in properly discussing and implementing plausible conditions for the Hadean Earth. A big thank you goes to all the developers of the open-source software Blender, a 3D computer graphics tool, Inkscape, a vector graphics editor, and GIMP, a raster graphics editor, as well as countless creators of tutorials on YouTube for allowing K.P. to render Figure~\ref{fig:hadean_earth_scheme}. We thank the referee for a constructive report that helped to clarify the manuscript. K.P. acknowledges financial support by the Deutsche Forschungsgemeinschaft (DFG, German Research Foundation) under Germany's Excellence Strategy EXC 2181/1 – 390900948 (the Heidelberg STRUCTURES Excellence Cluster). K.P. is a fellow of the International Max Planck Research School for Astronomy and Cosmic Physics at the University of Heidelberg (IMPRS-HD). T.K.H. and D.A.S. acknowledge financial support by the European Research Council under the Horizon 2020 Framework Program via the ERC Advanced Grant Origins 83 24 28. B.K.D.P. is supported by the 51 Pegasi b Postdoctoral Fellowship. R.E.P. is supported by an NSERC Discovery Grant. R.E.P. also gratefully acknowledges the Max Planck Institute for Astronomy for partial support of his sabbatical leave in 2022/23 when this work was started. Funding from Vector Stiftung is greatly acknowledged under project ID P2023-0152.

\clearpage

\appendix
\restartappendixnumbering

\section{Background: Early Earth's Mantle}\label{sec:mantle}

The structure, composition, and mechanism active in the Hadean Earth's mantle and crust are mostly unknown, as there is only little in the rock record that survived until today. The rocks embedded in the oldest cratons have been metamorphosed and lost over time, but the highly resistant mineral zircon has been preserved from the Hadean to the present \citep{Harrison2009}. Nevertheless, the conclusions drawn from their investigation can vastly differ. For example, when trying to understand if plate tectonics was active on the Hadean Earth, studies on zircons differing in their methodologies conclude either in favor of active tectonics \citep{Chowdhury2023} or the opposite \citep{Tarduno2023}. The evidence for active plate tectonics in the Hadean eon ($>$\SI{4}{\giga\year} ago) seems scarce and elusive, and only in the Eoarchean (\SIrange{4}{3.6}{\giga\year} ago) it seems certain that subduction was operational \citep{Hastie2023}.

\citet{Miyazaki2022} predict a Hadean Earth with a heterogeneous mantle composition and a thin crust as a consequence of mantle differentiation during magma ocean solidification. Because of the high solubility of water in magma, the primordial mantle was wet, possibly resulting in low melt viscosity and fractional crystallization of the cooling magma ocean \citep{Miyazaki2019,Dorn2021,Luo2024}. During magma ocean solidification, \ce{Mg} and \ce{SiO2}-rich material would accumulate in the lower mantle, whereas \ce{Fe}-rich and denser blobs would accumulate in the upper mantle and crust. Since the scale length of re-mixing is shorter than \SI{100}{\kilo\meter} \citep{Miyazaki2019}, this leads to a heterogeneous structure with a mostly depleted mantle and creates a thin crust and lithosphere. Conversely, a more homogeneous (pyrolitic) composition would lead to a very thick crust and lithosphere, decreasing the likelihood of active plate tectonics.

The heterogeneous Hadean mantle predicted by \citet{Miyazaki2022} allows for active plate tectonics and rapid crustal subduction with velocities around \SI{50}{\centi\meter\per\year} \citep[see also][it should be noted that, in contrast, the new models by \citet{Guo2025} employ a linearly decreasing plate velocity throughout the Hadean]{Sleep2001,Zahnle2007,Sleep2014}. In the case of today's Earth, crustal velocities measured as the spreading rates of mid-ocean ridges are much slower, with average values of $\sim\SIrange{2}{5}{\centi\meter\per\year}$ \citep{Parsons1982,Cogne2004}. Subduction allows the efficient sequestration of large amounts of carbonates, flushing \ce{CO2} out of the atmosphere. This might also be necessary to remove a \ce{CO2} atmosphere of \SIrange{110}{290}{\bar} at the beginning of the Hadean (after the Moon-forming impact). This massive atmosphere might have been outgassed from the global magma ocean with an initial \ce{CO2} mantle concentration of \SIrange{200}{500}{ppm}, based on present-day volatile budgets \citep{Hirschmann2009,Korenaga2017}. Converting this earliest oxidizing atmosphere to the more reducing conditions necessary for an effective prebiotic synthesis might require fast subduction rates, as in the heterogeneous mantle model by \citet{Miyazaki2022}. This scenario of a thick \ce{CO2} atmosphere in the earliest Hadean, flushed into the mantle within the first $\sim\SI{100}{\mega\year}$, is also supported in the review by \citet{Zahnle2007}.

The iron-enriched Hadean crust formed more olivine than the present-day oceanic crust, which in turn gives rise to a higher potential for the occurrence of the serpentinization reaction. This is a reaction of rocks in contact with water within hydrothermal vent systems. Olivine \ce{(Mg,Fe)2SiO4} reacts with water to form the minerals serpentine \ce{(Mg,Fe)_{3-Fe$^{3+}$}(Si,Fe)_{2-Fe$^{3+}$}O5(OH)4}, magnetite \ce{Fe3O4}, brucite \ce{(Mg,Fe)(OH)2}, and releases \ce{H2}. The ratio between the product minerals depends on the iron content of the olivine as well as the temperature, and \citet{Klein2013} describe it with the following general reaction:
\begin{equation}\label{eq:serp_general}
    \begin{aligned}
        &\ce{(Mg,Fe)2SiO4 + vH2O -> w(Mg,Fe)(OH)2}\\
        &\ce{+ x(Mg,Fe)_{3-Fe$^{3+}$}(Si,Fe)_{2-Fe$^{3+}$}O5(OH)4}\\
        &\ce{+ yFe3O4 + zH2_{\mathrm{(aq)}}},
    \end{aligned}
\end{equation}
with the generalized stoichiometric coefficients v, w, x, y, and z. Effectively, this reaction oxidizes ferrous (\ce{Fe^{2+}}) to ferric (\ce{Fe^{3+}}) iron, and reduces water to \ce{H2}. According to \citet{Klein2013}, the effective reaction can be summarized as
\begin{equation}\label{eq:serp_simple}
    \ce{2Fe^{2+}O + H2O -> Fe^{3+}2O3 + H2_{(\mathrm{aq})}}.
\end{equation}

\citet{Miyazaki2022} used the following simplification of Equation~\ref{eq:serp_general} in their calculation of the \ce{H2} output from the serpentinization reaction:
\begin{equation}\label{eq:serp_Yoshi}
    \begin{aligned}
        \ce{3Fe2SiO4 + 2H2O -> 2Fe3O4 + 3SiO2 + 2H2}, \\
        \ce{3Mg2SiO4 + SiO2 + 4H2O -> 2Mg3Si2O5(OH)4}.
    \end{aligned}
\end{equation}

The \ce{H2} resulting from the serpentinization reaction can further react to \ce{CH4} by combining it with \ce{CO2}. The methane is produced in a combination of the reverse water-gas shift reaction
\begin{equation}\label{eq:water_gas_shift}
    \ce{CO2 + H2 -> CO + H2O}
\end{equation}
and the Fischer-Tropsch reaction
\begin{equation}\label{eq:FT}
    \ce{CO + 3H2 -> CH4 + H2O},
\end{equation}
effectively following \citep{McCollom2007}:
\begin{equation}\label{eq:FT_eff}
    \ce{CO2 + 4H2 -> CH4 + 2H2O}.
\end{equation}

Another possibility is to form \ce{CH4} by direct reduction of \ce{CO2} with \ce{H2}, facilitated by mineral catalysts in the so-called Sabatier (methanation) reaction, a special case of the Fischer-Tropsch reaction \citep{Holm2015}. The chemical reaction is identical to Equation~\ref{eq:FT_eff}, except for the additional participation of catalysts, e.g. the minerals awaruite and chromite \citep{Bradley2016}.

These reactions only operate within the hydrothermal system at high pressures and temperatures inside the smokers. Therefore, the \ce{CO2} involved in the reaction must be be supplied to the reaction inside the hydrothermal vent system, coming directly from the crust and mantle, not from the atmosphere. This mantle \ce{CO2} is either left over from the original inventory in the forming mantle or has been added by subduction from the atmosphere.

The prediction is to have smokers, active undersea volcanoes, ``at most of the seafloor'' in the early Hadean \citep{Miyazaki2022}. This possibly led to a reduced Hadean atmosphere due to the outgassed reaction products in the serpentinization and Fischer-Tropsch reactions (Equations~\ref{eq:serp_simple} and \ref{eq:FT_eff}).

Another model of serpentinization and subsequent \ce{H2} and methane formation was simulated by \citet{Guzman-Marmolejo2013}. Their study used present-day crustal spreading rates and \ce{FeO} oceanic crustal content to calculate the \ce{H2} production from serpentinization. They identified the abundance of \ce{CO2} as a limiting reactant due to its limited abundance in aqueous hydrothermal vent systems on present-day Earth. The resulting \ce{CH4} flux from \citet{Guzman-Marmolejo2013} was used by \citet{Pearce2022}. Since the Hadean Earth's mantle and crust might have been very different from the present-day case represented in the study by \citet{Guzman-Marmolejo2013}, \citet{Pearce2022} might have underestimated the geological influence on the Hadean Earth's atmosphere. The ultramafic composition rich in fayalite (\ce{Fe^{2+}}-rich olivine group) of the Hadean Earth's crust in both homogeneous and heterogeneous cases by \citet{Miyazaki2022} might be more representative of the situation prevailing on the young Earth.

\citet{Klein2013} showed in simulations that olivine-rich rocks (peridotite) produce higher amounts of \ce{H2_{(aq)}} during serpentinization than pyroxene-rich rocks (pyroxenite). Therefore, the iron-rich Hadean mantle predicted by \citet{Miyazaki2022} entails a higher potential in reducing the Hadean atmosphere than the present-day scenario by \citet{Guzman-Marmolejo2013}.

\citet{McCollom2007} reviewed \textit{in situ} measurements of hydrocarbons including \ce{CH4} in hydrothermal vent fluids. They concluded that isotopic studies of $\delta\ce{^{13}C}$ and $\delta\ce{^{2}H}$ for \ce{CH4} in combination with compositional studies allow distinguishing sites with abiotic emissions from those dominated by methanogenic organisms. Unsedimented mid-ocean ridges such as ``Lost City'' or ``Rainbow'' are assumed to be free of any biological communities or sedimented biological material. Otherwise, these organisms and their remains emit biotic \ce{CH4} resulting from metabolic activity or thermogenic decomposition of biological material. These sites are hosted in serpentinites and are most likely dominated by abiotic \ce{CH4} emission due to serpentinization. Smokers with and without biological activity are clearly distinguishable from one another. This allows getting \textit{in situ} measured emission rates of abiotic \ce{CH4} in the natural environment, which might be representative of smokers active on the Hadean Earth prior to abiogenesis. \citet{Cannat2010} conclude that the most feasible value for the \ce{CH4}/\ce{H2} ratio is $\sim\SI{15}{\percent}$ as derived from \textit{in situ} measurements in fluids emitted from hydrothermal vents off-axis to the Mid-Atlantic Ridge \citep{Charlou2002,Kelley2005}, representing these uninhabited smokers.

\citet{Thompson2022} have compiled an impressive list of geological methane production processes and corresponding studies comparing various abiotic sources. We refer the reader to this comprehensive compilation, but note that their own calculations, and many of those cited, assumed present-day Earth conditions.  The focus of \citet{Thompson2022} was on methane as a biosignature on exoplanets, not on the evolutionary state of the Hadean Earth. Nevertheless, some of the cited studies make predictions for the Hadean Earth and were also used in the discussion Section above for comparison with the present work.

The prebiotic synthesis of biomolecules in WLPs depends on the presence of land masses, which depend on the growth of continental crust and sea level in the Hadean. There is evidence that already in the Hadean the first volcanic islands might have formed near subduction zones between oceanic plates, and these were pushed together by plate tectonics to form the first continental crust \citep{McCulloch1993,Menneken2007,Keller2018,McCoy-West2019,Guo2020,Chowdhury2023,Guo2023}. The question is whether or not these volcanic islands and continents rose to the surface above sea level. This depends on the volume of the oceans, which depends on the de- and regassing of water from the mantle over time. Evidence from zircons to constrain the behavior of the Hadean mantle and its interaction with water is scarce and limited to a single site (Jack Hills in Western Australia), so one must rely mostly on theoretical models. However, some proposed scenarios include at least temporary volcanic islands \citep{Bada2018} or possibly even continents \citep{Korenaga2021} above sea level in the Hadean, which might be sufficient for the presence of Hadean WLPs. Conversely, \citet{Russell2021} argues that the Hadean Earth was a water world.

It must be acknowledged that there is still no consensus found in the geoscientific community regarding the exact characteristics and course of magma ocean solidification, the onset of plate tectonics, plate velocities, water budgets within the mantle, water degassing, and continental growth in the Hadean \citep[and refs.\ therein]{Korenaga2021}. The rock record provides reliable evidence only for the Archean, which might or might not have been very different from the Hadean. Theoretical models of the Hadean Earth depend on many assumptions, leading to large variations within a model, and can lead to very different results between models.

\section{Background: Late Veneer/Late Heavy Bombardment/Late Accretion}\label{sec:late_veneer}

During its formation, the Earth's interior was molten and differentiated. During this differentiation process, metallic and siderophile elements sink to the core, while lithophile elements remain in the mantle. Therefore, HSEs should be entirely absent in the Earth's mantle, following the metal-silicate equilibrium in the Earth's interior \citep{Becker2006,Brenan2009}. However, the HSE concentration in the upper mantle is much higher than predicted by their partitioning behavior. Since experiments have shown that differentiation and equilibration in the Earth's interior cannot explain this HSE excess \citep{Mann2012}, external delivery by impactors is proposed to explain this excess. It is assumed that impactors delivered these HSEs to the upper crust and mantle in the Hadean and Archean. In particular, chondritic material, which unlike Earth is undifferentiated and therefore siderophile-rich, could have composed the impactors, which remained in the upper mantle to explain this HSE excess.

In the geosciences, the term ``late veneer'' is commonly used to refer to the late delivery of chondritic material to explain this HSE excess in the Earth's mantle, while in planet formation theory, the terms ``late accretion'' or ``late heavy bombardment'' refer to a period of impacts distinct from the formation of the Earth \citep{Morbidelli2015}, sometimes between $\sim\SIrange{4.5}{3.5}{Gyr}$ ago \citep{Chyba1990,Chyba1992}. Based on the lunar cratering record, the term ``late heavy bombardment'' or ``lunar cataclysm'' was coined to address the possibility that this period of impacts could have been as short as $\sim\SI{150}{Myr}$ or less \citep{Strom2005} around $\sim\SIrange{3.8}{3.9}{Gyr}$, but this is debated \citep{Zahnle2007}. A combination of a more continuous ``late veneer'' together with a distinct ``late heavy bombardment'' phase is also considered \citep{Morbidelli2015}. Whether or not this later addition of material is distinct from the Moon-forming impact is unclear \citep{Morbidelli2015,Hopp2020}, but the term ``late accretion'' is defined as all the material added after the Moon-forming impact.

After a long period in which the scientific community favored a carbonaceous late veneer \citep[see, e.g.,][]{Chou1983}, many studies concluded in a late veneer with a composition similar to enstatite chondrites \citep[see, e.g.,][]{Fischer-Godde2017,Dauphas2017,Bermingham2017}. Enstatite chondrites are rich in iron, poor in volatiles and carbon and might originate from the innermost region of the solar system. An estimate for the amount of accreted enstatite material necessary to explain the HSE excess in the Earth's mantle is about $\SI{0.34}{\percent}M_{\oplus}$ \citep{Walker2009}.

Conversely, recent isotope ratio measurements of \ce{Ru} in Eoarchean rocks support a significant contribution of carbonaceous chondrites to the late veneer \citep{Fischer-Godde2020}. These rocks are from southwest Greenland and might comprise pre-late veneer material of $>\SI{3.7}{Gyr}$ age, which might not have fully equilibrated with the rest of the upper mantle. The combination of this pre-late veneer mantle material with carbonaceous chondrites is consistent with the composition of the modern mantle. To explain the full HSE excess in the modern mantle, an upper bound of $\SI{0.3}{\percent}M_{\oplus}$ of carbonaceous chondritic material of class CM is sufficient, without the need for enstatite impactors. \citet{Varas-Reus2019} come to the same conclusion of a potentially carbonaceous late veneer, but looking at \ce{Se} isotopes. They give upper bounds of \SI{0.15}{\percent} $M_{\oplus}$ for CI, and \SI{0.26}{\percent} $M_{\oplus}$ for CM chondritic material to explain the full HSE excess. Further studies considering the relative abundances of \ce{Se}, \ce{Te}, and \ce{S} also concluded in a carbonaceous chondrite-like late veneer \citep[CI and CM chondrites,][]{Wang2013,Braukmüller2019}.

Finally, a combined \ce{Mo}-\ce{Ru} isotope analysis favors a heterogeneous late veneer with a contribution from both reservoirs \citep{Hopp2020}. In particular, the examination of the \ce{Mo} isotope \ce{^{94}Mo} across all meteorite populations leads to the conclusion that i) the late veneer had a mixed composition and/or ii) a giant impactor, possibly the Moon-forming one, was carbonaceous and/or iii) of mixed enstatite-carbonaceous composition \citet{Hopp2020}. All three scenarios are consistent with the \ce{Mo}-\ce{Ru} isotope signatures found. \citet{Hopp2020} ruled out a pure carbonaceous chondrite-like late veneer due to the $\varepsilon\ce{^{100}Ru} \approx 0$ isotope excess of the bulk silicate Earth (BSE) and $\varepsilon\ce{^{100}Ru} < 0$ of carbonaceous chondrites. However, the newly found positive $\varepsilon\ce{^{100}Ru}$ of the Eoarchean mantle by \citet{Fischer-Godde2020} relaxes this constraint, as this positive Eoarchean mantle signature might have been mixed with the negative signature of carbonaceous chondrites in the modern BSE mantle. Therefore, a purely carbonaceous late veneer is a possibility. \citet{Bermingham2025} also cannot exclude a contribution from carbonaceous material.

\subsection{Enstatite Iron-Rich Impactors}

If the late veneer was composed of a significant fraction of iron-rich impactors, the idea is that the exogenously introduced iron is able to reduce oxidized atmospheric gases, especially water. The reduced products of the reaction, mainly \ce{H2}, are outgassed and are able to convert the atmosphere to a more reduced state, which is beneficial for many prebiotic synthesis mechanisms.

A simple representation of this reduction reaction, where iron reduces water to hydrogen, according to \citet{Zahnle2020}, is given by
\begin{equation}\label{eq:iron_red}
    \ce{Fe + H2O -> FeO + H2}.
\end{equation}
In combination with the water-gas shift and the Fisher-Tropsch reactions in Equations~\ref{eq:water_gas_shift}~and~\ref{eq:FT}, this allows for the synthesis of methane if the temperature is high enough. These reactions are inhibited at low temperatures in the gas phase \citep{Zahnle2020}, but the energy dissipated at impact might allow for efficient synthesis of methane. \citet{Sekine2003} suggest that the impactor material leaves the atmosphere after impact and might re-enter it in the form of iron and nickel condensates. On the huge collective surface of these fine-grained condensates, a catalyzed version of the Fisher-Tropsch reaction might increase the yield of outgassed \ce{CH4}. \citet{Peters2023} showed in experiments that iron meteorites and iron-rich chondritic meteorites as well as volcanic ash can also catalyze these reactions.

\subsection{Carbonaceous Impactors}\label{sec:cc_impactors}

\citet{Kurosawa2013} showed in laboratory experiments that \ce{HCN} is synthesized during the impact of carbonaceous meteorites. They used hypervelocity impacts of a polycarbonate projectile fired from a gas gun and laser ablation experiments on graphite to study \ce{HCN} synthesis in an atmosphere of \ce{N2}, \ce{H2O}, and \ce{CO2} in varying mixing ratios at about \ce{1}{bar} total pressure. As the carbon is vaporized, it reacts to form short hydrocarbons and \ce{CN} radicals, which recombine to form several nitrile compounds and \ce{HCN}, according to the simplified reaction equation 
\begin{equation}\label{eq:cc_hcn}
    \ce{C + N2 + H2O -> HCN + nitriles + hydrocarbons}.
\end{equation}
As the \ce{CO2} mixing ratio is increased, the synthesis of these products becomes more and more suppressed, but is still productive as long as the molar mixing ratio $\ce{N2}/\ce{CO2}$ is greater than one. As soon as $\ce{N2}/\ce{CO2} < 1$, \ce{HCN} synthesis stops completely. This can be seen in the data by \citet[Table~3]{Kurosawa2013}, where at a partial pressure of \SI{500}{\milli\bar} of \ce{N2} and \SI{530}{\milli\bar} of \ce{CO2} ($\SI{500}{\milli\bar}/\SI{530}{\milli\bar} < 1$) no \ce{HCN} could be detected anymore. Therefore, this process is only feasible for impactors entering a \ce{CO2}-poor atmosphere. For $\ce{N2}/\ce{CO2} > 35.4$, up to \SI{2.8}{mol\percent} of the carbon is converted into \ce{HCN}, making it an interesting source term for a carbon-rich bombardment. This threshold of 35.4 results from experimental runs with partial pressures of \SI{920}{\milli\bar} of \ce{N2} and \SI{26}{\milli\bar} of \ce{CO2} ($\SI{920}{\milli\bar}/\SI{26}{\milli\bar} = 35.4$), approaching the maximum carbon to \ce{HCN} conversion as seen in experiments without any \ce{CO2} in the experiment \citep[Table~3]{Kurosawa2013}.

\section{Model: Timeline}\label{sec:time}

The formation history and composition of the early mantle determined the composition and temperature of the first primordial atmosphere in the early Hadean. The main greenhouse gases of interest are \ce{CO2} and water vapor, which depend on their partitioning between the atmosphere and the mantle. There seems to be a consensus that at the beginning of Earth's history, enormous amounts of \ce{CO2} were released from the mantle, resulting in several \SI{100}{\bar} in the atmosphere \citep{Zahnle2007,Miyazaki2022}. However, this thick \ce{CO2} atmosphere might not have lasted long, depending on how quickly and efficiently it was deposited as carbonates and subducted into the mantle by plate tectonics. If and when plate tectonics was active in the Hadean is hotly debated, as noted in Appendix~\ref{sec:mantle} \citep{Chowdhury2023,Tarduno2023}. If active, all the \ce{CO2} could have been flushed into the mantle within \SIrange{10}{200}{\mega\year} \citep{Zahnle2007, Miyazaki2022}, resulting in a strongly reduced atmosphere dominated by \ce{H2} at $\sim\SI{4.4}{\giga\year}$.

This \ce{H2} might be the remnant of the first primordial atmosphere accreted from the solar nebula that gave birth to the solar system. Serpentinization of ultramafic material upwelling in the mantle or the reduction by exogenously delivered chondrites further supplied this \ce{H2} atmosphere. This atmosphere might have been further reduced by the emission of \ce{CH4} in hydrothermal vents (see Equation~\ref{eq:FT_eff}) and the synthesis of \ce{HCN} by the carbonaceous portion of the late veneer (see Equation~\ref{eq:cc_hcn}), which is always productive as the ratio $\ce{N2}/\ce{CO2}>1$.

There seems to be no consensus on whether water was efficiently degassed as the magma ocean solidified on the surface of the nascent Earth. For example, \citet{Zahnle2007} suggested that most of the water was partitioned into surface reservoirs of the magma ocean and degassed from the freezing mantle into the atmosphere. This is based on the assumption that the solidified region of the mantle contained very little hydrated minerals. Oxygen isotope signatures in zircons formed at $\sim\SI{4.4}{\giga\year}$ and younger confirm that they were chemically altered by liquid water \citep{Wilde2001,Mojzsis2001,Valley2005,Cavosie2005}. Whether this evidence means that the mantle is mostly dry is not clear.

On the other hand, \citet{Miyazaki2022} draw another scenario where most of the water remained in the mantle, assuming that the percolation of volatiles through the porous melt of the freezing magma was too slow and trapped most of them \citep{Hier-Majumder2017}. This has strong implications for the viscosity of the Hadean mantle. Low viscosity and thus early active plate tectonics promoted very fast renewal of the Hadean Earth's crust, which enhanced the ability of serpentinization to reduce surface water to \ce{H2} and \ce{CH4} (in combination with \ce{CO2} in the mantle), as well as the rapid sequestration of atmospheric \ce{CO2} into the Hadean mantle.

\citet{Pearce2022} constructed a corresponding atmospheric composition under the assumption of maintaining habitability at the surface (\SIrange{0}{100}{\celsius}). To achieve this, a P-T profile was generated with the 1D radiative transfer code petitRADTRANS \citep{Guillot2010,Molliere2019}. The corresponding set of parameters for a habitable atmosphere with an exemplary temperature of \SI{78}{\celsius} in the MH at \SI{4.4}{\giga\year} is summarized in Table~\ref{tab:epochs}.

Later, toward the end of the Hadean, the rock record indicates more oxidizing conditions again, although it does not date all the way back to \SI{4.0}{\giga\year} ago. For example, redox-sensitive elements in the greenstone belt at \SI{3.8}{\giga\year} seem to indicate that the mantle was as oxidized as it is today \citep{Aulbach2016}, resulting in volcanic emission of mostly oxidizing gases \citep{Holland1984,Catling2017,Wogan2020a,Wogan2023}. However, Archean metamorphosed mid-ocean ridge basalts and picrites (up to \SI{3.0}{\giga\year} old) show lower oxidation states \citep{Aulbach2016}. Thermodynamic calculations also show a decreasing reducing power of serpentinization reactions in upwelling ultramafic rocks over the Archean (\SI{3.5}{\giga\year} and younger), setting the stage for the Great Oxidation Event \citep{Leong2021}.  Constraints on atmospheric \ce{H2} levels in detrital magnetites in Archean riverbeds (\SI{3.0}{\giga\year}) indicate pressures of $<\SI{e-2}{\bar}$ \citep{Kadoya2019}.

Zircons dated back to \SI{4.35}{\giga\year} show oxygen fugacities, indicating that they crystallized in magmatic melts with oxidation states similar to present-day conditions \citep{Trail2011}. This state of the Hadean mantle would be consistent with the quartz-fayalite-magnetite mineral buffer, suggesting that volcanic outgassing would contribute mainly oxidized gases. The assumption that volcanic outgassing strongly influenced the Hadean atmosphere leads to the conclusion that the Hadean atmosphere was oxidized as early as \SI{4.35}{\giga\year} \citep{Trail2011}. However, it is questionable whether fluxes of oxidizing gases from volcanic outgassing might not be outcompeted by secondary geologic processes such as serpentinization. Hadean zircons crystallize at temperatures above \SI{600}{\celsius} \citep{Harrison2007} and therefore probe conditions deep in the Earth's mantle. However, serpentinization of rocks in the near-surface crust is a chemical process that operates independently of the redox state of these deeper regions. Outgassing from hydrothermal vent fields off-axis to the mid-ocean ridges, driven by serpentinization, might result in a Hadean atmosphere out of equilibrium with the Hadean mantle. The present work aims to explore this scenario.

An extrapolation of the rock record to the end of the Hadean, combined with the redox state of the late Hadean mantle, motivates the consideration of an atmosphere in an initially oxidizing state. This might describe a potential scenario of the late Hadean Earth atmosphere in an oxidized state, assuming that serpentinization and the late veneer were not yet able to produce fluxes of reducing gases that significantly affected the Earth's atmosphere. After this point in time, about \SI{4.0}{\giga\year} ago, these sources of reducing gases might have finally begun to contribute significantly, potentially altering this initial state toward more reducing conditions. Starting the simulation with this initially oxidized atmosphere allows us to test whether or not our considered source terms of reducing gases (serpentinization and impact degassing) are capable of transforming the atmosphere into a sufficiently reducing state favorable for the synthesis of \ce{HCN} and other key precursors of prebiotic organics. This might lead to a reduced Hadean atmosphere out of equilibrium with the more oxidized state of the mantle. Even if this reducing atmosphere begins to equilibrate with the upper parts of the crust and mantle, near-surface zircon crystals might be too inert to be affected. This might explain why \SI{4.35}{\giga\year} old Hadean zircons do not reflect this in their oxygen fugacities. Thus, these oxidized zircons do not necessarily rule out a reduced atmosphere throughout the late Hadean.

A corresponding set of parameters for an initially oxidized and habitable atmosphere with a temperature of \SI{51}{\celsius} in the EH at \SI{4.0}{\giga\year} from \citet{Pearce2022} is given in Table~\ref{tab:epochs}. We build on these two epochs of potential Hadean atmospheres from the previous study by \citet{Pearce2022} and introduce our newly determined source terms of reducing gases plausible in these epochs. The oxidizing atmosphere in the EH starts with a ratio of $\ce{N2}/\ce{CO2} \ll 1$. Therefore, the direct synthesis of \ce{HCN} by the carbonaceous portion of the late veneer (see Equation~\ref{eq:cc_hcn}) is not operational until the composition of the atmosphere is not inverted by the supply of reducing gases from other considered sources (serpentinization and enstatite bombardment) or by photochemistry in the atmosphere.

\section{Model: Sources of Atmospheric Gases}\label{sec:sources}

\begin{deluxetable*}{cDDcDD}[t]
    \tablecaption{Atmospheric source fluxes in the Hadean eon due to geological processes and impacts.}
    \label{tab:source_terms}
        \tablehead{
            \colhead{} & \multicolumn{9}{c}{Flux [\unit{\per\centi\meter\squared\per\second}]} \\
            \cmidrule{2-10}
             & \multicolumn{4}{c}{Smokers/Mantle} &  & \multicolumn{4}{c}{Impacts} \\
            \cmidrule{2-5}\cmidrule{7-10} 
            \colhead{Gas} & \twocolhead{Hadean} & \twocolhead{Today} & \colhead{} & \twocolhead{Enstatite} & \twocolhead{Carbonaceous} 
        } 
        \decimals
        \startdata
            \ce{H2} & 1.87$\times10^{10}$~(\SI{0.5}{\kilo\meter})\tablenotemark{a,b} & 6.25$\times10^{7}$\tablenotemark{b} &  & 2.30$\times10^{11}$~(\SI{4.4}{\giga\year})\tablenotemark{f} & . \\
            & 3.74$\times10^{10}$~(\SI{1.0}{\kilo\meter})\tablenotemark{a,b} & 8.6\phn$\times10^{9}$\tablenotemark{c} &  & 2.30$\times10^{10}$~(\SI{4.0}{\giga\year})\tablenotemark{f} & . \\
            & 7.48$\times10^{10}$~(\SI{2.0}{\kilo\meter})\tablenotemark{a,b} & 2.10$\times10^{9}$\tablenotemark{d} &  & . & . \\
            \hline
            \ce{CO2} & 5.17$\times10^{10}$~(\SI{0.5}{\kilo\meter})\tablenotemark{a,b} & 7 \phd\phn\phn\:$\times\:10^{-1}$\tablenotemark{c} &  & . & . \\
             & 4.89$\times10^{10}$~(\SI{1.0}{\kilo\meter})\tablenotemark{a,b} & 3.00$\times10^{11}$\tablenotemark{e} &  & . & . \\
             & 4.33$\times10^{10}$~(\SI{2.0}{\kilo\meter})\tablenotemark{a,b} & . &  & . & . \\
            \hline
            \ce{CH4} & 2.77$\times10^{9\phn}$~(\SI{0.5}{\kilo\meter})\tablenotemark{a,b} & 9.35$\times10^{7}$\tablenotemark{b} &  & . & . \\
             & 5.61$\times10^{9\phn}$~(\SI{1.0}{\kilo\meter})\tablenotemark{a,b} & 6.8\phn$\times10^{8}$\tablenotemark{c} &  & . & . \\
             & 1.12$\times10^{10}$~(\SI{2.0}{\kilo\meter})\tablenotemark{a,b} & . &  & . & . \\
            \hline
            \ce{HCN} & . & . &  & . & 3.35$\times10^{9}$(\SI{4.4}{\giga\year})\tablenotemark{g} \\
             & . & . &  & . & 3.35$\times10^{8}$(\SI{4.0}{\giga\year})\tablenotemark{g} \\
        \enddata
    \tablenotetext{a}{\citet{Miyazaki2022}: Extended model with varying hydrothermal circulation depth (HCD, see Figure~\ref{fig:H2_mantle_flux}).}
    \tablenotetext{b}{\citet{Charlou2002,Kelley2005,Cannat2010}: Used methane/hydrogen ratio measured in hydrothermal fluids ($\approx\SI{15}{\percent}$).}
    \tablenotetext{c}{\citet{Guzman-Marmolejo2013}}
    \tablenotetext{d}{\citet{Liu2023}: Globally up-scaled from smokers at North Atlantic mid-ocean ridge.}
    \tablenotetext{e}{\citet{Hu2012}}
    \tablenotetext{f}{\citet{Pearce2017,Pearce2022}: Complete oxidation of incoming iron \citep[\ce{Fe + H2O -> FeO + H2},][]{Zahnle2020} based on a model of the late veneer using the lunar cratering record.}
    \tablenotetext{g}{\citet{Kurosawa2013}: For $\ce{N2}/\ce{CO2}>35.4$.}
\end{deluxetable*}

\vspace{-24pt}
An overview of the source terms for atmospheric gases, including \ce{H2}, \ce{CO2}, \ce{CH4}, and \ce{HCN}, is summarized in Table~\ref{tab:source_terms}. These include emissions from hydrothermal vents on a global scale in the Hadean, impact degassing from enstatite impacts, and \ce{HCN} synthesis from carbonaceous impacts. Each source is discussed in more detail in the following Sections.

\subsection{Hadean Earth's Mantle Model}\label{sec:mantle_hcd}

The geophysical Hadean mantle model developed by \citet{Miyazaki2022} examines different regimes of mantle convection, as well as the dissolution and degassing of volatiles. The overall budgets of volatiles such as water and \ce{CO2} follow present-day values, but their initial values are based on partitioning between the atmosphere and magma ocean based on solubilities to the silicate melt \citep[see also][]{Dorn2021,Luo2024}. 

Overall, the model results in a flux of \ce{H2} to the atmosphere due to serpentinization in hydrothermal vent systems. Upwelling ferrous iron comes into contact with water penetrating the crust through fissures around these submarine volcanic systems. During this process, the iron is oxidized to ferric iron and the water is reduced to \ce{H2}. The main parameter that determines how much \ce{H2} is released is the HCD. The deeper the water can penetrate the crust, the more iron can be oxidized. 

The most direct way to determine the HCD is to look at the serpentinites left in the crust after contact with water. In particular, the depth of serpentinization in young oceanic crust formed from mid-oceanic ridges is of interest, as we expect the hydrothermal circulation to be prevalent in this environment on the Hadean Earth. Serpentinite formed in subduction zones beneath continental crust, as found on the present Earth, might not be representative of the Hadean Earth, where subduction might be active but massive continental plates are not yet present.

\citet{Lissenberg2024} drilled \SI{1269}{\meter} into the Mid-Atlantic Ridge and found serpentinization of the recovered peridotite over the full depth of the drill core. Changes in seismic wave velocities measured in the oceanic crust around slow to ultra-slow spreading ridges indicate the presence of serpentinites to at least \SIrange{3}{4}{\kilo\meter} below the seafloor \citep{Guillot2015}. The degree of serpentinization varies somewhere between $>\SI{97}{\percent}$ near the surface and $\sim\SI{20}{\percent}$ in deep regions. Along cracks in the crust, it cannot be excluded that serpentinization might locally extend as far as \SI{8}{\kilo\meter} below the seafloor, as has been found by seismic measurements of microearthquakes along the Mid-Atlantic Ridge \citep{Toomey1988,Tilmann2004,Guillot2015}. Computational modeling suggests that thermal cracking of oceanic crust could lead to hydration and subsequent serpentinization of rocks down to depths of \SIrange{30}{50}{\kilo\meter} \citep{Korenaga2007}. In addition, this might have ruptured the first stagnant lid crust formed on the Hadean Earth, initiating the first plate tectonic convection in the early history of the Earth.

Another way to determine how deep the ocean water penetrates the oceanic crust is to look at carbonate deposition. It indicates the alteration of rocks by hydrothermal fluids that enter the cracks and carry dissolved \ce{CO2}, which is deposited as carbonates in the cracks. Carbonate deposits can be observed in drill cores taken from the ocean floor. Alteration of the oceanic crust occurs up to \SI{5}{\kilo\meter} below the seafloor, but the degree of alteration is significantly lower below \SI{500}{\meter} and is almost negligible below \SI{2}{\kilo\meter} \citep[see, e.g.,][]{Staudigel1981,Alt1999}.

\subsubsection{Mantle Model Assumptions}\label{sec:mantle_model_assumptions}

\begin{figure}[htb]
    \centering
    \includegraphics[width=\columnwidth]{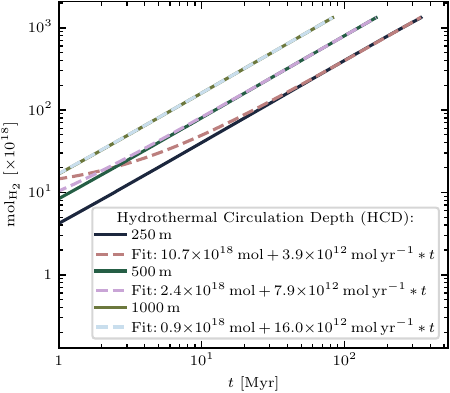}
    \caption{Global emissions of \ce{H2} over time from serpentinization in white smokers, hydrothermal vents located off-axis to mid-ocean ridges, in the Hadean. The hydrothermal circulation depth (HCD) was varied in extended calculations of the Hadean mantle model by \citet{Miyazaki2022}. Linear fitting allows \ce{H2} surface fluxes to be derived in units of \si{\mole\per\year}. A doubling of the HCD results in a doubling of the \ce{H2} flux, as more upwelling ferrous iron comes in contact with seawater, promoting the serpentinization reaction.}
    \label{fig:H2_mantle_flux}
\end{figure}

\citet{Miyazaki2022}'s original model used an HCD of \SI{500}{\meter} as a conservative estimate based on the drilling records of carbonate deposits. We used this existing model and extend it to include HCDs of 1 and \SI{2}{\kilo\meter}. Some exemplary data are shown in Figure~\ref{fig:H2_mantle_flux}. This is motivated by the fact that an HCD of \SI{2}{\kilo\meter} is consistent with still rather conservative constraints from both seismic measurements of serpentinization and carbonate deposits in drill cores as described above.

All resulting \ce{H2} surface fluxes were introduced into the photochemical atmosphere model to investigate the impact of HCD on prebiotic synthesis. This approach might better represent Hadean conditions and is supported by the fact that hydrothermal alteration can occur at these depths today. What distinguishes the present model from others is that it is specifically tailored to the Hadean and extends beyond the present-day situation commonly assumed as the basis for calculations in similar models \citep[cf., e.g.,][]{Guzman-Marmolejo2013,Thompson2022}.

In addition to the \ce{H2} fluxes, the geophysical mantle model also provides a surface degassing flux of \SI{6.53e11}{\kilo\gram\per\year} of \ce{CO2} \citep{Miyazaki2022}, based on the mantle processing rate. In addition, under the high temperatures and pressures of hydrothermal fluids, \ce{H2} and \ce{CO2} can react to form \ce{CH4} according to Equation~\ref{eq:FT_eff}. For the most realistic representation of this in our model, we have used the measurements from submersible missions that have sampled the fluid composition emitted from uninhabited hydrothermal vents \textit{in situ} \citep{Charlou2002,Kelley2005,Cannat2010}. The \ce{CH4}/\ce{H2} ratio is about $\SI{15}{\percent}$.

This ratio can now be combined with Equation~\ref{eq:FT_eff}, which states that the production of one mole of \ce{CH4} consumes one mole of \ce{CO2} and four moles of \ce{H2}. The model by \citet{Miyazaki2022} provides the initial fluxes $\Phi_\text{\ce{H2}}$ and $\Phi_\text{\ce{CO2}}$ emitted by the mantle for \ce{H2} and \ce{CO2}, respectively, prior to this reaction. As a result, the effectively emitted gas fluxes after this reaction is complete are
\begin{align}
\Phi_\text{\ce{H2},eff} &= 0.625*\Phi_\text{\ce{H2}},\\ \Phi_\text{\ce{CH4},eff} &= 0.09375*\Phi_\text{\ce{H2}},\\
\Phi_\text{\ce{CO2},eff} &= \Phi_\text{\ce{CO2}} - \Phi_\text{\ce{CH4}}.
\end{align}

These effective surface fluxes are listed in the second column of Table~\ref{tab:source_terms}. The third column lists surface fluxes predicted or extrapolated from current Earth measurements that differ from those representative of the Hadean.

\subsection{Bombardment Model}\label{sec:bombardment_model}

\begin{figure}[tb]
    \centering
    \includegraphics[width=\columnwidth]{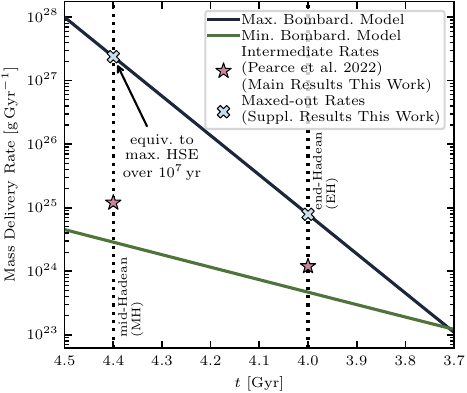}
    \caption{Early Earth bombardment models by \citet{Pearce2017} derived from exponentially decaying fits to the lunar cratering record \citep{Chyba1990}. The red stars represent the intermediate mass delivery rates in the mid-Hadean (MH) at \SI{4.4}{\giga\year} and the end-Hadean (EH) at \SI{4.0}{\giga\year} used by \citet{Pearce2022} and in the main results of this work (Section~\ref{sec:results}). The maxed-out rates marked with blue crosses were used in the supplementary results of this work (Section~\ref{sec:supp_results_high_bomb_rate}). The maxed-out rate at \SI{4.4}{\giga\year} roughly corresponds to the total excess of highly siderophile elements (HSEs) of the late veneer spread over a time span of 10 million years, the maximum duration of our atmospheric models. This is equivalent to a single impactor of \SI{2e25}{\gram} hitting Earth during the late veneer \citep{Zahnle2020,Wogan2023}.}
    \label{fig:bomb_rate}
\end{figure}

To quantify the contribution of meteorite impacts in reducing the atmosphere and allowing for prebiotic synthesis during the late veneer, it is first necessary to estimate the rates of meteorite bombardment in the Hadean. For the two scenarios at \SI{4.4}{\giga\year} and \SI{4.0}{\giga\year}, we applied the bombardment rates determined by \citet{Pearce2022} of \SI{1.2e25}{\gram\per\giga\year} and \SI{1.2e24}{\gram\per\giga\year}, respectively, based on exponentially decaying fits to the lunar cratering record analyzed by the Apollo program \citep[see Figure~\ref{fig:bomb_rate}]{Pearce2017,Chyba1990}. Comparable bombardment models have also been used in other studies of prebiotic synthesis and the favorable conditions that set the stage for the origins of life on early Earth \citep{Chyba1992,Laneuville2018,Kadoya2020}.

Red stars in Figure~\ref{fig:bomb_rate} mark these intermediate rates, which were used in the main results of this work in Section~\ref{sec:results}. We assume that these rates do not change substantially over the maximum duration of 10 million years in our atmospheric models and are therefore assumed to be constant in the respective epochs MH or EH, which simplifies the implementation of our simulations.

Blue crosses mark the maxed-out rates in the MH and EH of \SI{2.4e27}{\gram\per\giga\year} and \SI{7.9e24}{\gram\per\giga\year}, respectively, which were used in the supplementary results in Section~\ref{sec:supp_results_high_bomb_rate}. The maxed-out rate at \SI{4.4}{\giga\year} roughly corresponds to the total excess of HSEs of the late veneer spread over a time span of 10 million years, the maximum duration of our atmospheric models. This is equivalent to a single impactor of \SI{2e25}{\gram} (\SI{2300}{\kilo\meter} diameter) hitting Earth during the late veneer \citep[$\SI{2e25}{\gram} / \SI{e7}{\year} = \SI{2e27}{\gram\per\giga\year} \approx \SI{2.4e27}{\gram\per\giga\year}$]{Zahnle2020,Wogan2023}.

We then combined this with the rates of reducing gases emitted per impactor mass. For enstatite meteorites, equilibrium chemistry calculations of Equation~\ref{eq:iron_red} by \citet{Zahnle2020} resulted in a rate of $\sim$\SI{e-21}{\mole\HTwo\per\centi\meter\squared\per\gram} impactor.

Carbonaceous impacts could have facilitated the direct synthesis of \ce{HCN} in a \ce{N2}-\ce{H2O} atmosphere. This hypothesis is supported by experiments performed by \citet{Kurosawa2013} and summarized in Equation~\ref{eq:cc_hcn}. The experiments showed that up to \SI{2.8}{mol\percent} of the impacting carbon is converted into \ce{HCN} when the \ce{N2}/\ce{CO2} ratio exceeds 35.4 (as described in Appendix~\ref{sec:cc_impactors}). For the carbonaceous component of the bombardment we assumed a carbon content of \SI{3.2}{wt\%}, similar to CI chondrites \citep{Wasson1988}. This amount corresponds to a carbon content of \SI{2.66e-3}{\mole\carbon\per\gram}. Combined this gives a maximum possible \ce{HCN} production rate of \SI{7.45e-5}{\mole\HCN\per\gram} impactor.

\begin{figure}[tb]
    \centering
    \includegraphics[width=\columnwidth]{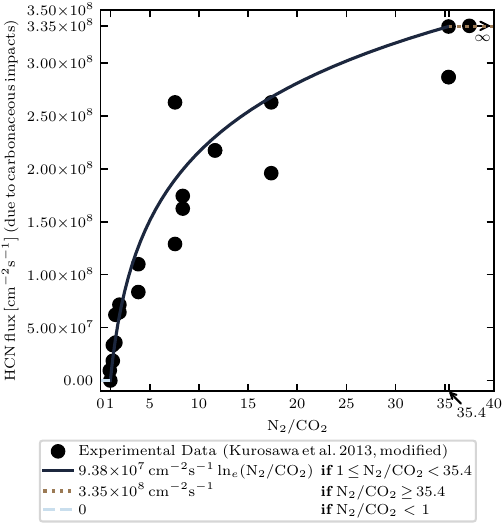}
    \caption{Impact degassing of \ce{HCN} by impacts of carbonaceous chondrites in the EH. The yields of carbon to \ce{HCN} conversion in the experimental data of \citet[Table~3]{Kurosawa2013} were multiplied by the bombardment rate determined by \citet{Pearce2022}. The partial pressures of \ce{N2} and \ce{CO2} used in the experiments were divided, and are represented by the plotted ratio \ce{N2}/\ce{CO2}. By fitting the functions given in the figure legend to the respective ranges of \ce{N2}/\ce{CO2}, \ce{HCN} surface fluxes can be derived in units of \si{\molecules\HCN\per\centi\meter\squared\per\second}. The data point in the upper right corner represents a measurement without any \ce{CO2} in the atmosphere employed in the experiment. Therefore, this point would have to be at an infinite ratio of \ce{N2}/\ce{CO2}, since the amount of \ce{CO2} in this measurement was zero.}
    \label{fig:cc_hcn}
\end{figure}

Figure~\ref{fig:cc_hcn} shows the performed fits of the experimental data by \citet[Table~3]{Kurosawa2013} used to determine the surface flux of \ce{HCN} generated by impact degassing of carbonaceous chondrites in the Hadean. If the atmosphere contains more \ce{CO2} than \ce{N2} ($\ce{N2}/\ce{CO2} < 1$), no \ce{HCN} formation was detected in the experiments. As soon as more \ce{N2} then \ce{CO2} is in the atmosphere ($\ce{N2}/\ce{CO2} > 1$), \ce{HCN} production during impact starts to increase logarithmically. It reaches a maximum as soon as the amount of \ce{N2} is more than 35.4 times the amount of \ce{CO2} in the surrounding atmosphere.

Overall, this allowed us to infer the global degassing rates of these reducing gases injected into the photochemical atmosphere model. To accomplish this, we combined the bombardment rates of exogenous iron and carbon delivery at a given epoch with the synthesis rates of \ce{H2} and \ce{HCN}. The results are given in Table~\ref{tab:source_terms} in the two rightmost columns. The different bombardment rates in the different epochs have been taken into account. In the EH, the initial \ce{CO2} level in the atmosphere was too high to allow \ce{HCN} synthesis by carbonaceous impacts. Therefore, the \ce{HCN} flux is not valid until the \ce{N2}/\ce{CO2} ratio exceeds 35.4 as the atmospheric composition evolves in the model. To account for different possible compositions of the late veneer (enstatite vs.~carbonaceous, see Appendix~\ref{sec:late_veneer}), we will consider several cases with either a pure enstatite, a pure carbonaceous, or a mixed bombardment.

\section{Model: Hadean Earth Scenarios}\label{sec:scenarios}

In a parameter study, we explore different scenarios and their interplay in multiple cases, which serve as the basis for several model runs. The goal is to explore as much of the parameter space as possible.

\begin{deluxetable*}{lDDDD}[htp]
    \tablecaption{Different scenarios considered for the mid-Hadean Earth (MH) in the initially \textit{reducing} scenario at \SI{4.4}{\giga\year}.}
    \label{tab:cases_red}
        \tablehead{
            \colhead{} & \multicolumn{8}{c}{Flux [\unit{\per\centi\meter\squared\per\second}]} \\
            \cmidrule{2-9}
            \colhead{Case} & \twocolhead{\ce{H2}} & \twocolhead{\ce{CO2}} & \twocolhead{\ce{CH4}} & \twocolhead{\ce{HCN}}
        }
        \decimals
        \startdata
        MH\_G: Geology only & . & . & . & . \\
        \makebox[1.4cm][l]{MH\_G0.5:} \SI{0.5}{\kilo\meter} HCD\tablenotemark{a} & 1.85$\times10^{10}$ & 5.17$\times10^{10}$ & 2.77$\times10^{9}$ & . \\
        \makebox[1.4cm][l]{MH\_G1:} \SI{1.0}{\kilo\meter} HCD\tablenotemark{a} & 3.74$\times10^{10}$ & 4.89$\times10^{10}$ & 5.61$\times10^{9}$ & . \\
        \makebox[1.4cm][l]{MH\_G2:} \SI{2.0}{\kilo\meter} HCD\tablenotemark{a} & 7.48$\times10^{10}$ & 4.33$\times10^{10}$ & 1.12$\times10^{10}$ & . \\
        MH\_E: \SI{100}{\percent} enstatite bombardment only & 2.30$\times10^{11}$ & 4.33$\times10^{10}$ & 0 & . \\
        MH\_C: \SI{100}{\percent} carbonaceous bombardment only & 0 & 4.33$\times10^{10}$ & 0 & 3.35$\times10^{9}$ \\
        MH\_G2E: Geology (\SI{2.0}{\kilo\meter}) + \SI{100}{\percent} enstatite & 3.05$\times10^{11}$ & 4.33$\times10^{10}$ & 1.12$\times10^{10}$ & . \\
        MH\_G2C: Geology (\SI{2.0}{\kilo\meter}) + \SI{100}{\percent} carbonaceous & 7.48$\times10^{10}$ & 4.33$\times10^{10}$ & 1.12$\times10^{10}$ & 3.35$\times10^{9}$ \\
        MH\_G2EC: Geology (\SI{2.0}{\kilo\meter}) + \SI{50}{\percent} enstatite + \SI{50}{\percent} carbonaceous & 1.90$\times10^{11}$ & 4.33$\times10^{10}$ & 1.12$\times10^{10}$ & 1.68$\times10^{9}$ \\
       \enddata
   \tablenotetext{a}{HCD: hydrothermal circulation depth} 
\end{deluxetable*}

\begin{deluxetable*}{lDDDD}[htbp]
    \tablecaption{Different scenarios considered for the end-Hadean Earth (EH) in the initially \textit{oxidizing} scenario at \SI{4.0}{\giga\year}.}
    \label{tab:cases_ox}
        \tablehead{
            \colhead{} & \multicolumn{8}{c}{Flux [\unit{\per\centi\meter\squared\per\second}]} \\
            \cmidrule{2-9}
            \colhead{Case} & \twocolhead{\ce{H2}} & \twocolhead{\ce{CO2}} & \twocolhead{\ce{CH4}} & \twocolhead{\ce{HCN}}
        }
        \decimals
        \startdata
        EH\_G: Geology only & . & . & . & . \\
        \makebox[1.4cm][l]{EH\_G0.5:} \SI{0.5}{\kilo\meter} HCD\tablenotemark{a} & 1.85$\times10^{10}$ & 5.17$\times10^{10}$ & 2.77$\times10^{9}$ & . \\
        \makebox[1.4cm][l]{EH\_G1:} \SI{1.0}{\kilo\meter} HCD\tablenotemark{a} & 3.74$\times10^{10}$ & 4.89$\times10^{10}$ & 5.61$\times10^{9}$ & . \\
        \makebox[1.4cm][l]{EH\_G2:} \SI{2.0}{\kilo\meter} HCD\tablenotemark{a} & 7.48$\times10^{10}$ & 4.33$\times10^{10}$ & 1.12$\times10^{10}$ & . \\
        EH\_E: \SI{100}{\percent} enstatite bombardment only & 2.30$\times10^{10}$ & 4.33$\times10^{10}$ & 0 & . \\
        EH\_C: \SI{100}{\percent} carbonaceous bombardment only & 0 & 4.33$\times10^{10}$ & 0 & 3.35$\times10^{8}$ \\
        EH\_G2E: Geology (\SI{2.0}{\kilo\meter}) + \SI{100}{\percent} enstatite & 9.78$\times10^{10}$ & 4.33$\times10^{10}$ & 1.12$\times10^{10}$ & . \\
        EH\_G2C: Geology (\SI{2.0}{\kilo\meter}) + \SI{100}{\percent} carbonaceous & 7.48$\times10^{10}$ & 4.33$\times10^{10}$ & 1.12$\times10^{10}$ & 3.35$\times10^{8}$ \\
        EH\_G2EC: Geology (\SI{2.0}{\kilo\meter}) + \SI{50}{\percent} enstatite + \SI{50}{\percent} carbonaceous & 8.63$\times10^{10}$ & 4.33$\times10^{10}$ & 1.12$\times10^{10}$ & 1.68$\times10^{8}$ \\
       \enddata
    \tablenotetext{a}{HCD: hydrothermal circulation depth} 
\end{deluxetable*}

\vspace{-24pt}\vspace{-24pt}
Table~\ref{tab:cases_red} presents the different cases analyzed in the reducing scenario during the MH at \SI{4.4}{\giga\year}, while Table~\ref{tab:cases_ox} represents the oxidizing scenario in the EH, \SI{4.0}{\giga\year} ago. Each case listed represents a set of initial parameters that we used in a separate run of the atmosphere model. The parameters are combinations of the source fluxes specified in Table~\ref{tab:source_terms}.

First, in both the reducing and oxidizing cases, we examine how the HCD affects the injection of \ce{H2}, \ce{CO2}, and methane by the geophysical mantle processes of the early Earth. We will refer to this as the ``geology'' case G. We examined three specific sub-cases, each with a different HCD of \SI{0.5}{\kilo\meter}, \SI{1}{\kilo\meter}, and \SI{2}{\kilo\meter}, to explore the range that is potentially feasible for the Hadean (see Appendix~\ref{sec:mantle_hcd}). These sub-cases are designated G0.5, G1, and G2, respectively.

Additionally, we explore the effects of either a purely enstatite or a purely carbonaceous bombardment in cases E and C, respectively. The geological contributions of \ce{H2} and \ce{CH4} are deactivated to examine the reducing potential of the late veneer alone, but the \ce{CO2} emitted by the mantle remains in the model. The purpose is to investigate whether the bombardment is capable of reducing the atmosphere while counteracting the oxidizing gases emitted by volcanism. Numerous models have examined the same interplay between the reducing effect of an exogenous enstatite bombardment and the oxidizing effect of endogenous volcanism in competition with each other \citep{Wogan2023,Zahnle2020,Itcovitz2022,Citron2022}. Case E will be used to compare our results with these previous studies. We have assumed an HCD of \SI{2}{\kilo\meter}, which results in the lowest net \ce{CO2} release from the geological model by \citet{Miyazaki2022}. Therefore, this represents the best case scenario for the successful reduction of the atmosphere by the late veneer. The same holds for case C, where a purely carbonaceous bombardment and \ce{HCN} synthesis competes with the geological \ce{CO2} emission from volcanoes.

Finally, we examined the combination of these source terms by combining the geology with an HCD of \SI{2}{\kilo\meter} with the enstatite-only bombardment in case G2E, with the carbonaceous-only bombardment in case G2C, and perhaps the most agnostic assumption of a mixed bombardment of half and half composition in case G2EC.

\begingroup
\setlength{\tabcolsep}{3pt}
\begin{splitdeluxetable*}{lDDcDDcDDcDDBlDDcDDcDDBlDDcDDcDD}
    \tablecaption{Maximum yields of prebiotic organic molecules in warm little ponds (WLPs) with turned off seepage after \SI{10000}{\year}.}
    \label{tab:results_WLP_no_seepage}
    \tabletypesize{\small}
        \tablehead{
            \colhead{} & \multicolumn{19}{c}{Max.~Warm Little Pond Concentration [\unit{\micro\Molar}]} & \colhead{} & \multicolumn{14}{c}{Max.~Warm Little Pond Concentration [\unit{\micro\Molar}]} & \colhead{} & \multicolumn{14}{c}{Max.~Warm Little Pond Concentration [\unit{\micro\Molar}]} \\
            \cmidrule{2-20}\cmidrule{22-35}\cmidrule{37-50}
            \colhead{} & \multicolumn{4}{c}{\ce{HCN} from Rain-out} & \colhead{} & \multicolumn{4}{c}{\ce{H2CO} from Rain-out} & \colhead{} & \multicolumn{4}{c}{\ce{H2CO} from Aqueous Synth.\tablenotemark{a}} & \colhead{} & \multicolumn{4}{c}{Ribose} & \colhead{} & \multicolumn{4}{c}{2-Aminooxazole} & \colhead{} & \multicolumn{4}{c}{Adenine} & \colhead{} & \multicolumn{4}{c}{Guanine}& \colhead{} & \multicolumn{4}{c}{Cytosine} & \colhead{} & \multicolumn{4}{c}{Uracil} & \colhead{} & \multicolumn{4}{c}{Thymine} \\
            \cmidrule{2-5}\cmidrule{7-10}\cmidrule{12-15}\cmidrule{17-20}\cmidrule{22-25}\cmidrule{27-30}\cmidrule{32-35}\cmidrule{37-40}\cmidrule{42-45}\cmidrule{47-50}
            \colhead{Case} & \twocolhead{MH (red.)} & \twocolhead{EH (ox.)} & \colhead{} & \twocolhead{MH (red.)} & \twocolhead{EH (ox.)} & \colhead{} & \twocolhead{MH (red.)} & \twocolhead{EH (ox.)} & \colhead{} & \twocolhead{MH (red.)} & \twocolhead{EH (ox.)} & \colhead{Case} & \twocolhead{MH (red.)} & \twocolhead{EH (ox.)} & \colhead{} & \twocolhead{MH (red.)} & \twocolhead{EH (ox.)} & \colhead{} & \twocolhead{MH (red.)} & \twocolhead{EH (ox.)} & \colhead{Case} & \twocolhead{MH (red.)} & \twocolhead{EH (ox.)} & \colhead{} & \twocolhead{MH (red.)} & \twocolhead{EH (ox.)} & \colhead{} & \twocolhead{MH (red.)} & \twocolhead{EH (ox.)}
        }
        \decimals
        \startdata
        G0.5 & 4.22$\times 10^{3}$ & 1.21$\times 10^{-2}$ && 2.87$\times 10^{-1}$ & 5.04 && 1.52$\times 10^{2}$ & 4.37$\times 10^{-4}$ && 1.86$\times 10^{-1}$\tablenotemark{b} & 6.14$\times 10^{-3}$\tablenotemark{c} & G0.5 & 4.65 & 1.33$\times 10^{-5}$ && 4.85 & 1.39$\times 10^{-5}$ && 5.59 & 1.61$\times 10^{-5}$ & G0.5 & 4.10$\times 10^{-2}$ & 1.18$\times 10^{-7}$ && 1.72$\times 10^{1}$ & 4.95$\times 10^{-5}$ && 3.38$\times 10^{1}$ & 9.71$\times 10^{-5}$ \\ 
        G1 & 9.78$\times 10^{2}$ & 5.37$\times 10^{-2}$ && 1.14$\times 10^{-3}$ & 2.32 && 3.52$\times 10^{1}$ & 1.93$\times 10^{-3}$ && 4.30$\times 10^{-2}$\tablenotemark{b} & 2.83$\times 10^{-3}$\tablenotemark{c} & G1 & 1.08 & 5.91$\times 10^{-5}$ && 1.12 & 6.17$\times 10^{-5}$ && 1.29 & 7.11$\times 10^{-5}$ & G1 & 9.50$\times 10^{-3}$ & 5.22$\times 10^{-7}$ && 3.99 & 2.19$\times 10^{-4}$ && 7.83 & 4.30$\times 10^{-4}$ \\ 
        G2 & 7.04$\times 10^{3}$ & 5.95$\times 10^{4}$ && 7.87$\times 10^{-6}$ & 5.01 && 2.54$\times 10^{2}$ & 2.14$\times 10^{3}$ && 3.09$\times 10^{-1}$\tablenotemark{b} & 2.62\tablenotemark{b} & G2 & 7.75 & 6.55$\times 10^{1}$ && 8.09 & 6.84$\times 10^{1}$ && 9.32 & 7.88$\times 10^{1}$ & G2 & 6.84$\times 10^{-2}$ & 5.78$\times 10^{-1}$ && 2.88$\times 10^{1}$ & 2.43$\times 10^{2}$ && 5.64$\times 10^{1}$ & 4.76$\times 10^{2}$ \\
        E & 2.58$\times 10^{-2}$ & 1.17$\times 10^{-6}$ && 6.46$\times 10^{-10}$ & 2.14$\times 10^{-12}$ && 9.29$\times 10^{-4}$ & 4.23$\times 10^{-8}$ && 1.13$\times 10^{-6}$\tablenotemark{b} & 5.16$\times 10^{-11}$\tablenotemark{b} & E & 2.84$\times 10^{-5}$ & 1.29$\times 10^{-9}$ && 2.97$\times 10^{-5}$ & 1.35$\times 10^{-9}$ && 3.42$\times 10^{-5}$ & 1.56$\times 10^{-9}$ & E & 2.51$\times 10^{-7}$ & 1.14$\times 10^{-11}$ && 1.05$\times 10^{-4}$ & 4.80$\times 10^{-9}$ && 2.07$\times 10^{-4}$ & 9.40$\times 10^{-9}$ \\ 
        C & 2.19$\times 10^{5}$ & 1.34$\times 10^{-6}$ && 6.48$\times 10^{-10}$ & 1.32$\times 10^{-12}$ && 7.88$\times 10^{3}$ & 4.83$\times 10^{-8}$ && 9.61\tablenotemark{b} & 5.89$\times 10^{-11}$\tablenotemark{b} & C & 2.41$\times 10^{2}$ & 1.48$\times 10^{-9}$ && 2.51$\times 10^{2}$ & 1.54$\times 10^{-9}$ && 2.90$\times 10^{2}$ & 1.78$\times 10^{-9}$ & C & 2.13 & 1.30$\times 10^{-11}$ && 8.94$\times 10^{2}$ & 5.48$\times 10^{-9}$ && 1.75$\times 10^{3}$ & 1.07$\times 10^{-8}$ \\ 
        G2E & 1.88$\times 10^{4}$ & 5.44$\times 10^{4}$ && 8.78$\times 10^{-4}$ & 4.00 && 6.77$\times 10^{2}$ & 1.96$\times 10^{3}$ && 8.26$\times 10^{-1}$\tablenotemark{b} & 2.39\tablenotemark{b} & G2E & 2.07$\times 10^{1}$ & 5.99$\times 10^{1}$ && 2.16$\times 10^{1}$ & 6.25$\times 10^{1}$ && 2.49$\times 10^{1}$ & 7.20$\times 10^{1}$ & G2E & 1.83$\times 10^{-1}$ & 5.29$\times 10^{-1}$ && 7.68$\times 10^{1}$ & 2.22$\times 10^{2}$ && 1.50$\times 10^{2}$ & 4.35$\times 10^{2}$ \\
        G2C & 1.34$\times 10^{5}$ & 3.39$\times 10^{2}$ && 7.86$\times 10^{-6}$ & 1.42$\times 10^{2}$ && 4.82$\times 10^{3}$ & 1.22$\times 10^{1}$ && 5.89\tablenotemark{b} & 1.88$\times 10^{-1}$\tablenotemark{c} & G2C & 1.47$\times 10^{2}$ & 3.73$\times 10^{-1}$ && 1.54$\times 10^{2}$ & 3.89$\times 10^{-1}$ && 1.77$\times 10^{2}$ & 4.49$\times 10^{-1}$ & G2C & 1.30 & 3.29$\times 10^{-3}$ && 5.47$\times 10^{2}$ & 1.38 && 1.07$\times 10^{3}$ & 2.71 \\
        G2EC & 7.73$\times 10^{4}$ & 1.19$\times 10^{4}$ && 6.31$\times 10^{-3}$ & 1.02$\times 10^{1}$ && 2.78$\times 10^{3}$ & 4.29$\times 10^{2}$ && 3.40\tablenotemark{b} & 5.36$\times 10^{-1}$\tablenotemark{b} & G2EC & 8.50$\times 10^{1}$ & 1.31$\times 10^{1}$ && 8.88$\times 10^{1}$ & 1.37$\times 10^{1}$ && 1.02$\times 10^{2}$ & 1.58$\times 10^{1}$ & G2EC & 7.51$\times 10^{-1}$ & 1.16$\times 10^{-1}$ && 3.16$\times 10^{2}$ & 4.87$\times 10^{1}$ && 6.19$\times 10^{2}$ & 9.55$\times 10^{1}$ \\
       \enddata
    \tablenotetext{a}{Formaldehyde synthesized aqueously from rained-out \ce{HCN}.}
    \tablenotetext{b}{Most ribose synthesized in formose reaction starting from formaldehyde, which in turn was aqueously synthesized from rained-out \ce{HCN}.} 
    \tablenotetext{c}{Most ribose synthesized in formose reaction starting from formaldehyde rained-out directly from the atmosphere.}
\end{splitdeluxetable*}
\endgroup

The same cases were considered for the oxidizing scenario summarized in Table~\ref{tab:cases_ox} in the EH at \SI{4.0}{\giga\year}. The main difference is a reduced bombardment rate due to the decline of the late veneer over time, while the geologic mantle fluxes remain unchanged. Since the oxidizing scenario starts with a ratio $\ce{N2}/\ce{CO2}<1$ (see Table~\ref{tab:epochs}), the source fluxes of reducing gases due to carbonaceous impacts are initially inactive (see Appendix~\ref{sec:cc_impactors}). They only become active when this ratio exceeds 1, due to the effects of geology and enstatite bombardment, as well as photochemistry changing the atmospheric composition. This highlights the need to study all these processes and their interactions in combined models, as done in the present study.

\section{Supplementary Results: No Seepage}\label{sec:supp_results}

Table~\ref{tab:results_WLP_no_seepage} shows the maximum yields of prebiotic organic molecules reached in the simulation of WLPs in the \textit{absence} of seepage after a simulation period of \SI{10000}{\year}. This could be a reasonable scenario if the pores at the bottom of the pond are clogged due to adsorption of biomolecules on the mineral surfaces or deposition of amphiphiles and mineral gels \citep{Hazen2010,Deamer2017,Damer2020}.

\section{Supplementary Results: Increased Impact Rates}\label{sec:supp_results_high_bomb_rate}

\begin{figure*}[htp]
    \centering
    \includegraphics[width=\textwidth]{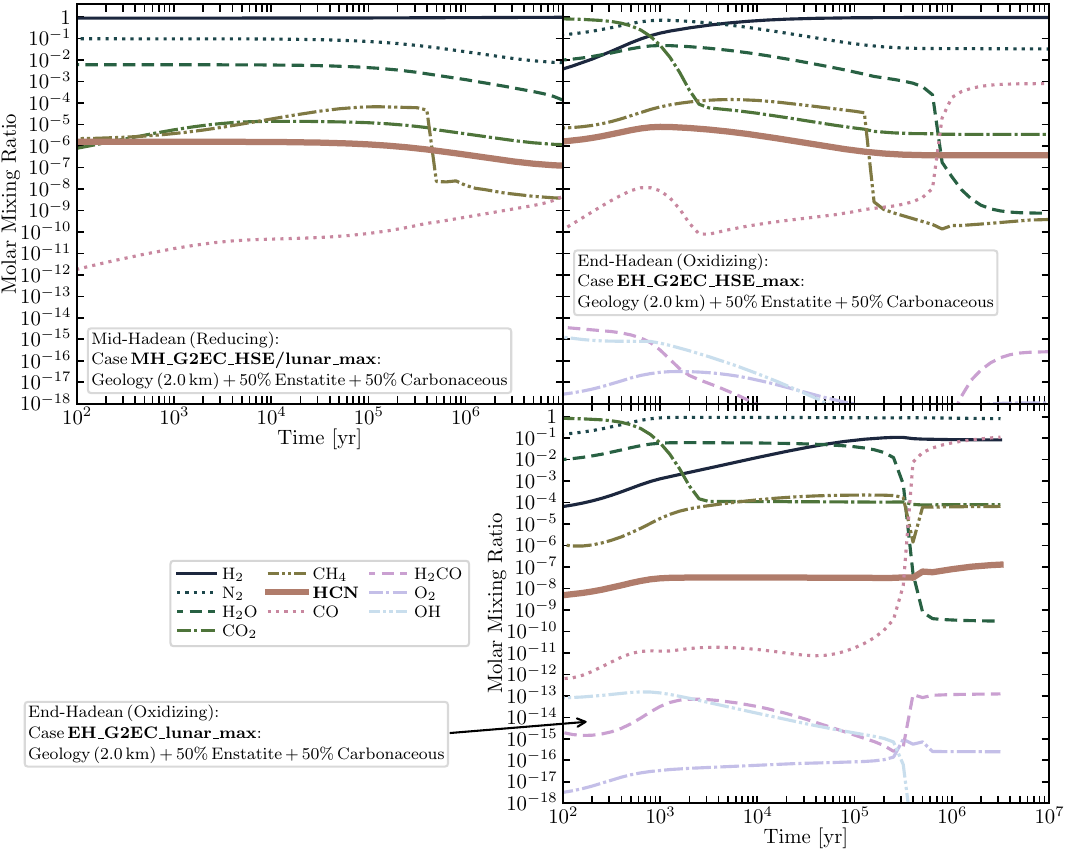}
    \caption{Exploring the effect of maxed-out exogenous chondritic bombardment. Shown is the simulated atmospheric composition of key species in the layer closest to the surface as a function of time. All cases consider the combination of geologic and late veneer source terms of atmospheric gases (referred to as \textit{G2EC}), with serpentinization driven by an hydrothermal circulation depth (HCD) of \SI{2}{\kilo\meter}, and maxed-out bombardment with a composition split half-half between enstatite and carbonaceous chondrites (see Table~\ref{tab:cases_high_bomb_rate}). The two epochs of the mid-Hadean (MH) at \SI{4.4}{\giga\year} with reducing initial conditions and the end-Hadean (EH) at \SI{4.0}{\giga\year} with oxidizing initial conditions are compared (left vs.~right column). The initial conditions for the reducing and oxidizing models are summarized in Table~\ref{tab:epochs}, closely following the established atmosphere models developed by \citet{Pearce2022}. \textit{HSE\_max} refers to a bombardment that delivered the entire excess of highly siderophile elements (HSEs) to the Hadean Earth spread over a time span of only 10 million years, the maximum duration of our atmospheric models. This is equivalent to the maximum rate of impacting material at \SI{4.4}{\giga\year} derived from the lunar cratering record, termed \textit{lunar\_max}. Panel \textbf{MH\_G2EC\_HSE/lunar\_max} shows the effect of a maxed-out bombardment, comprising the whole late veneer, on the initially reducing atmosphere in the MH. Panel \textbf{EH\_G2EC\_HSE\_max} shows the effect of this maxed-out bombardment on the initially oxidized atmosphere in the EH. And panel \textbf{EH\_G2EC\_lunar\_max} shows the effect of the maximum bombardment rate derived from the lunar cratering record at this time (\SI{4.0}{\giga\year}.}\label{fig:photochem_high_bomb_rate}
\end{figure*}

Table~\ref{tab:cases_high_bomb_rate} presents three additional models that explore what happens when the mass delivery rates in Figure~\ref{fig:bomb_rate} are at their maximum possible values, in contrast to all the models presented before using intermediate rates of impacting material.

\begin{deluxetable*}{lDDDD}[htp]
    \tablecaption{Additional scenarios considered with maximum impact rates (see Figure~\ref{fig:bomb_rate}) in the mixed case G2EC (serpentinization driven by an hydrothermal circulation depth (HCD) of \SI{2}{\kilo\meter}, \SI{50}{\percent} enstatite, \SI{50}{\percent} carbonaceous bombardment) for both the mid-Hadean (MH) and end-Hadean (EH).}
    \label{tab:cases_high_bomb_rate}
        \tablehead{
            \colhead{} & \multicolumn{8}{c}{Flux [\unit{\per\centi\meter\squared\per\second}]} \\
            \cmidrule{2-9}
            \colhead{Case} & \twocolhead{\ce{H2}} & \twocolhead{\ce{CO2}} & \twocolhead{\ce{CH4}} & \twocolhead{\ce{HCN}}
        }
        \decimals
        \startdata
        MH\_G2EC\_HSE/lunar\_max @ \SI{4.4}{\giga\year}: & 2.29$\times10^{13}$ & 4.33$\times10^{10}$ & 1.12$\times10^{10}$ & 3.35$\times10^{11}$ \\
        EH\_G2EC\_HSE\_max (equiv.\ to lunar\_max @ \SI{4.4}{\giga\year}): & 2.29$\times10^{13}$ & 4.33$\times10^{10}$ & 1.12$\times10^{10}$ & 3.35$\times10^{11}$ \\
        EH\_G2EC\_lunar\_max @ \SI{4.0}{\giga\year}: & 1.50$\times10^{11}$ & 4.33$\times10^{10}$ & 1.12$\times10^{10}$ & 1.10$\times10^{9}$ \\
       \enddata
\end{deluxetable*}

\vspace{-24pt}
In naming these three cases, \textit{G2EC} refers to a mixed scenario in which serpentinization driven by an HCD of \SI{2}{\kilo\meter}, \SI{50}{\percent} enstatite, and \SI{50}{\percent} carbonaceous bombardment are active at the same time. \textit{HSE\_max} refers to a bombardment that delivered the entire excess of HSEs to the Hadean Earth spread over a time span of only 10 million years, the maximum duration of our atmospheric models. This means that the late veneer was deposited in only this relatively short period of time in truly cataclysmic events of dwarf planet-sized impacts (\SI{2e25}{\gram}, \SI{2300}{\kilo\meter} diameter), as suggested by \citet{Zahnle2020,Wogan2023}. This is equivalent to the maximum rate of impacting material at \SI{4.4}{\giga\year} ago, derived from the lunar cratering record, and fitted with an exponentially decaying bombardment model spanning from \SI{4.5}{\giga\year} to \SI{3.9}{\giga\year} ago \citep[see Figure~\ref{fig:bomb_rate}]{Chyba1990,Pearce2017}, denoted \textit{lunar\_max}. Therefore, case MH\_G2EC\_HSE/lunar\_max got its name due to this equivalence.

Figures~\ref{fig:photochem_high_bomb_rate}(MH\_G2EC\_HSE/lunar\_max) and (EH\_G2EC\_HSE\_max) explore what would happen to the atmosphere if the entire late veneer was delivered to the Hadean Earth during the simulated time, either to an initially reducing or oxidizing atmosphere, respectively. In both cases, atmospheric \ce{HCN} concentrations are mainly driven by direct synthesis during carbonaceous impacts, and their levels reach the \num{e-6} range, two orders of magnitude higher than in the most productive case MH\_C (see Table~\ref{tab:results_atm}) using intermediate mass delivery rates in the main results Section~\ref{sec:results}. In case EH\_G2EC\_lunar\_max, \ce{HCN} levels reach the \num{e-8} range, comparable to the most productive cases in the main results. This means that even without assuming that dwarf planet-sized impactors struck Earth, the maximum bombardment model based on the lunar cratering record \SI{4.0}{\giga\year} ago (see Figure~\ref{fig:bomb_rate}) is capable of producing enough atmospheric \ce{HCN} to motivate nucleotide synthesis in WLPs, despite the initially \ce{CO2}-rich atmosphere.

\begin{deluxetable*}{lDDD}[t]
    \tablecaption{Maximum resulting rain-out rates of prebiotic precursors with maxed-out bombardment rates.}
    \label{tab:results_rain_high_bomb_rate}
        \tablehead{
            \colhead{} & \multicolumn{6}{c}{Max.~Rain-out Rate [\unit{\kilo\gram\per\meter\squared\per\year}]} \\
            \cmidrule{2-7}
            \colhead{Case} & \twocolhead{\ce{CO2}} & \twocolhead{\ce{HCN}} & \twocolhead{\ce{H2CO}}
        }
        \decimals
        \startdata
        MH\_G2EC\_HSE/lunar\_max & 1.55$\times 10^{-3}$ & 7.23$\times 10^{-3}$ & 1.35$\times 10^{-10}$ \\
        EH\_G2EC\_HSE\_max & 3.47$\times 10^{-3}$ & 1.40$\times 10^{-2}$ & 1.88$\times 10^{-11}$ \\ 
        EH\_G2EC\_lunar\_max & 1.15$\times 10^{-2}$ & 5.73$\times 10^{-4}$ & 8.68$\times 10^{-9}$ \\ 
        \enddata
\end{deluxetable*}

\vspace{-24pt}
Table~\ref{tab:results_rain_high_bomb_rate} presents the maximum rain-out rates for \ce{CO2}, \ce{HCN}, and \ce{H2CO} with maxed-out bombardment rates. In particular, the \ce{HCN} rain-out in case EH\_G2EC\_HSE\_max is very high, reaching the \SI{e-2}{\kilo\gram\per\meter\squared\per\year} range, more than an order of magnitude higher than in the most productive case MH\_C (see Table~\ref{tab:results_rain}) using intermediate mass delivery rates in the main results Section~\ref{sec:results}. This shows that if the late veneer was formed in a very short time in a concentrated bombardment in the EH, it might have been able to transform the initially \ce{CO2}-rich atmosphere into a reducing state. This is a very promising result to justify a robust prebiotic synthesis throughout the Hadean.

\begin{splitdeluxetable*}{lDDDDBlDDDDDD}
    \tablecaption{Maximum yields of prebiotic organic molecules in warm little ponds (WLPs) with maxed-out bombardment rates.}
    \label{tab:results_WLP_high_bomb_rate}
        \tablehead{
            \colhead{} & \multicolumn{8}{c}{Max.~Warm Little Pond Concentration [\unit{\micro\Molar}]} & \colhead{} & \multicolumn{12}{c}{Max.~Warm Little Pond Concentration [\unit{\micro\Molar}]} \\
            \cmidrule{2-9}\cmidrule{11-22}
            \colhead{Case} & \twocolhead{\ce{HCN} from Rain-out} & \twocolhead{\ce{H2CO} from Rain-out} & \twocolhead{\ce{H2CO} from Aqueous Synth.\tablenotemark{a}} & \twocolhead{Ribose} & \colhead{Case} & \twocolhead{2-Aminooxazole} & \twocolhead{Adenine} & \twocolhead{Guanine} & \twocolhead{Cytosine} & \twocolhead{Uracil} & \twocolhead{Thymine}
        }
        \decimals
        \startdata
        MH\_G2EC\_HSE/lunar\_max & 9.22$\times 10^{4}$ & 4.01$\times 10^{-5}$ & 3.32$\times 10^{3}$ & 4.05\tablenotemark{b} & MH\_G2EC\_HSE/lunar\_max & 1.01$\times 10^{2}$ & 1.66$\times 10^{4}$ & 1.84$\times 10^{4}$ & 3.32$\times 10^{3}$ & 1.66$\times 10^{3}$ & 1.11$\times 10^{3}$ \\ 
        EH\_G2EC\_HSE\_max & 1.79$\times 10^{5}$ & 5.60$\times 10^{-6}$ & 6.43$\times 10^{3}$ & 7.84\tablenotemark{b} & EH\_G2EC\_HSE\_max & 1.96$\times 10^{2}$ & 3.21$\times 10^{4}$ & 3.57$\times 10^{4}$ & 6.42$\times 10^{3}$ & 3.21$\times 10^{3}$ & 2.14$\times 10^{3}$ \\ 
        EH\_G2EC\_lunar\_max & 6.56$\times 10^{3}$ & 2.59$\times 10^{-3}$ & 2.36$\times 10^{2}$ & 2.88$\times 10^{-1}$\tablenotemark{b} & EH\_G2EC\_lunar\_max & 7.21 & 1.18$\times 10^{3}$ & 1.31$\times 10^{3}$ & 2.36$\times 10^{2}$ & 1.18$\times 10^{2}$ & 7.87$\times 10^{1}$ \\ 
        \enddata
    \tablenotetext{a}{Formaldehyde synthesized aqueously from rained-out \ce{HCN}.}
    \tablenotetext{b}{Most ribose synthesized in formose reaction starting from formaldehyde, which in turn was aqueously synthesized from rained-out \ce{HCN}.} 
\end{splitdeluxetable*}

Table~\ref{tab:results_WLP_high_bomb_rate} summarizes the maximum concentrations of prebiotic molecules in WLPs resulting from these three additional cases with maxed-out impact rates, and Table~\ref{tab:results_WLP_high_bomb_rate_no_seepage} shows the corresponding WLP concentrations with seepage turned off due to clogged pores at the bottom of the pond. With seepage, purine concentrations reach the \SI{10}{\milli\Molar} range, pyrimidine concentrations reach the \unit{\milli\Molar} range, and ribose concentrations reach the \unit{\micro\Molar} range. Without seepage, pyrimidine concentrations also reach the \SI{10}{\milli\Molar} range, and ribose concentrations reach the \SI{100}{\micro\Molar} range. These nucleobase concentrations are about an order of magnitude higher than the intermediate impact rates (see Tables~\ref{tab:results_WLP} and \ref{tab:results_WLP_no_seepage}) and are sufficient for nucleotide synthesis. Even with maxed-out bombardment rates and without seepage, ribose concentrations are close to, but still fall short of, the \unit{\milli\Molar} concentrations required for nucleotide synthesis in aqueous solution as performed in laboratory experiments \citep{Ponnamperuma1963,Fuller1972,Nam2018,Powner2009}.
 
\begin{splitdeluxetable*}{lDDDDBlDDDDDD}
    \tablecaption{Maximum yields of prebiotic organic molecules in warm little ponds (WLPs) with maxed-out bombardment rates and with turned off seepage after \SI{10000}{\year}.}
    \label{tab:results_WLP_high_bomb_rate_no_seepage}
        \tablehead{
            \colhead{} & \multicolumn{8}{c}{Max.~Warm Little Pond Concentration [\unit{\micro\Molar}]} & \colhead{} & \multicolumn{12}{c}{Max.~Warm Little Pond Concentration [\unit{\micro\Molar}]} \\
            \cmidrule{2-9}\cmidrule{11-22}
            \colhead{Case} & \twocolhead{\ce{HCN} from Rain-out} & \twocolhead{\ce{H2CO} from Rain-out} & \twocolhead{\ce{H2CO} from Aqueous Synth.\tablenotemark{a}} & \twocolhead{Ribose} & \colhead{Case} & \twocolhead{2-Aminooxazole} & \twocolhead{Adenine} & \twocolhead{Guanine} & \twocolhead{Cytosine} & \twocolhead{Uracil} & \twocolhead{Thymine}
        }
        \decimals
        \startdata
        MH\_G2EC\_HSE/lunar\_max & 2.75$\times 10^{6}$ & 4.61$\times 10^{-2}$ & 9.90$\times 10^{4}$ & 1.21$\times 10^{2}$\tablenotemark{b} & MH\_G2EC\_HSE/lunar\_max & 3.03$\times 10^{3}$ & 3.16$\times 10^{3}$ & 3.64$\times 10^{3}$ & 2.67$\times 10^{1}$ & 1.12$\times 10^{4}$ & 2.20$\times 10^{4}$ \\ 
        EH\_G2EC\_HSE\_max & 5.31$\times 10^{6}$ & 6.44$\times 10^{-3}$ & 1.91$\times 10^{5}$ & 2.33$\times 10^{2}$\tablenotemark{b} & EH\_G2EC\_HSE\_max & 5.84$\times 10^{3}$ & 6.10$\times 10^{3}$ & 7.02$\times 10^{3}$ & 5.15$\times 10^{1}$ & 2.17$\times 10^{4}$ & 4.25$\times 10^{4}$ \\ 
        EH\_G2EC\_lunar\_max & 2.18$\times 10^{5}$ & 2.97$\times 10^{0}$ & 7.84$\times 10^{3}$ & 9.57$\times 10^{0}$\tablenotemark{b} & EH\_G2EC\_lunar\_max & 2.40$\times 10^{2}$ & 2.50$\times 10^{2}$ & 2.88$\times 10^{2}$ & 2.12 & 8.89$\times 10^{2}$ & 1.74$\times 10^{3}$ \\ 
        \enddata
    \tablenotetext{a}{Formaldehyde synthesized aqueously from rained-out \ce{HCN}.}
    \tablenotetext{b}{Most ribose synthesized in formose reaction starting from formaldehyde, which in turn was aqueously synthesized from rained-out \ce{HCN}.} 
\end{splitdeluxetable*}
 
\FloatBarrier

\bibliography{main}{}

\begin{thebibliography}{}
\expandafter\ifx\csname natexlab\endcsname\relax\def\natexlab#1{#1}\fi
\providecommand{\url}[1]{\href{#1}{#1}}
\providecommand{\dodoi}[1]{doi:~\href{http://doi.org/#1}{\nolinkurl{#1}}}
\providecommand{\doeprint}[1]{\href{http://ascl.net/#1}{\nolinkurl{http://ascl.net/#1}}}
\providecommand{\doarXiv}[1]{\href{https://arxiv.org/abs/#1}{\nolinkurl{https://arxiv.org/abs/#1}}}

\bibitem[{Abe(1993)}]{Abe1993}
Abe, Y. 1993, Lithos, 30, 223, \dodoi{10.1016/0024-4937(93)90037-D}

\bibitem[{Adler \& Gu(2024)}]{Adler2024}
Adler, R.~F., \& Gu, G. 2024, Atmosphere, 15, 535, \dodoi{10.3390/atmos15050535}

\bibitem[{Alt \& Teagle(1999)}]{Alt1999}
Alt, J.~C., \& Teagle, D. A.~H. 1999, Geochimica et Cosmochimica Acta, 63, 1527, \dodoi{10.1016/S0016-7037(99)00123-4}

\bibitem[{Attwater {et~al.}(2018)Attwater, Raguram, Morgunov, Gianni, \& Holliger}]{Attwater2018}
Attwater, J., Raguram, A., Morgunov, A.~S., Gianni, E., \& Holliger, P. 2018, eLife, 7, e35255, \dodoi{10.7554/eLife.35255}

\bibitem[{Aulbach \& Stagno(2016)}]{Aulbach2016}
Aulbach, S., \& Stagno, V. 2016, Geology, 44, 751, \dodoi{10.1130/G38070.1}

\bibitem[{Bada(2013)}]{Bada2013}
Bada, J.~L. 2013, Chemical Society Reviews, 42, 2186, \dodoi{10.1039/C3CS35433D}

\bibitem[{Bada \& Korenaga(2018)}]{Bada2018}
Bada, J.~L., \& Korenaga, J. 2018, Life, 8, 55, \dodoi{10.3390/life8040055}

\bibitem[{Becker {et~al.}(2006)Becker, Horan, Walker, Gao, Lorand, \& Rudnick}]{Becker2006}
Becker, H., Horan, M., Walker, R., {et~al.} 2006, Geochimica et Cosmochimica Acta, 70, 4528, \dodoi{10.1016/j.gca.2006.06.004}

\bibitem[{Becker {et~al.}(2018)Becker, Schneider, Okamura, Crisp, Amatov, Dejmek, \& Carell}]{Becker2018}
Becker, S., Schneider, C., Okamura, H., {et~al.} 2018, NatCo, 9, 1, \dodoi{10.1038/s41467-017-02639-1}

\bibitem[{Becker {et~al.}(2016)Becker, Thoma, Deutsch, Gehrke, Mayer, Zipse, \& Carell}]{Becker2016}
Becker, S., Thoma, I., Deutsch, A., {et~al.} 2016, Science, 352, 833, \dodoi{10.1126/science.aad2808}

\bibitem[{Benner {et~al.}(2019{\natexlab{a}})Benner, Kim, \& Biondi}]{Benner2019b}
Benner, S.~A., Kim, H.-J., \& Biondi, E. 2019{\natexlab{a}}, Life, 9, 84, \dodoi{10.3390/life9040084}

\bibitem[{Benner {et~al.}(2019{\natexlab{b}})Benner, Bell, Biondi, Brasser, Carell, Kim, Mojzsis, Omran, Pasek, \& Trail}]{Benner2019a}
Benner, S.~A., Bell, E.~A., Biondi, E., {et~al.} 2019{\natexlab{b}}, ChemSystemsChem, 2, e1900035, \dodoi{10.1002/syst.201900035}

\bibitem[{Bermingham \& Walker(2017)}]{Bermingham2017}
Bermingham, K., \& Walker, R. 2017, Earth and Planetary Science Letters, 474, 466, \dodoi{10.1016/j.epsl.2017.06.052}

\bibitem[{Bermingham {et~al.}(2025)Bermingham, Tornabene, Walker, Godfrey, Meyer, Piccoli, \& Mojzsis}]{Bermingham2025}
Bermingham, K.~R., Tornabene, H.~A., Walker, R.~J., {et~al.} 2025, Geochimica et Cosmochimica Acta, 392, 38, \dodoi{10.1016/j.gca.2024.11.005}

\bibitem[{Bradley(2016)}]{Bradley2016}
Bradley, A.~S. 2016, Proceedings of the National Academy of Sciences, 113, 13944, \dodoi{10.1073/pnas.1617103113}

\bibitem[{Braukmüller {et~al.}(2019)Braukmüller, Wombacher, Funk, \& Münker}]{Braukmüller2019}
Braukmüller, N., Wombacher, F., Funk, C., \& Münker, C. 2019, Nature Geoscience, 12, 564, \dodoi{10.1038/s41561-019-0375-x}

\bibitem[{Brenan \& McDonough(2009)}]{Brenan2009}
Brenan, J.~M., \& McDonough, W.~F. 2009, Nature Geoscience, 2, 798, \dodoi{10.1038/ngeo658}

\bibitem[{Breslow(1959)}]{Breslow1959}
Breslow, R. 1959, Tetrahedron Letters, 1, 22, \dodoi{10.1016/S0040-4039(01)99487-0}

\bibitem[{Budde {et~al.}(2019)Budde, Burkhardt, \& Kleine}]{Budde2019}
Budde, G., Burkhardt, C., \& Kleine, T. 2019, Nature Astronomy, 3, 736, \dodoi{10.1038/s41550-019-0779-y}

\bibitem[{Butlerow(1861)}]{Butlerow1861}
Butlerow, A. 1861, Justus Liebigs Annalen der Chemie, 120, 295, \dodoi{10.1002/jlac.18611200308}

\bibitem[{Callahan {et~al.}(2011)Callahan, Smith, Cleaves, Ruzicka, Stern, Glavin, House, \& Dworkin}]{Callahan2011}
Callahan, M.~P., Smith, K.~E., Cleaves, H.~J., {et~al.} 2011, PNAS, 108, 13995, \dodoi{10.1073/pnas.1106493108}

\bibitem[{Cannat {et~al.}(2010)Cannat, Fontaine, \& EscartíN}]{Cannat2010}
Cannat, M., Fontaine, F., \& EscartíN, J. 2010, Serpentinization and Associated Hydrogen And Methane Fluxes at Slow Spreading Ridges (American Geophysical Union (AGU)), 241--264, \dodoi{10.1029/2008GM000760}

\bibitem[{Catling \& Kasting(2017)}]{Catling2017}
Catling, D.~C., \& Kasting, J.~F. 2017, Atmospheric {{Evolution}} on {{Inhabited}} and {{Lifeless Worlds}} ({Cambridge}: {Cambridge University Press}), \dodoi{10.1017/9781139020558}

\bibitem[{Cavosie {et~al.}(2005)Cavosie, Valley, Wilde, \& {E.i.m.f.}}]{Cavosie2005}
Cavosie, A.~J., Valley, J.~W., Wilde, S.~A., \& {E.i.m.f.} 2005, Earth and Planetary Science Letters, 235, 663, \dodoi{10.1016/j.epsl.2005.04.028}

\bibitem[{Cech(1986)}]{Cech1986}
Cech, T.~R. 1986, PNAS, 83, 4360, \dodoi{10.1073/pnas.83.12.4360}

\bibitem[{Charlou {et~al.}(2002)Charlou, Donval, Fouquet, {Jean-Baptiste}, \& Holm}]{Charlou2002}
Charlou, J.~L., Donval, J.~P., Fouquet, Y., {Jean-Baptiste}, P., \& Holm, N. 2002, Chemical Geology, 191, 345, \dodoi{10.1016/S0009-2541(02)00134-1}

\bibitem[{Chou {et~al.}(1983)Chou, Shaw, \& Crocket}]{Chou1983}
Chou, C.-L., Shaw, D.~M., \& Crocket, J.~H. 1983, Journal of Geophysical Research: Solid Earth, 88, A507, \dodoi{10.1029/JB088iS02p0A507}

\bibitem[{Chowdhury {et~al.}(2023)Chowdhury, Trail, Miller, \& Savage}]{Chowdhury2023}
Chowdhury, W., Trail, D., Miller, M., \& Savage, P. 2023, Nature Communications, 14, 1140, \dodoi{10.1038/s41467-023-36538-5}

\bibitem[{Chyba(1990)}]{Chyba1990}
Chyba, C.~F. 1990, Nature, 343, 129, \dodoi{10.1038/343129a0}

\bibitem[{Chyba \& Sagan(1992)}]{Chyba1992}
Chyba, C.~F., \& Sagan, C. 1992, Nature, 355, 125, \dodoi{10.1038/355125a0}

\bibitem[{Citron \& Stewart(2022)}]{Citron2022}
Citron, R.~I., \& Stewart, S.~T. 2022, The Planetary Science Journal, 3, 116, \dodoi{10.3847/PSJ/ac66e8}

\bibitem[{Cleaves {et~al.}(2008)Cleaves, Chalmers, Lazcano, Miller, \& Bada}]{Cleaves2008}
Cleaves, H.~J., Chalmers, J.~H., Lazcano, A., Miller, S.~L., \& Bada, J.~L. 2008, Origins of Life and Evolution of Biospheres, 38, 105, \dodoi{10.1007/s11084-007-9120-3}

\bibitem[{Cleaves(2015)}]{Cleaves2015}
Cleaves, H. J.~J. 2015, in Encyclopedia of {{Astrobiology}}, ed. M.~Gargaud, W.~M. Irvine, R.~Amils, H.~J.~J. Cleaves, D.~L. Pinti, J.~C. Quintanilla, D.~Rouan, T.~Spohn, S.~Tirard, \& M.~Viso (Berlin, Heidelberg: Springer), 877--884, \dodoi{10.1007/978-3-662-44185-5_587}

\bibitem[{Cogn{\'e} \& Humler(2004)}]{Cogne2004}
Cogn{\'e}, J.-P., \& Humler, E. 2004, Earth and Planetary Science Letters, 227, 427, \dodoi{10.1016/j.epsl.2004.09.002}

\bibitem[{Cojocaru \& Unrau(2021)}]{Cojocaru2021}
Cojocaru, R., \& Unrau, P.~J. 2021, Science, 371, 1225, \dodoi{10.1126/science.abd9191}

\bibitem[{Da~Silva {et~al.}(2015)Da~Silva, Maurel, \& Deamer}]{DaSilva2015}
Da~Silva, L., Maurel, M.~C., \& Deamer, D. 2015, JMolE, 80, 86, \dodoi{10.1007/s00239-014-9661-9}

\bibitem[{Damer \& Deamer(2020)}]{Damer2020}
Damer, B., \& Deamer, D. 2020, AsBio, 20, 429, \dodoi{10.1089/ast.2019.2045}

\bibitem[{Dauphas(2017)}]{Dauphas2017}
Dauphas, N. 2017, Nature, 541, 521, \dodoi{10.1038/nature20830}

\bibitem[{Deamer(2017)}]{Deamer2017}
Deamer, D. 2017, Life, 7, 5, \dodoi{10.3390/life7010005}

\bibitem[{Debaille {et~al.}(2013)Debaille, O'Neill, Brandon, Haenecour, Yin, Mattielli, \& Treiman}]{Debaille2013}
Debaille, V., O'Neill, C., Brandon, A.~D., {et~al.} 2013, Earth and Planetary Science Letters, 373, 83, \dodoi{10.1016/j.epsl.2013.04.016}

\bibitem[{Di~Giulio(1997)}]{DiGiulio1997}
Di~Giulio, M. 1997, Journal of Molecular Evolution, 45, 571, \dodoi{10.1007/PL00006261}

\bibitem[{Dirscherl {et~al.}(2023)Dirscherl, Ianeselli, Tetiker, Matreux, Queener, Mast, \& Braun}]{Dirscherl2023}
Dirscherl, C.~F., Ianeselli, A., Tetiker, D., {et~al.} 2023, Physical Chemistry Chemical Physics, 25, 3375, \dodoi{10.1039/D2CP04538A}

\bibitem[{Dorn \& Lichtenberg(2021)}]{Dorn2021}
Dorn, C., \& Lichtenberg, T. 2021, The Astrophysical Journal Letters, 922, L4, \dodoi{10.3847/2041-8213/ac33af}

\bibitem[{Ferus {et~al.}(2019)Ferus, Pietrucci, Saitta, Ivanek, Knizek, Kubel{\'i}k, Krus, Juha, Dudzak, Dost{\'a}l, Pastorek, Petera, Hrncirova, Saeidfirozeh, Shestivsk{\'a}, Sponer, Sponer, Rimmer, Civi{\v s}, \& Cassone}]{Ferus2019}
Ferus, M., Pietrucci, F., Saitta, A.~M., {et~al.} 2019, Astronomy \& Astrophysics, 626, A52, \dodoi{10.1051/0004-6361/201935435}

\bibitem[{Fiebig {et~al.}(2007)Fiebig, Woodland, Spangenberg, \& Oschmann}]{Fiebig2007}
Fiebig, J., Woodland, A.~B., Spangenberg, J., \& Oschmann, W. 2007, Geochimica et Cosmochimica Acta, 71, 3028, \dodoi{10.1016/j.gca.2007.04.010}

\bibitem[{Fischer-G{\"{o}}dde \& Kleine(2017)}]{Fischer-Godde2017}
Fischer-G{\"{o}}dde, M., \& Kleine, T. 2017, \nat, 541, 525, \dodoi{10.1038/nature21045}

\bibitem[{Fischer-G{\"{o}}dde {et~al.}(2020)Fischer-G{\"{o}}dde, Elfers, M{\"{u}}nker, Szilas, Maier, Messling, Morishita, {Van Kranendonk}, \& Smithies}]{Fischer-Godde2020}
Fischer-G{\"{o}}dde, M., Elfers, B.~M., M{\"{u}}nker, C., {et~al.} 2020, \nat, 579, 240, \dodoi{10.1038/s41586-020-2069-3}

\bibitem[{Fuller {et~al.}(1972)Fuller, Sanchez, \& Orgel}]{Fuller1972}
Fuller, W.~D., Sanchez, R.~A., \& Orgel, L.~E. 1972, Journal of Molecular Biology, 67, 25, \dodoi{10.1016/0022-2836(72)90383-X}

\bibitem[{Furukawa {et~al.}(2019)Furukawa, Chikaraishi, Ohkouchi, Ogawa, Glavin, Dworkin, Abe, \& Nakamura}]{Furukawa2019}
Furukawa, Y., Chikaraishi, Y., Ohkouchi, N., {et~al.} 2019, PNAS, 116, 24440, \dodoi{10.1073/pnas.1907169116}

\bibitem[{Genda {et~al.}(2017{\natexlab{a}})Genda, Brasser, \& Mojzsis}]{Genda2017b}
Genda, H., Brasser, R., \& Mojzsis, S.~J. 2017{\natexlab{a}}, Earth and Planetary Science Letters, 480, 25, \dodoi{10.1016/j.epsl.2017.09.041}

\bibitem[{Genda {et~al.}(2017{\natexlab{b}})Genda, Iizuka, Sasaki, Ueno, \& Ikoma}]{Genda2017a}
Genda, H., Iizuka, T., Sasaki, T., Ueno, Y., \& Ikoma, M. 2017{\natexlab{b}}, Earth and Planetary Science Letters, 470, 87, \dodoi{10.1016/j.epsl.2017.04.035}

\bibitem[{Gilbert(1986)}]{Gilbert1986}
Gilbert, W. 1986, \nat, 319, 618, \dodoi{10.1038/319618a0}

\bibitem[{Gilmour(2003)}]{Gilmour2003}
Gilmour, I. 2003, in Treatise on Geochemistry, Vol.~1 (Elsevier Inc.), 269--290, \dodoi{10.1016/b0-08-043751-6/01146-4}

\bibitem[{Guerrier-Takada \& Altman(1984)}]{Guerrier-Takada1984}
Guerrier-Takada, C., \& Altman, S. 1984, Science, 223, 285, \dodoi{10.1126/science.6199841}

\bibitem[{Guerrier-Takada {et~al.}(1983)Guerrier-Takada, Gardiner, Marsh, Pace, \& Altman}]{Guerrier-Takada1983}
Guerrier-Takada, C., Gardiner, K., Marsh, T., Pace, N., \& Altman, S. 1983, Cell, 35, 849, \dodoi{10.1016/0092-8674(83)90117-4}

\bibitem[{Guillot {et~al.}(2015)Guillot, Schwartz, Reynard, Agard, \& Prigent}]{Guillot2015}
Guillot, S., Schwartz, S., Reynard, B., Agard, P., \& Prigent, C. 2015, Tectonophysics, 646, 1, \dodoi{10.1016/j.tecto.2015.01.020}

\bibitem[{Guillot(2010)}]{Guillot2010}
Guillot, T. 2010, Astronomy \& Astrophysics, 520, A27, \dodoi{10.1051/0004-6361/200913396}

\bibitem[{Guo \& Korenaga(2020)}]{Guo2020}
Guo, M., \& Korenaga, J. 2020, Science Advances, 6, eaaz6234, \dodoi{10.1126/sciadv.aaz6234}

\bibitem[{Guo \& Korenaga(2023)}]{Guo2023}
---. 2023, Science Advances, 9, eade2711, \dodoi{10.1126/sciadv.ade2711}

\bibitem[{Guo \& Korenaga(2024)}]{Guo2024}
---. 2024, Rapidly Evolving Ocean {{pH}} in the Early {{Earth}}: {{Insights}} from Global Carbon Cycle Coupled with Ocean Chemistry,  Research Square, \dodoi{10.21203/rs.3.rs-4247090/v1}

\bibitem[{Guo \& Korenaga(2025)}]{Guo2025}
---. 2025, Nature Geoscience, 1, \dodoi{10.1038/s41561-025-01649-9}

\bibitem[{Guzm\'{a}n-Marmolejo {et~al.}(2013)Guzm\'{a}n-Marmolejo, Segura, \& Escobar-Briones}]{Guzman-Marmolejo2013}
Guzm\'{a}n-Marmolejo, A., Segura, A., \& Escobar-Briones, E. 2013, Astrobiology, 13, 550, \dodoi{10.1089/ast.2012.0817}

\bibitem[{Haldane(1929)}]{Haldane1929}
Haldane, J. B.~S. 1929, Rationalist Annual, 12

\bibitem[{Harrison(2009)}]{Harrison2009}
Harrison, T.~M. 2009, Annual Review of Earth and Planetary Sciences, 37, 479, \dodoi{10.1146/annurev.earth.031208.100151}

\bibitem[{Harrison {et~al.}(2007)Harrison, Watson, \& Aikman}]{Harrison2007}
Harrison, T.~M., Watson, E.~B., \& Aikman, A.~B. 2007, Geology, 35, 635, \dodoi{10.1130/G23505A.1}

\bibitem[{Hashimoto {et~al.}(2007)Hashimoto, Abe, \& Sugita}]{Hashimoto2007}
Hashimoto, G.~L., Abe, Y., \& Sugita, S. 2007, Journal of Geophysical Research: Planets, 112, \dodoi{10.1029/2006JE002844}

\bibitem[{Hastie {et~al.}(2023)Hastie, Law, Bromiley, Fitton, Harley, \& Muir}]{Hastie2023}
Hastie, A.~R., Law, S., Bromiley, G.~D., {et~al.} 2023, Nature Geoscience, 16, 816, \dodoi{10.1038/s41561-023-01249-5}

\bibitem[{Hazen \& Sverjensky(2010)}]{Hazen2010}
Hazen, R.~M., \& Sverjensky, D.~A. 2010, Cold Spring Harbor Perspectives in Biology, 2, a002162, \dodoi{10.1101/cshperspect.a002162}

\bibitem[{{Hier-Majumder} \& Hirschmann(2017)}]{Hier-Majumder2017}
{Hier-Majumder}, S., \& Hirschmann, M.~M. 2017, Geochemistry, Geophysics, Geosystems, 18, 3078, \dodoi{10.1002/2017GC006937}

\bibitem[{Hill \& Orgel(2002)}]{Hill2002}
Hill, A., \& Orgel, L.~E. 2002, Origins of life and evolution of the biosphere, 32, 99, \dodoi{10.1023/A:1016070723772}

\bibitem[{Hirschmann \& Dasgupta(2009)}]{Hirschmann2009}
Hirschmann, M.~M., \& Dasgupta, R. 2009, Chemical Geology, 262, 4, \dodoi{10.1016/j.chemgeo.2009.02.008}

\bibitem[{Holland(1984)}]{Holland1984}
Holland, H.~D. 1984, The Chemical Evolution of the Atmosphere and Oceans. (Princeton University Press)

\bibitem[{Holm {et~al.}(2015)Holm, Oze, Mousis, Waite, \& {Guilbert-Lepoutre}}]{Holm2015}
Holm, N., Oze, C., Mousis, O., Waite, J., \& {Guilbert-Lepoutre}, A. 2015, Astrobiology, 15, 587, \dodoi{10.1089/ast.2014.1188}

\bibitem[{Hopp {et~al.}(2020)Hopp, Budde, \& Kleine}]{Hopp2020}
Hopp, T., Budde, G., \& Kleine, T. 2020, Earth and Planetary Science Letters, 534, 116065, \dodoi{10.1016/j.epsl.2020.116065}

\bibitem[{Hu {et~al.}(2012)Hu, Seager, \& Bains}]{Hu2012}
Hu, R., Seager, S., \& Bains, W. 2012, The Astrophysical Journal, 761, 166, \dodoi{10.1088/0004-637X/761/2/166}

\bibitem[{Itcovitz {et~al.}(2022)Itcovitz, Rae, Citron, Stewart, Sinclair, Rimmer, \& Shorttle}]{Itcovitz2022}
Itcovitz, J.~P., Rae, A. S.~P., Citron, R.~I., {et~al.} 2022, The Planetary Science Journal, 3, 115, \dodoi{10.3847/PSJ/ac67a9}

\bibitem[{Jackson {et~al.}(2020)Jackson, Saunois, Bousquet, Canadell, Poulter, Stavert, Bergamaschi, Niwa, Segers, \& Tsuruta}]{Jackson2020}
Jackson, R.~B., Saunois, M., Bousquet, P., {et~al.} 2020, Environmental Research Letters, 15, 071002, \dodoi{10.1088/1748-9326/ab9ed2}

\bibitem[{Johansen {et~al.}(2024)Johansen, Camprubi, {van Kooten}, \& Hoeijmakers}]{Johansen2024}
Johansen, A., Camprubi, E., {van Kooten}, E., \& Hoeijmakers, H.~J. 2024, Astrobiology, 24, 856, \dodoi{10.1089/ast.2023.0104}

\bibitem[{Johansen {et~al.}(2023)Johansen, Ronnet, Schiller, Deng, \& Bizzarro}]{Johansen2023}
Johansen, A., Ronnet, T., Schiller, M., Deng, Z., \& Bizzarro, M. 2023, Astronomy \& Astrophysics, 671, A76, \dodoi{10.1051/0004-6361/202142143}

\bibitem[{Johnson {et~al.}(2008)Johnson, Cleaves, Dworkin, Glavin, Lazcano, \& Bada}]{Johnson2008}
Johnson, A.~P., Cleaves, H.~J., Dworkin, J.~P., {et~al.} 2008, Science, 322, 404, \dodoi{10.1126/science.1161527}

\bibitem[{Johnston {et~al.}(2001)Johnston, Unrau, Lawrence, Glasner, \& Bartel}]{Johnston2001}
Johnston, W.~K., Unrau, P.~J., Lawrence, M.~S., Glasner, M.~E., \& Bartel, D.~P. 2001, Science, 292, 1319, \dodoi{10.1126/science.1060786}

\bibitem[{Kadoya \& Catling(2019)}]{Kadoya2019}
Kadoya, S., \& Catling, D.~C. 2019, Geochimica et Cosmochimica Acta, 262, 207, \dodoi{10.1016/j.gca.2019.07.041}

\bibitem[{Kadoya {et~al.}(2020)Kadoya, {Krissansen-Totton}, \& Catling}]{Kadoya2020}
Kadoya, S., {Krissansen-Totton}, J., \& Catling, D.~C. 2020, Geochemistry, Geophysics, Geosystems, 21, e2019GC008734, \dodoi{10.1029/2019GC008734}

\bibitem[{Kasting(1990)}]{Kasting1990}
Kasting, J.~F. 1990, Origins of life and evolution of the biosphere, 20, 199, \dodoi{10.1007/BF01808105}

\bibitem[{Kasting(2005)}]{Kasting2005}
---. 2005, Precambrian Research, 137, 119, \dodoi{10.1016/j.precamres.2005.03.002}

\bibitem[{Keller \& Schoene(2018)}]{Keller2018}
Keller, B., \& Schoene, B. 2018, Earth and Planetary Science Letters, 481, 290, \dodoi{10.1016/j.epsl.2017.10.031}

\bibitem[{Kelley {et~al.}(2005)Kelley, Karson, {Fr{\"u}h-Green}, Yoerger, Shank, Butterfield, Hayes, Schrenk, Olson, Proskurowski, Jakuba, Bradley, Larson, Ludwig, Glickson, Buckman, Bradley, Brazelton, Roe, Elend, Delacour, Bernasconi, Lilley, Baross, Summons, \& Sylva}]{Kelley2005}
Kelley, D.~S., Karson, J.~A., {Fr{\"u}h-Green}, G.~L., {et~al.} 2005, Science, 307, 1428, \dodoi{10.1126/science.1102556}

\bibitem[{Klein {et~al.}(2013)Klein, Bach, \& McCollom}]{Klein2013}
Klein, F., Bach, W., \& McCollom, T.~M. 2013, Lithos, 178, 55, \dodoi{10.1016/j.lithos.2013.03.008}

\bibitem[{Korenaga(2007)}]{Korenaga2007}
Korenaga, J. 2007, Journal of Geophysical Research: Solid Earth, 112, \dodoi{10.1029/2006JB004502}

\bibitem[{Korenaga(2021)}]{Korenaga2021}
---. 2021, Life, 11, 1142, \dodoi{10.3390/life11111142}

\bibitem[{Korenaga {et~al.}(2017)Korenaga, Planavsky, \& Evans}]{Korenaga2017}
Korenaga, J., Planavsky, N.~J., \& Evans, D. A.~D. 2017, Philosophical Transactions of the Royal Society A: Mathematical, Physical and Engineering Sciences, 375, 20150393, \dodoi{10.1098/rsta.2015.0393}

\bibitem[{Kristoffersen {et~al.}(2022)Kristoffersen, Burman, Noy, \& Holliger}]{Kristoffersen2022}
Kristoffersen, E.~L., Burman, M., Noy, A., \& Holliger, P. 2022, eLife, 11, e75186, \dodoi{10.7554/eLife.75186}

\bibitem[{Kruger {et~al.}(1982)Kruger, Grabowski, Zaug, Sands, Gottschling, \& Cech}]{Kruger1982}
Kruger, K., Grabowski, P.~J., Zaug, A.~J., {et~al.} 1982, Cell, 31, 147, \dodoi{10.1016/0092-8674(82)90414-7}

\bibitem[{Kurosawa {et~al.}(2013)Kurosawa, Sugita, Ishibashi, Hasegawa, Sekine, Ogawa, Kadono, Ohno, Ohkouchi, Nagaoka, \& Matsui}]{Kurosawa2013}
Kurosawa, K., Sugita, S., Ishibashi, K., {et~al.} 2013, Origins of Life and Evolution of Biospheres, 43, 221, \dodoi{10.1007/s11084-013-9339-0}

\bibitem[{Kuwahara \& Sugita(2015)}]{Kuwahara2015}
Kuwahara, H., \& Sugita, S. 2015, Icarus, 257, 290, \dodoi{10.1016/j.icarus.2015.05.007}

\bibitem[{Laneuville {et~al.}(2018)Laneuville, Kameya, \& Cleaves}]{Laneuville2018}
Laneuville, M., Kameya, M., \& Cleaves, H.~J. 2018, Astrobiology, 18, 897, \dodoi{10.1089/ast.2017.1700}

\bibitem[{LaRowe \& Regnier(2008)}]{LaRowe2008}
LaRowe, D.~E., \& Regnier, P. 2008, Origins of Life and Evolution of Biospheres, 38, 383, \dodoi{10.1007/s11084-008-9137-2}

\bibitem[{Leong {et~al.}(2021)Leong, Ely, \& Shock}]{Leong2021}
Leong, J. A.~M., Ely, T., \& Shock, E.~L. 2021, Nature Communications, 12, 7341, \dodoi{10.1038/s41467-021-27589-7}

\bibitem[{Li(2022)}]{Li2022}
Li, C.-H. 2022, Acta Geochimica, 41, 650, \dodoi{10.1007/s11631-021-00517-8}

\bibitem[{Lissenberg {et~al.}(2024)Lissenberg, McCaig, Lang, Blum, Abe, Brazelton, Coltat, Deans, Dickerson, Godard, John, Klein, Kuehn, Lin, Liu, Lopes, Nozaka, Parsons, Pathak, Reagan, Robare, Savov, Schwarzenbach, Sissmann, Southam, Wang, Wheat, Anderson, \& Treadwell}]{Lissenberg2024}
Lissenberg, C.~J., McCaig, A.~M., Lang, S.~Q., {et~al.} 2024, Science, 385, 623, \dodoi{10.1126/science.adp1058}

\bibitem[{Liu {et~al.}(2023)Liu, Perez-Gussinye, García-Pintado, Mezri, \& Bach}]{Liu2023}
Liu, Z., Perez-Gussinye, M., García-Pintado, J., Mezri, L., \& Bach, W. 2023, Geology, 51, 284, \dodoi{10.1130/G50722.1}

\bibitem[{Luo {et~al.}(2024)Luo, Dorn, \& Deng}]{Luo2024}
Luo, H., Dorn, C., \& Deng, J. 2024, Nature Astronomy, 1, \dodoi{10.1038/s41550-024-02347-z}

\bibitem[{Mann {et~al.}(2012)Mann, Frost, Rubie, Becker, \& Audétat}]{Mann2012}
Mann, U., Frost, D.~J., Rubie, D.~C., Becker, H., \& Audétat, A. 2012, Geochimica et Cosmochimica Acta, 84, 593, \dodoi{10.1016/j.gca.2012.01.026}

\bibitem[{Martin \& Russell(2006)}]{Martin2006}
Martin, W., \& Russell, M.~J. 2006, Philosophical Transactions of the Royal Society B: Biological Sciences, 362, 1887, \dodoi{10.1098/rstb.2006.1881}

\bibitem[{Matreux {et~al.}(2024)Matreux, Aikkila, Scheu, Braun, \& Mast}]{Matreux2024}
Matreux, T., Aikkila, P., Scheu, B., Braun, D., \& Mast, C.~B. 2024, Nature, 628, 110, \dodoi{10.1038/s41586-024-07193-7}

\bibitem[{McCollom \& Seewald(2007)}]{McCollom2007}
McCollom, T.~M., \& Seewald, J.~S. 2007, Chemical Reviews, 107, 382, \dodoi{10.1021/cr0503660}

\bibitem[{{McCoy-West} {et~al.}(2019){McCoy-West}, Chowdhury, Burton, Sossi, Nowell, Fitton, Kerr, Cawood, \& Williams}]{McCoy-West2019}
{McCoy-West}, A.~J., Chowdhury, P., Burton, K.~W., {et~al.} 2019, Nature Geoscience, 12, 946, \dodoi{10.1038/s41561-019-0451-2}

\bibitem[{McCulloch \& Bennett(1993)}]{McCulloch1993}
McCulloch, M.~T., \& Bennett, V.~C. 1993, Lithos, 30, 237, \dodoi{10.1016/0024-4937(93)90038-E}

\bibitem[{Menneken {et~al.}(2007)Menneken, Nemchin, Geisler, Pidgeon, \& Wilde}]{Menneken2007}
Menneken, M., Nemchin, A.~A., Geisler, T., Pidgeon, R.~T., \& Wilde, S.~A. 2007, Nature, 448, 917, \dodoi{10.1038/nature06083}

\bibitem[{Miller(1953)}]{Miller1953}
Miller, S.~L. 1953, Science, 117, 528, \dodoi{10.1126/science.117.3046.528}

\bibitem[{Miller(1955)}]{Miller1955}
---. 1955, Journal of the American Chemical Society, 77, 2351, \dodoi{10.1021/ja01614a001}

\bibitem[{Miller(1957{\natexlab{a}})}]{Miller1957a}
---. 1957{\natexlab{a}}, Annals of the New York Academy of Sciences, 69, 260, \dodoi{10.1111/j.1749-6632.1957.tb49662.x}

\bibitem[{Miller(1957{\natexlab{b}})}]{Miller1957b}
---. 1957{\natexlab{b}}, Biochimica et Biophysica Acta, 23, 480, \dodoi{10.1016/0006-3002(57)90366-9}

\bibitem[{Miller \& Urey(1959)}]{Miller1959}
Miller, S.~L., \& Urey, H.~C. 1959, Science, 130, 245, \dodoi{10.1126/science.130.3370.245}

\bibitem[{Miyakawa {et~al.}(2002)Miyakawa, Yamanashi, Kobayashi, Cleaves, \& Miller}]{Miyakawa2002}
Miyakawa, S., Yamanashi, H., Kobayashi, K., Cleaves, H.~J., \& Miller, S.~L. 2002, Proceedings of the National Academy of Sciences, 99, 14628, \dodoi{10.1073/pnas.192568299}

\bibitem[{Miyazaki \& Korenaga(2019)}]{Miyazaki2019}
Miyazaki, Y., \& Korenaga, J. 2019, Journal of Geophysical Research: Solid Earth, 124, 3399, \dodoi{10.1029/2018JB016928}

\bibitem[{Miyazaki \& Korenaga(2022)}]{Miyazaki2022}
---. 2022, \nat, 603, 86, \dodoi{10.1038/s41586-021-04371-9}

\bibitem[{Mojzsis {et~al.}(2001)Mojzsis, Harrison, \& Pidgeon}]{Mojzsis2001}
Mojzsis, S.~J., Harrison, T.~M., \& Pidgeon, R.~T. 2001, Nature, 409, 178, \dodoi{10.1038/35051557}

\bibitem[{Molaverdikhani {et~al.}(2020)Molaverdikhani, Helling, Lew, MacDonald, Samra, Iro, Woitke, \& Parmentier}]{Molaverdikhani2020}
Molaverdikhani, K., Helling, C., Lew, B. W.~P., {et~al.} 2020, Astronomy \& Astrophysics, 635, A31, \dodoi{10.1051/0004-6361/201937044}

\bibitem[{Molaverdikhani {et~al.}(2019)Molaverdikhani, Henning, \& Molli{\`e}re}]{Molaverdikhani2019}
Molaverdikhani, K., Henning, T., \& Molli{\`e}re, P. 2019, The Astrophysical Journal, 883, 194, \dodoi{10.3847/1538-4357/ab3e30}

\bibitem[{Molli{\`e}re {et~al.}(2019)Molli{\`e}re, Wardenier, van Boekel, Henning, Molaverdikhani, \& Snellen}]{Molliere2019}
Molli{\`e}re, P., Wardenier, J.~P., van Boekel, R., {et~al.} 2019, Astronomy \& Astrophysics, 627, A67, \dodoi{10.1051/0004-6361/201935470}

\bibitem[{Morbidelli \& Wood(2015)}]{Morbidelli2015}
Morbidelli, A., \& Wood, B.~J. 2015, in The {{Early Earth}} ({American Geophysical Union (AGU)}), 71--82, \dodoi{10.1002/9781118860359.ch4}

\bibitem[{M{\"u}ller {et~al.}(2022)M{\"u}ller, Escobar, Xu, W{\k e}grzyn, Nainyt{\.e}, Amatov, Chan, Pichler, \& Carell}]{Muller2022}
M{\"u}ller, F., Escobar, L., Xu, F., {et~al.} 2022, Nature, 605, 279, \dodoi{10.1038/s41586-022-04676-3}

\bibitem[{Nam {et~al.}(2018)Nam, Nam, \& Zare}]{Nam2018}
Nam, I., Nam, H.~G., \& Zare, R.~N. 2018, Proceedings of the National Academy of Sciences, 115, 36, \dodoi{10.1073/pnas.1718559115}

\bibitem[{Nisbet \& Sleep(2001)}]{Nisbet2001}
Nisbet, E.~G., \& Sleep, N.~H. 2001, Nature, 409, 1083, \dodoi{10.1038/35059210}

\bibitem[{Oba {et~al.}(2022)Oba, Takano, Furukawa, Koga, Glavin, Dworkin, \& Naraoka}]{Oba2022}
Oba, Y., Takano, Y., Furukawa, Y., {et~al.} 2022, NatCo, 13, 2008, \dodoi{10.1038/s41467-022-29612-x}

\bibitem[{Oba {et~al.}(2023)Oba, Koga, Takano, Ogawa, Ohkouchi, Sasaki, Sato, Glavin, Dworkin, Naraoka, Tachibana, Yurimoto, Nakamura, Noguchi, Okazaki, Yabuta, Sakamoto, Yada, Nishimura, Nakato, Miyazaki, Yogata, Abe, Okada, Usui, Yoshikawa, Saiki, Tanaka, Terui, Nakazawa, Watanabe, \& Tsuda}]{Oba2023}
Oba, Y., Koga, T., Takano, Y., {et~al.} 2023, Nature Communications, 14, 1292, \dodoi{10.1038/s41467-023-36904-3}

\bibitem[{Ogino {et~al.}(2016)Ogino, Yamanaka, Mori, \& Matsumoto}]{Ogino2016}
Ogino, S.-Y., Yamanaka, M.~D., Mori, S., \& Matsumoto, J. 2016, Journal of Climate, 29, 1231, \dodoi{10.1175/JCLI-D-15-0484.1}

\bibitem[{Oparin(1924)}]{Oparin1938}
Oparin, A.~I. 1924, The Origin of Life [Russian: Proiskhozhdenie zhizni] (Moscow: Izd. Moskovskii Rabochii)

\bibitem[{Or{\'o}(1961)}]{Oro1961}
Or{\'o}, J. 1961, Nature, 191, 1193, \dodoi{10.1038/1911193a0}

\bibitem[{Or{\'o} \& Kamat(1961)}]{Oro1961a}
Or{\'o}, J., \& Kamat, S.~S. 1961, Nature, 190, 442, \dodoi{10.1038/190442a0}

\bibitem[{Or{\'o} {et~al.}(1990)Or{\'o}, Miller, \& Lazcano}]{Oro1990}
Or{\'o}, J., Miller, S.~L., \& Lazcano, A. 1990, Annual Review of Earth and Planetary Sciences, 18, 317, \dodoi{10.1146/annurev.ea.18.050190.001533}

\bibitem[{Owen \& Wu(2017)}]{Owen2017}
Owen, J.~E., \& Wu, Y. 2017, The Astrophysical Journal, 847, 29, \dodoi{10.3847/1538-4357/aa890a}

\bibitem[{Parsons(1982)}]{Parsons1982}
Parsons, B. 1982, Journal of Geophysical Research: Solid Earth, 87, 289, \dodoi{10.1029/JB087iB01p00289}

\bibitem[{Paschek {et~al.}(2022)Paschek, Kohler, Pearce, Lange, Henning, Trapp, Pudritz, \& Semenov}]{Paschek2022}
Paschek, K., Kohler, K., Pearce, B. K.~D., {et~al.} 2022, Life, 12, 404, \dodoi{10.3390/life12030404}

\bibitem[{Paschek {et~al.}(2024)Paschek, Lee, Semenov, \& Henning}]{Paschek2024}
Paschek, K., Lee, M., Semenov, D.~A., \& Henning, T.~K. 2024, ChemPlusChem, 89, e202300508, \dodoi{10.1002/cplu.202300508}

\bibitem[{Paschek {et~al.}(2023)Paschek, Semenov, Pearce, Lange, Henning, \& Pudritz}]{Paschek2023}
Paschek, K., Semenov, D.~A., Pearce, B. K.~D., {et~al.} 2023, The Astrophysical Journal, 942, 50, \dodoi{10.3847/1538-4357/aca27e}

\bibitem[{Pearce {et~al.}(2019)Pearce, Ayers, \& Pudritz}]{Pearce2019}
Pearce, B. K.~D., Ayers, P.~W., \& Pudritz, R.~E. 2019, The Journal of Physical Chemistry A, 123, 1861, \dodoi{10.1021/acs.jpca.8b11323}

\bibitem[{Pearce {et~al.}(2020{\natexlab{a}})Pearce, Ayers, \& Pudritz}]{Pearce2020}
---. 2020{\natexlab{a}}, The Journal of Physical Chemistry A, 124, 8594, \dodoi{10.1021/acs.jpca.0c06804}

\bibitem[{Pearce {et~al.}(2024)Pearce, H{\"o}rst, Sebree, \& He}]{Pearce2024}
Pearce, B. K.~D., H{\"o}rst, S.~M., Sebree, J.~A., \& He, C. 2024, The Planetary Science Journal, 5, 23, \dodoi{10.3847/PSJ/ad17bd}

\bibitem[{Pearce {et~al.}(2022)Pearce, Molaverdikhani, Pudritz, Henning, \& Cerrillo}]{Pearce2022}
Pearce, B. K.~D., Molaverdikhani, K., Pudritz, R.~E., Henning, T., \& Cerrillo, K.~E. 2022, The Astrophysical Journal, 932, 9, \dodoi{10.3847/1538-4357/ac47a1}

\bibitem[{Pearce {et~al.}(2020{\natexlab{b}})Pearce, Molaverdikhani, Pudritz, Henning, \& H{\'e}brard}]{Pearce2020a}
Pearce, B. K.~D., Molaverdikhani, K., Pudritz, R.~E., Henning, T., \& H{\'e}brard, E. 2020{\natexlab{b}}, The Astrophysical Journal, 901, 110, \dodoi{10.3847/1538-4357/abae5c}

\bibitem[{Pearce {et~al.}(2017)Pearce, Pudritz, Semenov, \& Henning}]{Pearce2017}
Pearce, B. K.~D., Pudritz, R.~E., Semenov, D.~A., \& Henning, T.~K. 2017, Proceedings of the National Academy of Sciences, 114, 11327, \dodoi{10.1073/pnas.1710339114}

\bibitem[{Peters {et~al.}(2023)Peters, Semenov, Hochleitner, \& Trapp}]{Peters2023}
Peters, S., Semenov, D.~A., Hochleitner, R., \& Trapp, O. 2023, Scientific Reports, 13, 6843, \dodoi{10.1038/s41598-023-33741-8}

\bibitem[{Pinto {et~al.}(1980)Pinto, Gladstone, \& Yung}]{Pinto1980}
Pinto, J.~P., Gladstone, G.~R., \& Yung, Y.~L. 1980, Science, 210, 183, \dodoi{10.1126/science.210.4466.183}

\bibitem[{Pizzarello {et~al.}(2006)Pizzarello, Cooper, \& Flynn}]{Pizzarello2006}
Pizzarello, S., Cooper, G.~W., \& Flynn, G.~J. 2006, in Meteorites and the Early Solar System II (Tucson: University of Arizona Press), 625--651

\bibitem[{Poch {et~al.}(2015)Poch, Jaber, Stalport, Nowak, Georgelin, Lambert, Szopa, \& Coll}]{Poch2015}
Poch, O., Jaber, M., Stalport, F., {et~al.} 2015, Astrobiology, 15, 221, \dodoi{10.1089/ast.2014.1230}

\bibitem[{Ponnamperuma {et~al.}(1963)Ponnamperuma, Sagan, \& Mariner}]{Ponnamperuma1963}
Ponnamperuma, C., Sagan, C., \& Mariner, R. 1963, Nature, 199, 222, \dodoi{10.1038/199222a0}

\bibitem[{Powner {et~al.}(2009)Powner, Gerland, \& Sutherland}]{Powner2009}
Powner, M.~W., Gerland, B., \& Sutherland, J.~D. 2009, Nature, 459, 239, \dodoi{10.1038/nature08013}

\bibitem[{Preiner {et~al.}(2018)Preiner, Xavier, Sousa, Zimorski, Neubeck, Lang, Greenwell, Kleinermanns, T{\"u}ys{\"u}z, McCollom, Holm, \& Martin}]{Preiner2018}
Preiner, M., Xavier, J.~C., Sousa, F.~L., {et~al.} 2018, Life, 8, 41, \dodoi{10.3390/life8040041}

\bibitem[{Reynard(2013)}]{Reynard2013}
Reynard, B. 2013, Lithos, 178, 171, \dodoi{10.1016/j.lithos.2012.10.012}

\bibitem[{Rich(1962)}]{Rich1962}
Rich, A. 1962, in Horizons in Biochemistry, ed. M.~Kasha \& B.~Pullman (Academic Press: New York, NY, USA), 103--126

\bibitem[{Russell(2021)}]{Russell2021}
Russell, M.~J. 2021, Life, 11, 429, \dodoi{10.3390/life11050429}

\bibitem[{Russell {et~al.}(2010)Russell, Hall, \& Martin}]{Russell2010}
Russell, M.~J., Hall, A.~J., \& Martin, W. 2010, Geobiology, 8, 355, \dodoi{10.1111/j.1472-4669.2010.00249.x}

\bibitem[{Saladino {et~al.}(2017)Saladino, Bizzarri, Botta, {\v S}poner, {\v S}poner, Georgelin, Jaber, Rigaud, Kapralov, Timoshenko, Rozanov, Krasavin, Timperio, \& Mauro}]{Saladino2017}
Saladino, R., Bizzarri, B.~M., Botta, L., {et~al.} 2017, Scientific Reports, 7, 14709, \dodoi{10.1038/s41598-017-15392-8}

\bibitem[{Schaefer \& Fegley(2007)}]{Schaefer2007}
Schaefer, L., \& Fegley, B. 2007, Icarus, 186, 462, \dodoi{10.1016/j.icarus.2006.09.002}

\bibitem[{Schaefer \& Fegley(2010)}]{Schaefer2010}
---. 2010, Icarus, 208, 438, \dodoi{10.1016/j.icarus.2010.01.026}

\bibitem[{Schaefer \& Fegley(2017)}]{Schaefer2017}
---. 2017, The Astrophysical Journal, 843, 120, \dodoi{10.3847/1538-4357/aa784f}

\bibitem[{Schlesinger \& Miller(1983)}]{Schlesinger1983}
Schlesinger, G., \& Miller, S.~L. 1983, Journal of Molecular Evolution, 19, 376, \dodoi{10.1007/BF02101642}

\bibitem[{Sekine {et~al.}(2003)Sekine, Sugita, Kadono, \& Matsui}]{Sekine2003}
Sekine, Y., Sugita, S., Kadono, T., \& Matsui, T. 2003, Journal of Geophysical Research: Planets, 108, 5070, \dodoi{10.1029/2002JE002034}

\bibitem[{Shimoyama {et~al.}(1990)Shimoyama, Hagishita, \& Harada}]{Shimoyama1990}
Shimoyama, A., Hagishita, S., \& Harada, K. 1990, GeocJ, 24, 343, \dodoi{10.2343/geochemj.24.343}

\bibitem[{Sleep {et~al.}(2001)Sleep, Zahnle, \& Neuhoff}]{Sleep2001}
Sleep, N.~H., Zahnle, K., \& Neuhoff, P.~S. 2001, Proceedings of the National Academy of Sciences, 98, 3666, \dodoi{10.1073/pnas.071045698}

\bibitem[{Sleep {et~al.}(2014)Sleep, Zahnle, \& Lupu}]{Sleep2014}
Sleep, N.~H., Zahnle, K.~J., \& Lupu, R.~E. 2014, Philosophical Transactions of the Royal Society A: Mathematical, Physical and Engineering Sciences, 372, 20130172, \dodoi{10.1098/rsta.2013.0172}

\bibitem[{Smith {et~al.}(2014)Smith, Callahan, Gerakines, Dworkin, \& House}]{Smith2014}
Smith, K.~E., Callahan, M.~P., Gerakines, P.~A., Dworkin, J.~P., \& House, C.~H. 2014, Geochimica et Cosmochimica Acta, 136, 1, \dodoi{10.1016/j.gca.2014.04.001}

\bibitem[{Staudigel {et~al.}(1981)Staudigel, Hart, \& Richardson}]{Staudigel1981}
Staudigel, H., Hart, S.~R., \& Richardson, S.~H. 1981, Earth and Planetary Science Letters, 52, 311, \dodoi{10.1016/0012-821X(81)90186-2}

\bibitem[{Stoks \& Schwartz(1979)}]{Stoks1979}
Stoks, P.~G., \& Schwartz, A.~W. 1979, \nat, 282, 709, \dodoi{10.1038/282709a0}

\bibitem[{Stoks \& Schwartz(1981)}]{Stoks1981}
---. 1981, \gca, 45, 563, \dodoi{10.1016/0016-7037(81)90189-7}

\bibitem[{Stribling \& Miller(1987)}]{Stribling1987}
Stribling, R., \& Miller, S.~L. 1987, Origins of life and evolution of the biosphere, 17, 261, \dodoi{10.1007/BF02386466}

\bibitem[{Strom {et~al.}(2005)Strom, Malhotra, Ito, Yoshida, \& Kring}]{Strom2005}
Strom, R.~G., Malhotra, R., Ito, T., Yoshida, F., \& Kring, D.~A. 2005, Science, 309, 1847, \dodoi{10.1126/science.1113544}

\bibitem[{Sutherland(2016)}]{Sutherland2016}
Sutherland, J.~D. 2016, Angewandte Chemie International Edition, 55, 104, \dodoi{10.1002/anie.201506585}

\bibitem[{Tarduno {et~al.}(2023)Tarduno, Cottrell, Bono, Rayner, Davis, Zhou, Nimmo, Hofmann, Jodder, {Iba{\~n}ez-Mejia}, Watkeys, Oda, \& Mitra}]{Tarduno2023}
Tarduno, J.~A., Cottrell, R.~D., Bono, R.~K., {et~al.} 2023, Nature, 618, 531, \dodoi{10.1038/s41586-023-06024-5}

\bibitem[{Teichert {et~al.}(2019)Teichert, Kruse, \& Trapp}]{Teichert2019}
Teichert, J.~S., Kruse, F.~M., \& Trapp, O. 2019, Angewandte Chemie International Edition, 58, 9944, \dodoi{10.1002/anie.201903400}

\bibitem[{Thompson {et~al.}(2022)Thompson, {Krissansen-Totton}, Wogan, Telus, \& Fortney}]{Thompson2022}
Thompson, M.~A., {Krissansen-Totton}, J., Wogan, N., Telus, M., \& Fortney, J.~J. 2022, Proceedings of the National Academy of Sciences, 119, e2117933119, \dodoi{10.1073/pnas.2117933119}

\bibitem[{Tilmann {et~al.}(2004)Tilmann, Flueh, Planert, Reston, \& Weinrebe}]{Tilmann2004}
Tilmann, F., Flueh, E., Planert, L., Reston, T., \& Weinrebe, W. 2004, Journal of Geophysical Research: Solid Earth, 109, \dodoi{10.1029/2003JB002827}

\bibitem[{Toomey {et~al.}(1988)Toomey, Solomon, \& Purdy}]{Toomey1988}
Toomey, D.~R., Solomon, S.~C., \& Purdy, G.~M. 1988, Journal of Geophysical Research: Solid Earth, 93, 9093, \dodoi{10.1029/JB093iB08p09093}

\bibitem[{Tosi {et~al.}(2017)Tosi, Godolt, Stracke, Ruedas, Grenfell, H{\"o}ning, Nikolaou, Plesa, Breuer, \& Spohn}]{Tosi2017}
Tosi, N., Godolt, M., Stracke, B., {et~al.} 2017, Astronomy \& Astrophysics, 605, A71, \dodoi{10.1051/0004-6361/201730728}

\bibitem[{Trail {et~al.}(2011)Trail, Watson, \& Tailby}]{Trail2011}
Trail, D., Watson, E.~B., \& Tailby, N.~D. 2011, Nature, 480, 79, \dodoi{10.1038/nature10655}

\bibitem[{Urey(1951)}]{Urey1951}
Urey, H.~C. 1951, Geochimica et Cosmochimica Acta, 1, 209, \dodoi{10.1016/0016-7037(51)90001-4}

\bibitem[{Urey(1952)}]{Urey1952}
---. 1952, Proceedings of the National Academy of Sciences, 38, 351, \dodoi{10.1073/pnas.38.4.351}

\bibitem[{Vaidya {et~al.}(2012)Vaidya, Manapat, Chen, Xulvi-Brunet, Hayden, \& Lehman}]{Vaidya2012}
Vaidya, N., Manapat, M.~L., Chen, I.~A., {et~al.} 2012, Nature, 491, 72, \dodoi{10.1038/nature11549}

\bibitem[{Valley {et~al.}(2005)Valley, Lackey, Cavosie, Clechenko, Spicuzza, Basei, Bindeman, Ferreira, Sial, King, Peck, Sinha, \& Wei}]{Valley2005}
Valley, J.~W., Lackey, J.~S., Cavosie, A.~J., {et~al.} 2005, Contributions to Mineralogy and Petrology, 150, 561, \dodoi{10.1007/s00410-005-0025-8}

\bibitem[{van~der Velden \& Schwartz(1977)}]{vanderVelden1977}
van~der Velden, W., \& Schwartz, A.~W. 1977, \gca, 41, 961, \dodoi{10.1016/0016-7037(77)90155-7}

\bibitem[{Varas-Reus {et~al.}(2019)Varas-Reus, König, Yierpan, Lorand, \& Schoenberg}]{Varas-Reus2019}
Varas-Reus, M.~I., König, S., Yierpan, A., Lorand, J.-P., \& Schoenberg, R. 2019, Nature Geoscience, 12, 779, \dodoi{10.1038/s41561-019-0414-7}

\bibitem[{Wakamatsu {et~al.}(1966)Wakamatsu, Yamada, Saito, Kumashiro, \& Takenishi}]{Wakamatsu1966}
Wakamatsu, H., Yamada, Y., Saito, T., Kumashiro, I., \& Takenishi, T. 1966, The Journal of Organic Chemistry, 31, 2035, \dodoi{10.1021/jo01344a545}

\bibitem[{Walker(2009)}]{Walker2009}
Walker, R.~J. 2009, Geochemistry, 69, 101, \dodoi{10.1016/j.chemer.2008.10.001}

\bibitem[{Wang \& Becker(2013)}]{Wang2013}
Wang, Z., \& Becker, H. 2013, Nature, 499, 328, \dodoi{10.1038/nature12285}

\bibitem[{Wasson {et~al.}(1988)Wasson, Kallemeyn, Runcorn, Turner, \& Woolfson}]{Wasson1988}
Wasson, J.~T., Kallemeyn, G.~W., Runcorn, S.~K., Turner, G., \& Woolfson, M.~M. 1988, Philosophical Transactions of the Royal Society of London. Series A, Mathematical and Physical Sciences, 325, 535, \dodoi{10.1098/rsta.1988.0066}

\bibitem[{Westall \& Brack(2018)}]{Westall2018}
Westall, F., \& Brack, A. 2018, Space Science Reviews, 214, 50, \dodoi{10.1007/s11214-018-0476-7}

\bibitem[{Wilde {et~al.}(2001)Wilde, Valley, Peck, \& Graham}]{Wilde2001}
Wilde, S.~A., Valley, J.~W., Peck, W.~H., \& Graham, C.~M. 2001, Nature, 409, 175, \dodoi{10.1038/35051550}

\bibitem[{Wogan {et~al.}(2020)Wogan, {Krissansen-Totton}, \& Catling}]{Wogan2020a}
Wogan, N., {Krissansen-Totton}, J., \& Catling, D.~C. 2020, The Planetary Science Journal, 1, 58, \dodoi{10.3847/PSJ/abb99e}

\bibitem[{Wogan {et~al.}(2023)Wogan, Catling, Zahnle, \& Lupu}]{Wogan2023}
Wogan, N.~F., Catling, D.~C., Zahnle, K.~J., \& Lupu, R. 2023, The Planetary Science Journal, 4, 169, \dodoi{10.3847/PSJ/aced83}

\bibitem[{Yadav {et~al.}(2020)Yadav, Kumar, \& Krishnamurthy}]{Yadav2020}
Yadav, M., Kumar, R., \& Krishnamurthy, R. 2020, Chemical Reviews, 120, 4766, \dodoi{10.1021/acs.chemrev.9b00546}

\bibitem[{Yi {et~al.}(2020)Yi, Tran, Ali, Yoda, Adam, Cleaves, \& Fahrenbach}]{Yi2020}
Yi, R., Tran, Q.~P., Ali, S., {et~al.} 2020, Proceedings of the National Academy of Sciences, 117, 13267, \dodoi{10.1073/pnas.1922139117}

\bibitem[{Young {et~al.}(2023)Young, Shahar, \& Schlichting}]{Young2023}
Young, E.~D., Shahar, A., \& Schlichting, H.~E. 2023, Nature, 616, 306, \dodoi{10.1038/s41586-023-05823-0}

\bibitem[{Zahnle {et~al.}(2007)Zahnle, Arndt, Cockell, Halliday, Nisbet, Selsis, \& Sleep}]{Zahnle2007}
Zahnle, K., Arndt, N., Cockell, C., {et~al.} 2007, Space Science Reviews, 129, 35, \dodoi{10.1007/s11214-007-9225-z}

\bibitem[{Zahnle \& Sleep(2006)}]{Zahnle2006}
Zahnle, K., \& Sleep, N.~H. 2006, in Comets and the {{Origin}} and {{Evolution}} of {{Life}}, ed. P.~J. Thomas, R.~D. Hicks, C.~F. Chyba, \& C.~P. McKay (Berlin, Heidelberg: Springer), 207--251, \dodoi{10.1007/3-540-33088-7_7}

\bibitem[{Zahnle {et~al.}(2019)Zahnle, Gacesa, \& Catling}]{Zahnle2019}
Zahnle, K.~J., Gacesa, M., \& Catling, D.~C. 2019, Geochimica et Cosmochimica Acta, 244, 56, \dodoi{10.1016/j.gca.2018.09.017}

\bibitem[{Zahnle {et~al.}(2020)Zahnle, Lupu, Catling, \& Wogan}]{Zahnle2020}
Zahnle, K.~J., Lupu, R., Catling, D.~C., \& Wogan, N. 2020, The Planetary Science Journal, 1, 11, \dodoi{10.3847/PSJ/ab7e2c}

\bibitem[{Zaug \& Cech(1986)}]{Zaug1986}
Zaug, A.~J., \& Cech, T.~R. 1986, Science, 231, 470, \dodoi{10.1126/science.3941911}

\end{thebibliography}
\bibliographystyle{aasjournal}

\end{document}